\documentclass[journal]{IEEEtran}
\usepackage{cite}
\usepackage{epsfig}
\usepackage{amssymb}
\usepackage{multirow}
\usepackage[cmex10]{amsmath, nccmath}
\usepackage{upgreek}
\usepackage{graphicx}
\usepackage{mathtools}
\usepackage{enumerate}
\usepackage{bm}
\usepackage{orcidlink}

\newtheorem{theorem}{Theorem}

\newtheorem{lemma}{Lemma}
\newtheorem{definition}{Definition}

\newcommand{\norm}[1]{\left\lVert#1\right\rVert}
\newcommand{\cB}[1]{\left\{#1\right\}}
\newcommand{\nB}[1]{\left(#1\right)}

\newcommand{\cX}{{\mathcal X}}
\newcommand{\cY}{{\mathcal Y}}

\newcommand{\Ez}[1]{\mathbb{E}_0[#1]}
\newcommand{\Eo}[1]{\mathbb{E}_1[#1]}
\newcommand{\Pz}[1]{\mathbb{P}_0[#1]}
\newcommand{\Po}[1]{\mathbb{P}_1[#1]}

\newcommand{\E}[1]{\mathbb E\left[#1\right]}
\newcommand{\Prob}[1]{\mathbb P\left[#1\right]}

\newcommand{\bs}[1]{\boldsymbol{#1}}
\newcommand{\mc}[1]{\mathcal{#1}}
\newcommand{\mb}[1]{\mathbf{#1}}
\newcommand{\ms}[1]{\mathsf{#1}}
\newcommand{\mr}[1]{\mathrm{#1}}
\newcommand{\bigo}[1]{O\left(#1\right)}

\newcommand{\Var}[1]{\mathrm{Var}\left[#1\right]}

\newcommand{\thmref}[1]{Theorem~\ref{#1}}

\newcommand{\secref}[1]{Section~\ref{#1}}
\newcommand{\lemref}[1]{Lemma~\ref{#1}}

\newcommand{\appref}[1]{Appendix~\ref{#1}}

\newcommand{\figref}[1]{Fig.~\ref{#1}}

\usepackage{hyperref}
\usepackage{url}
\usepackage{cases}
\hypersetup{colorlinks=true, linkcolor=blue, citecolor=blue, urlcolor = blue}
\usepackage{ifthen}

\newif\ifshowtodo
\showtodotrue
\renewcommand{\IEEEproofindentspace}{1\parindent}

\newcommand{\VersionLength}{long}
\providecommand{\ver}{\ifthenelse{\equal{\VersionLength}{long}}}

\allowdisplaybreaks
\begin{document}
\title{Variable-Length Sparse Feedback Codes for Point-to-Point, Multiple Access, and Random Access Channels}

\author{Recep~Can~Yavas,~\IEEEmembership{Member,~IEEE}, Victoria~Kostina,~\IEEEmembership{Senior~Member,~IEEE}, and Michelle~Effros,~\IEEEmembership{Fellow,~IEEE}
\thanks{Manuscript received January 31, 2023; revised October 9, 2023; accepted November 17, 2023.}
\thanks{When this work was completed, R. C. Yavas, V. Kostina, and M. Effros were all with the Department of Electrical Engineering, California
Institute of Technology, Pasadena, CA 91125, USA. R. C. Yavas is currently with CNRS@CREATE, 138602, Singapore (e-mail: vkostina,
effros@caltech.edu, recep.yavas@cnrsatcreate.sg). This work was supported in part by the National Science Foundation (NSF) under grant
CCF-1817241 and CCF-1956386. This paper was presented in part at ISIT 2021 \cite{yavas2021VLF} and at ITW 2021 \cite{nested}.}}
\date{\today}
\IEEEoverridecommandlockouts
\maketitle 
\begin{abstract}
This paper investigates variable-length stop-feedback codes for memoryless channels in point-to-point, multiple access, and random access communication scenarios. The proposed codes employ $L$ decoding times $n_1, n_2, \dots, n_L$ for the point-to-point and multiple access channels and $KL + 1$ decoding times for the random access channel with at most $K$ active transmitters. In the point-to-point and multiple access channels, the decoder uses the observed channel outputs to decide whether to decode at each of the allowed decoding times $n_1, \dots, n_L$, at each time telling the encoder whether or not to stop transmitting using a single bit of feedback. In the random access scenario, the decoder estimates the number of active transmitters at time $n_0$ and then chooses among decoding times $n_{k, 1}, \dots, n_{k, L}$ if it believes that there are $k$ active transmitters. In all cases, the choice of allowed decoding times is part of the code design; given fixed value $L$, allowed decoding times are chosen to minimize the expected decoding time for a given codebook size and target average error probability. The number $L$ in each scenario is assumed to be constant even when the blocklength is allowed to grow; the resulting code therefore requires only sparse feedback. 
The central results are asymptotic approximations of achievable rates as a function of the error probability, the expected decoding time, and the number of decoding times. A converse for variable-length stop-feedback codes with uniformly-spaced decoding times is included for the point-to-point channel.
\end{abstract}

\section{Introduction}
Although feedback does not increase the capacity of memoryless, point-to-point channels (PPCs) \cite{shannon1956zero}, feedback can simplify coding schemes and improve the speed of approach to capacity with blocklength. Examples that demonstrate this effect include Horstein's scheme for the binary symmetric channel (BSC) \cite{horstein}
and Schalkwijk and Kailath's scheme 
for the Gaussian channel \cite{schalkwijk}, both of which leverage full channel feedback to simplify coding in the fixed-length regime. Wagner \textit{et al.} \cite{wagner2020} show that feedback improves the second-order term in the achievable rate as a function of blocklength for fixed-rate coding over discrete, memoryless, point-to-point channels (DM-PPCs) that have multiple capacity-achieving input distributions giving distinct dispersions. 

\subsection{Literature Review on Variable-Length Feedback Codes}
The benefits of feedback increase for codes with multiple decoding times (called variable-length or rateless codes). In \cite{burnashev1976data}, Burnashev shows that feedback significantly improves the optimal error exponent of variable-length codes for DM-PPCs. In \cite{polyanskiy2011feedback}, Polyanskiy \textit{et al.} extend the work of Burnashev to the finite-length regime with non-vanishing error probabilities,
introducing variable-length feedback (VLF) codes and deriving achievability and converse bounds on their performance. Tchamkerten and Telatar  \cite{tchamkerten2006variable} show that Burnashev's optimal error exponent is achieved for a family of BSCs and $Z$ channels, where the cross-over probability of the channel is unknown.
For the BSC, 
Naghshvar \emph{et al.} \cite{naghshvar} propose a VLF coding scheme with a novel encoder called the small-enough-difference (SED) encoder and derive a non-asymptotic achievability bound. Their scheme is an alternative to Burnashev's scheme to achieve the optimal error exponent. Yang \emph{et al.} \cite{yangSED} extend the SED encoder to the binary asymmetric channel, of which the BSC is a special case, and derive refined non-asymptotic achievability bounds for the binary asymmetric channel. Guo and Kostina \cite{guo2021} propose an instantaneous SED code for a source whose symbols progressively arrive at the encoder in real time.

The feedback in VLF codes can be limited in its amount and frequency. Here, the amount refers to how much feedback is sent from the receiver at each time feedback is available; the frequency refers to how many times feedback is available throughout the communication epoch. The extreme cases in the frequency are no feedback and feedback after every channel use.  The extreme cases in the amount are full feedback and stop feedback. With full feedback, at time $n_i$, the receiver sends all symbols received until that time, $Y^{n_i}$, which can be used by the transmitter to encode the $(n_{i+1})$-th symbol. With stop feedback, the receiver sends a single bit of feedback to inform the transmitter whether or not to stop transmitting. Unlike full-feedback codes, variable-length stop-feedback (VLSF) codes employ codewords that are fixed when the code is designed; that is, feedback affects how much of a codeword is sent but does not affect the codeword's value. 

In \cite{polyanskiy2011feedback}, Polyanskiy \textit{et al.} define VLSF codes with feedback after every channel use. 
The result in \cite[Th.~2]{polyanskiy2011feedback} shows that variable-length coding improves the first-order term in the asymptotic expansion of the maximum achievable message set size from $N C$ to $\frac{N C}{1-\epsilon}$, where $C$ is the capacity of the DM-PPC, $N$ is the average decoding time (averaging is with respect to both the random message and the random noise), and $\epsilon$ is the average error probability. The second-order term achievable for VLF codes is $O(\log  N)$, which means that VLF codes have zero dispersion and that the convergence to the capacity is much faster than that achieved by the fixed-length codes \cite{polyanskiy2010Channel, fong2015feedbacknot}.
In \cite{altug2015probabilistic}, Altu\u{g} \textit{et al.} modify the VLSF coding paradigm by replacing the average decoding time constraint with a constraint on the probability that the decoding time exceeds a target value; the benefit in the first-order term does not appear under this probabilistic delay constraint, and the dispersion is no longer zero. A VLSF scenario with noisy feedback and a finite largest available decoding time is studied in \cite{ostman2021noisy}. For VLSF codes,  Forney \cite{forney68} shows an achievable error exponent  that is strictly better than that of fixed-length, no-feedback codes and is strictly worse than Burnashev's error exponent for variable-length full-feedback codes. Ginzach \emph{et al.} \cite{ginzach2017} derive the exact error exponent of VLSF codes for the BSC.

Bounds on the performance of VLSF codes that allow feedback after every channel use are derived for several network communication problems.
Truong and Tan \cite{truong2016gaussian, truong2018Journal} extend the results from \cite{polyanskiy2011feedback} to the Gaussian multiple access channel (MAC) under an average power constraint. Trillingsgaard \textit{et al.} \cite{trillingsgaard2018broadcast} study the VLSF scenario where a common message is transmitted across a $K$-user discrete memoryless broadcast channel. Heidari \emph{et al.} \cite{heidari2018} extend Burnashev's work from the DM-PPC to the DM-MAC, deriving lower and upper bounds on the error exponents of VLF codes for the DM-MAC. Bounds on the performance of VLSF codes for the DM-MAC with an unbounded number of decoding times appear in \cite{trillingsgaard2014VLF}. The achievability bounds for $K$-transmitter MAC in \cite{truong2018Journal} and \cite{trillingsgaard2014VLF} employ $2^K-1$ simultaneous information density threshold rules. 
 
While high rates of feedback are impractical for many applications --- especially wireless applications on half-duplex devices --- most prior work on VLSF codes (e.g., \cite{polyanskiy2011feedback, altug2015probabilistic, truong2016gaussian, truong2018Journal,  trillingsgaard2018broadcast, heidari2018, trillingsgaard2014VLF}) considers the densest feedback possible, using feedback at each of the (at most) $n_{\max}$ time steps before decoding, where $n_{\max}$ is the largest blocklength used by a given VLSF code. To consider more limited feedback scenarios, let $L$ denote the number of potential decoding times in a VLSF code, a number that we assume to be independent of the blocklength. We further assume that feedback is available only at the $L$ fixed decoding times $n_1, \dots, n_L$, which are fixed in the code design and known by the transmitter and receiver before the start of transmission. In \cite{kim2015VLF}, Kim \emph{et al.} choose the decoding time for each message from the set $\{d, 2d, \dots, Ld\}$ for some positive integer $d$ and $L < \infty$, In \cite{williamson2015VLConvolution}, Williamson \emph{et al.} numerically optimize the values of $L$ decoding times and employ punctured convolutional codes and a Viterbi algorithm. In \cite{vakilinia2016}, Vakilinia \emph{et al.} introduce a sequential differential optimization (SDO) algorithm to optimize the choices of the $L$ potential decoding times $n_1, \dots, n_L$, approximating the random decoding time $\tau$ by a Gaussian random variable. Vakilinia \emph{et al.} apply the SDO algorithm to non-binary low-density parity-check codes over binary-input, additive white Gaussian channels; the mean and variance of $\tau$ are determined through simulation.
Heidarzadeh \emph{et al.} \cite{heidarzadeh2019Systematic} extend \cite{vakilinia2016} to account for the feedback rate and apply the SDO algorithm to random linear codes over the binary erasure channel. In \cite{yavas2020Random}, we develop a communication strategy for a random access scenario with a total of $K$ transmitters; in this scenario, neither the transmitters nor the receiver knows the set of active transmitters, which can vary from one epoch to the next. The code in \cite{yavas2020Random} is a VLSF code with decoding times $n_0 < n_1 < \dots < n_K$. The decoder decodes messages only if it decides at that time that $k$ out of total $K$ transmitters are active at time $n_k$. It informs the transmitters about its decision by sending a one-bit signal at each time $n_i$ until the time at which it decodes. We show that our random access channel (RAC) code with sparse stop feedback achieves performance identical in the capacity and dispersion terms to that of the best-known code without feedback for a MAC in which the set of active transmitters is known a priori. An extension of \cite{yavas2020Random} to low-density parity-check codes appears in \cite{LiuE:20}. Building upon an earlier version of the present paper \cite{yavas2021VLF}, Yang \emph{et al.} \cite{yang2022} construct an integer program to minimize the upper bound on the average blocklength subject to constraints on the average error probability and the minimum gap between consecutive decoding times. By employing a combination of the Edgeworth expansion \cite[Sec. XVI.4]{feller1971introduction} and the Petrov expansion (\lemref{lem:moderate}), that paper develops an approximation to the cumulative distribution function of the information density random variable $\imath(X^n; Y^n)$; the numerical comparison of their approximation and the empirical cumulant distribution function shows that the approximation is tight even for small values of $n$. Their analysis uses this tight approximation to numerically evaluate the non-asymptotic achievability bound (\thmref{thm:nonasyPPC}, below) for the BSC, binary erasure channel, and binary-input Gaussian PPC for all $L \leq 32$. The resulting numerical results show performance that closely approaches Polyanskiy's VLSF achievability bound \cite{polyanskiy2011feedback} with a relatively small $L$. For the binary erasure channel, \cite{yang2022} also proposes a new zero-error code that employs systematic transmission followed
by random linear fountain coding; the proposed code  
outperforms Polyanskiy’s achievability
bound. 

Sparse feedback is known to achieve the optimal error exponent for VLF codes. Yamamoto and Itoh \cite{yamamoto1979} construct a two-phase scheme that achieves Burnashev's optimal error exponent \cite{burnashev1976data}. Although their scheme allows an unlimited number of feedback instances and decoding times, it is sparse in the sense that feedback is available only at times $\alpha n, n, (1+\alpha)n, 2n, \dots$ for some $\alpha \in (0, 1)$ and integer $n$. 
Lalitha and Javidi \cite{lalitha2020Almost} show that Burnashev's optimal error exponent can be achieved by only $L = 3$ decoding times by truncating the Yamamoto--Itoh scheme.

Decoding for VLSF codes can be accomplished by running a sequential hypothesis test (SHT) on each possible message. At each increasingly larger stopping times, the SHT compares a hypothesis $H_0$ corresponding to a particular transmitted message to the hypothesis $H_1$ corresponding to the marginal distribution of the channel output. In \cite{berlin2009conv}, Berlin \emph{et al.} derive a bound on the average stopping time of an SHT. They then use this bound to derive a non-asymptotic converse bound for VLF codes. This result is an alternative proof for the converse of Burnashev's error exponent \cite{burnashev1976data}. 

\subsection{Contributions of This Work}
Like \cite{vakilinia2016, williamson2015VLConvolution, heidarzadeh2019Systematic}, this paper studies VLSF codes under a finite constraint $L$ on the number of decoding times. While \cite{vakilinia2016, williamson2015VLConvolution, heidarzadeh2019Systematic} focus on practical coding and performance, our goal is to derive new achievability bounds on the asymptotic rate achievable by VLSF codes between $L = 1$ (the fixed-length regime analyzed in \cite{polyanskiy2010Channel, tan2015Third}) and $L = n_{
\max}$ (the classical variable-length regime defined in \cite[Def.~1]{polyanskiy2011feedback} where all decoding times 1, 2, $\dots, n_{\max}$ are available). 

Our contributions are summarized as follows.
\begin{enumerate}
    \item We derive second-order achievability bounds for VLSF codes over DM-PPCs, DM-MACs, DM-RACs, and the Gaussian PPC with maximal power constraints. These bounds are presented in Theorems~\ref{thm:mainPPC},~\ref{thm:MAC},~\ref{thm:RAC}, and~\ref{thm:Gaussian}, respectively. In our analysis for each problem, we consider the asymptotic regime where the number of decoding times $L$ is fixed while the average decoding time $N$ grows without bound, i.e., $L = O(1)$ with respect to $N$. Each of our asymptotic bounds follows from the corresponding non-asymptotic bound that employs an information-density threshold rule with a stop-at-time-zero procedure. Asymptotically optimizing the values of the $L$ decoding times yields the given results. By viewing the proposed decoder as a special case of SHT-based decoders, we show a more general non-asymptotic achievability bound; \thmref{thm:SHTnonasy} employs an arbitrary SHT to decide whether a message is transmitted. 

    \item Linking the error probability of any given VLSF code to that of an SHT, in \thmref{thm:metaconv}, we prove a converse bound in the spirit of the meta-converse bound from \cite[Th.~27]{polyanskiy2010Channel}. Analyzing the new bound with infinitely many uniformly-spaced decoding times over Cover--Thomas symmetric channels, in \thmref{thm:VLSFdN}, we prove a converse bound  for VLSF codes; the resulting bound is tight up to its second-order term. 
    Unfortunately, since analyzing our meta-converse bound is challenging in the general case of an arbitrary DM-PPC and an arbitrary number $L$ of decoding times (see \cite[Th.~3.2.3]{sequentialbook} for the structure of optimal SHTs with finitely many decoding times), whether or not the second-order term is tight in the general case remains an open question.

\end{enumerate}

\begin{table*}[htbp!]
 \renewcommand{\arraystretch}{1.8}
\caption{The performance of VLSF codes according to the number of decoding times $L$ and the channel type}
\label{table:T2}
\centering
\begin{tabular}{|c|c|c|cc|}
\hline
\multirow{2}{*}{\textbf{Number of decoding times}}                                                    & \multicolumn{1}{l|}{\multirow{2}{*}{\textbf{Channel type}}}                     & \multirow{2}{*}{\textbf{First-order term}} & \multicolumn{2}{c|}{\textbf{Second-order term}}                                                  \\ \cline{4-5} 
                                                                              & \multicolumn{1}{l|}{}                                                  &                                   & \multicolumn{1}{c|}{\textbf{Lower bound}}                       & \textbf{Upper bound}                    \\ \hline
\begin{tabular}[c]{@{}c@{}}Fixed-length, no-feedback \\ $(L = 1)$\end{tabular}                                                      & DM-PPC                                                                 & $N C$                             & \multicolumn{1}{c|}{$-\sqrt{N V} Q^{-1}(\epsilon)$ \quad \cite{polyanskiy2010Channel}}    & $-\sqrt{N V} Q^{-1}(\epsilon)$ \quad \cite{polyanskiy2010Channel} \\ \hline
\begin{tabular}[c]{@{}c@{}}Variable-length \\ $(1 < L < \infty)$\end{tabular} & DM-PPC                                                                 & $\frac{N C}{1-\epsilon}$          & \multicolumn{1}{c|}{$-\sqrt{N \log_{(L-1)}(N) \frac{V}{1-\epsilon}}$ \quad (\thmref{thm:mainPPC})}                                  & $+ O(1)$   
\quad \cite{polyanskiy2011feedback} \\ \hline
\begin{tabular}[c]{@{}c@{}}Variable-length \\ $(L = \infty)$\end{tabular}     & DM-PPC                                                                 & $\frac{N C}{1-\epsilon}$          & \multicolumn{1}{c|}{$-\log N + O(1)$ \quad \cite{polyanskiy2011feedback}}                   & $+ O(1)$ \quad \cite{polyanskiy2011feedback}                       \\ \hline
\begin{tabular}[c]{@{}c@{}}Fixed-length, no-feedback \\ $(L = 1)$\end{tabular}                                                      & DM-MAC                                                                 & $ N I_K $                         & \multicolumn{1}{c|}{$- \sqrt{N V_K} Q^{-1}(\epsilon)$ \quad \cite{tan2014dispersions}} & $+ O(\sqrt{N})$ \quad \cite{kosut}                \\ \hline
\begin{tabular}[c]{@{}c@{}}Variable-length \\ $(1 < L < \infty)$\end{tabular} & DM-MAC                                                                 & $ \frac{N I_K}{1-\epsilon}$       & \multicolumn{1}{c|}{$-\sqrt{N \log_{(L-1)}(N) \frac{V_K}{1-\epsilon}}$ \quad (\thmref{thm:MAC})}                                 & $+ O(1)$ \quad \cite{truong2018Journal}                             \\ \hline
\begin{tabular}[c]{@{}c@{}}Variable-length \\ $(L = \infty)$\end{tabular}     & DM-MAC                                                                 & $ \frac{N I_K}{1-\epsilon}$       & \multicolumn{1}{c|}{$-\log N + O(1)$ \quad eq. \eqref{eq:mainresultMACinf}}                                 & $+ O(1)$ \quad \cite{truong2018Journal}                             \\ \hline
\begin{tabular}[c]{@{}c@{}}Variable-length \\ $(L = \infty)$\end{tabular}     & \begin{tabular}[c]{@{}c@{}}Gaussian MAC\\ (average power)\end{tabular} &   \multicolumn{1}{c|}{$\frac{N C(KP)}{1-\epsilon}$}                                & \multicolumn{1}{c|}{$-O(\sqrt{N})$ \quad \cite{truong2018Journal}}                                 & $+O(1)$  \quad \cite{truong2018Journal}                           \\ \hline
$(L = 1)$                                                                     & DM-RAC                                                                 & $N_k I_k$                         & \multicolumn{1}{c|}{$-\sqrt{N_k V_k} Q^{-1}(\epsilon_k)$ \quad \cite{yavas2020Random}}                                 & $+ O(\sqrt{N_k})$   \quad \cite{kosut}           \\ \hline
$(L = 1)$                                                                     & \begin{tabular}[c]{@{}c@{}}Gaussian RAC \\ (maximal power)\end{tabular}                                                           & $N_k C(kP)$                       & \multicolumn{1}{c|}{$-\sqrt{N_k V_k(P)} Q^{-1}(\epsilon_k)$ \quad \cite{yavas2021Gaussian}}                                 & $+ O(\sqrt{N_k})$  \quad \cite{kosut}            \\ \hline
$(1 < L < \infty)$                                                            & DM-RAC                                                                 & $\frac{N_k I_k}{1-\epsilon_k}$    & \multicolumn{1}{c|}{$-\sqrt{N_k \log_{(L-1)}(N_k) \frac{V_k}{1-\epsilon_k}}$ \quad (\thmref{thm:RAC})}                                 & $+O(1)$ \quad \cite{truong2018Journal}                             \\ \hline
\begin{tabular}[c]{@{}c@{}}Variable-length \\ $(1 < L < \infty)$\end{tabular} & \begin{tabular}[c]{@{}c@{}}Gaussian PPC \\ (maximal power)\end{tabular}                                                           & $\frac{N C(P)}{1-\epsilon}$         & \multicolumn{1}{c|}{$-\sqrt{N \log_{(L-1)}(N) \frac{V(P)}{1-\epsilon}}$ \quad (\thmref{thm:Gaussian})}                                 & $+ O(1)$ \quad \cite{truong2016gaussian}                             \\ \hline
\begin{tabular}[c]{@{}c@{}}Variable-length\\ $(L = \infty)$\end{tabular}        & \begin{tabular}[c]{@{}c@{}}Gaussian PPC \\ (average power)\end{tabular}                                                           & $\frac{N C(P)}{1-\epsilon}$         & \multicolumn{1}{c|}{$- \log N + O(1)$ \quad \cite{truong2016gaussian} }                                 & $+O(1)$ \quad \cite{truong2016gaussian}                             \\ \hline
\end{tabular}
\end{table*}

Below, we detail these contributions. Our main result shows that for VLSF codes with $L = O(1) \geq 2$ decoding times over a DM-PPC, message set size $M$ satisfying
\begin{align}
    \log  M \approx \frac{N C}{1-\epsilon} - \sqrt{N \log _{(L-1)}(N) \frac{V}{1-\epsilon}} \label{eq:Mapprox}
\end{align}
is achievable. Here $\log _{(L)}(\cdot)$ denotes the $L$-fold nested logarithm, $N$ is the average decoding time, $\epsilon$ is the average error probability, and $C$ and $V$ are the capacity and dispersion of the DM-PPC, respectively. Similar formulas arise for the DM-MAC and DM-RAC, where $C$ and $V$ are replaced by the sum-rate mutual information and the sum-rate dispersion. The speed of convergence to $\frac{C}{1-\epsilon}$ depends on $L$. It is slower than the convergence to $C$ in the fixed-length scenario, which has second-order term $O(\sqrt{N})$ \cite{polyanskiy2010Channel}. The $L = 2$ case in \eqref{eq:Mapprox} recovers the rate of convergence for the variable-length scenario without feedback, which has second-order term $O(\sqrt{N \log  N})$ \cite[Proof~of~Th.~1]{polyanskiy2011feedback}; that rate is achieved with $n_1 = 0$. 
The nested logarithm term in \eqref{eq:Mapprox} arises because after writing the average decoding time as $\E{\tau} = n_1 + \sum_{i = 1}^{L-1} (n_{i + 1} - n_i) \Prob{\tau > n_i}$, the decoding time choices in \eqref{eq:decodingeq}, below, satisfy $(n_{i + 1} - n_i) \Prob{\tau > n_i} = o(\sqrt{n_1})$ for $i \in [L-1]$, making the effect of each decoding time on the average decoding time asymptotically similar. 
We then use the SDO algorithm introduced in \cite{vakilinia2016} to show that our particular choice of $n_1, \dots, n_L$ is second-order optimal (see \appref{app:optimalityproof}). Despite the order-wise dependence of the rate of convergence on $L$, \eqref{eq:Mapprox} grows so slowly with $L$ that it suggests little benefit to choosing a large $L$. For example, when $L = 4$, $\sqrt{N \log _{(L-1)}(N)}$ behaves very similarly to $O(\sqrt{N})$ for practical values of $N$ (e.g., $N \in [10^3, 10^5]$). Notice, however, that the given achievability result provides a \emph{lower} bound on the benefit of increasing $L$; bounding the benefit from above requires a converse result. We note, however, that the numerical results in \cite{yang2022} support our conclusion from the asymptotic achievability bound \eqref{eq:Mapprox} that indicates that the improvement over achievable $\log M$ from $L$ to $L+1$ decoding times diminishes as $L$ increases.

For the PPC and MAC, the feedback rate of our code is $\frac{\ell}{n_\ell}$ if the decoding time is $n_\ell$; for the RAC, that rate becomes $\frac{(k-1)L + \ell + 1}{n_{k, \ell}}$ if the decoding time is $n_{k, \ell}$. In both cases, our feedback rate approaches 0 as the decoding time grows. In contrast, VLSF codes like in \cite{polyanskiy2011feedback, forney68} use feedback rate 1 bit per channel use. In VLSF codes for the RAC, the decoder decodes at one of the available times $n_{k, 1}, n_{k, 2}, \dots, n_{k, L}$ if it estimates that the number of active transmitters is $k \neq 0$; we reserve a single decoding time $n_0$ for decoding the possibility that no active transmitters are active. \thmref{thm:RAC} extends the RAC code in \cite{yavas2020Random} from $L = 1$ to any $L \geq 2$.

The converse result in \thmref{thm:VLSFdN} shows that in order to achieve \eqref{eq:Mapprox} with evenly spaced decoding times, one needs at least $L = \Omega \left(\sqrt{\frac{N}{\log_{(L-1)}(N)}} \right)$ decoding times. In contrast, our optimized codes achieve \eqref{eq:Mapprox} with a finite $L$ that does not grow with the average decoding time $N$, which highlights the importance of optimizing the values of decoding times in a VLSF code.

Table~\ref{table:T2} summarizes the literature on VLSF codes and the new results from this work, showing how they vary with the number of decoding times and the channel type.

In what follows, \secref{sec:problemstatement} gives notation and definitions. Sections \ref{sec:PPC}--\ref{sec:GaussianPPC} introduce variable-length sparse stop-feedback codes for the DM-PPC, DM-MAC, DM-RAC, and the Gaussian PPC, respectively, and present our main theorems for those channel models; \secref{sec:conclusion} concludes the paper. The proofs appear in the Appendix.

\section{Preliminaries}\label{sec:problemstatement}
\subsection{Notation}
For any positive integers $k$ and $n$, $[k] \triangleq \{1, \dots, k\}$, $x^n \triangleq (x_1, \dots, x_n)$, and $x^{a:b} \triangleq (x_a, x_{a+1}, \dots, x_b)$. The collection of length-$n$ vectors from the transmitter index set $\mc{A}$ is denoted by $x_{\mc{A}}^n \triangleq (x_a^n \colon a \in \mc{A})$; we drop the superscript $n$ if $n = 1$, i.e., $x^1_{\mc{A}} = x_{\mc{A}}$. The collection of non-empty strict subsets of a set $\mc{A}$ is denoted by $\mc{P}(\mc{A}) \triangleq \{ \mc{B} \colon \mc{B} \subseteq \mc{A}, 0<|\mc{B}| < |\mc{A}|\}.$  
All-zero and all-one vectors are denoted by $\mathbf{0}$ and $\mathbf{1}$, respectively; dimension is determined from the context. The sets of positive integers and non-negative integers are denoted by $\mathbb{Z}_+$ and $\mathbb{Z}_\geq$, respectively. We write $x^n \stackrel{\pi}{=} y^n$ if there exists a permutation $\pi$ of $x^n$ such that $\pi(x^n) = y^n$, and $x^n \stackrel{\pi}{\neq} y^n$ if such a permutation does not exist. The identity matrix of dimension $n$ is denoted by $\mathsf{I}_n$. The Euclidean norm of vector $x^n$ is denoted by $\norm{x^n} \triangleq \sqrt{\sum_{i = 1}^n x_i^2}$.
Unless specified otherwise, all logarithms and exponents have base $e$. Information is measured in nats. The standard $O(\cdot)$, $o(\cdot)$, and $\Omega(\cdot)$ notations are defined as $f(n) = O(g(n))$ if $\limsup_{n \to \infty} |f(n) / g(n)| < \infty$, $f(n) = o(g(n))$ if $\lim_{n \to \infty} |f(n) / g(n)| = 0$, and $f(n) = \Omega(g(n))$ if $\lim_{n \to \infty} |f(n) / g(n)| > 0$. The distribution of a random variable $X$ is denoted by $P_X$; $\mathcal{N}(\bs{\mu}, \ms{V})$ denotes the Gaussian distribution with mean $\bs{\mu}$ and covariance matrix $\ms{V}$, $Q(\cdot)$ represents the complementary standard Gaussian cumulative distribution function $Q(x) \triangleq \frac{1}{\sqrt{2 \pi}} \int_{x}^{\infty} \exp\cB{-\frac{t^2}{2}} dt$, and $Q^{-1}(\cdot)$ is its functional inverse. We define the nested logarithm function
\begin{align}
        \log _{(L)}(x) \triangleq \begin{cases}
    \log (x) &\text{if } L = 1, \,\, x > 0 \\
    \log (\log _{(L-1)}(x)) &\text{if } L \geq 2, \,\, \log _{(L-1)}(x) > 0;
    \end{cases} \IEEEeqnarraynumspace
\end{align}
$\log _{(L)}(x)$ is undefined for all other $(L, x)$ pairs.

We denote the Radon-Nikodym derivative of distribution $P$ with respect to distribution $Q$ by $\frac{\mathrm{d}P}{\mathrm{d}Q}$. We denote the relative entropy and relative entropy variance between $P$ and $Q$ by $D(P \| Q) = \E{\log  \frac{\mathrm{d}P}{\mathrm{d}Q}(X)}$ and $V(P \| Q) = \mathrm{Var}\left[\log  \frac{\mathrm{d}P}{\mathrm{d}Q}(X) \right]$, respectively, where $X \sim P$. The $\sigma$-algebra generated by random variable $X$ is denoted by $\mc{F}(X)$. A random variable $X$ is called arithmetic if there exists some $d > 0$ such that $\Prob{X \in d \mathbb{Z}} = 1$. The largest $d$ that satisfies this condition is called the span. If such a $d$ does not exist, then the random variable is non-arithmetic. Denote $X^+ \triangleq \max\{0, X\}$ and $X^- \triangleq -\min\{0, X\}$ for any random variable $X$. 

\subsection{Discrete Memoryless Channel and Information Density}

A DM-PPC is defined by the triple $(\mc{X}, P_{Y|X}, \mc{Y})$, where $\mc{X}$ is the finite input alphabet, $P_{Y|X}$ is the channel transition kernel, and $\mc{Y}$ is the finite output alphabet. The $n$-letter input-output relationship of the channel is given by
$P_{Y^n|X^n}(y^n|x^n) = \prod_{i = 1}^n P_{Y|X}(y_i|x_i)$ for all $n$, $x^n$, and $y^n$.

The $n$-letter information density of a channel $P_{Y|X}$ under input distribution $P_{X^n}$ is defined as
\begin{align}
    \imath(x^n; y^n) \triangleq \log  \frac{P_{Y^n|X^n}(y^n|x^n)}{P_{Y^n}(y^n)},
 \end{align}
where $P_{Y^n}$ is the $Y^n$ marginal of $P_{X^n}P_{Y^n|X^n}$. If the inputs $X_1, X_2, \dots, X_n$ are independently and identically distributed (i.i.d.) according to $P_X$, then
\begin{align}
    \imath(x^n; y^n) = \sum_{i = 1}^n \imath(x_i; y_i),
 \end{align}
 where the single-letter information density is given by
 \begin{align}
     \imath(x; y) \triangleq \log  \frac{P_{Y|X}(y|x)}{P_{Y}(y)}, \quad x \in \mc{X}, y \in \mc{Y}.
 \end{align}
 The mutual information and dispersion are defined as
 \begin{align}
     I(X;Y) &\triangleq \E{\imath(X; Y)} \\
     V(X; Y) &\triangleq \Var{\imath(X; Y)},
 \end{align}
 respectively, where $(X, Y) \sim P_X P_{Y|X}$. 
 
Let $\mc{P}$ denote all distributions on the alphabet $\mc{X}$. The capacity of the DM-PPC is 
 \begin{align}
     C = \max_{P_X \in \mc{P}} I(X; Y),
 \end{align}
 and the dispersion of the DM-PPC is 
 \begin{align}
     V = \min_{P_X \in \mc{P} \colon I(X; Y) = C} V(X; Y). \label{eq:dispersion}
 \end{align}

\section{VLSF Codes for the DM-PPC} \label{sec:PPC}
\subsection{VLSF Codes with $L$ Decoding Times}

We consider VLSF codes with a finite number of potential decoding times
$n_1 < n_2 < \cdots < n_L$ over a DM-PPC. The receiver chooses to end the transmission at the first time $n_\ell \in \{n_1, \ldots, n_L\}$ that it is ready to decode. The transmitter learns of the receiver's decision via a single bit of feedback at each of times $n_1, \ldots, n_\ell$. Feedback bit ``0'' at time $n_i$ means that the receiver is not yet ready to decode, and transmission should continue; feedback bit ``1'' means that the receiver can decode at time $n_i$, which signals the transmitter to stop. Using this feedback, the transmitter and the receiver are synchronized and aware of the current state of the transmission at all times. Since $n_L$ is the last decoding time available, the receiver always makes a final decision if time $n_L$ is reached. Unlike \cite{burnashev1976data, yamamoto1979, williamson2015VLConvolution}, we do not allow re-transmission of the message after time $n_L$. Since the transmitter and the receiver both know the values of decoding times, the receiver does not need to send feedback at the last available time $n_L$. 
We assume that the transmitter and the receiver know the channel transition kernel $P_{Y|X}$. 
We employ average decoding time and average error probability constraints. Definition~\ref{def:gaussvlf}, below, formalizes our code description.
	\begin{definition} \label{def:gaussvlf}
	Fix $\epsilon \in (0, 1)$, positive integers $L$ and $M$, and a positive scalar $N$. An $(N, L, M, \epsilon)$-VLSF code for the DM-PPC comprises
	\begin{enumerate}
	\item non-negative integer-valued decoding times $n_1 < \ldots < n_L$,
	\item a finite alphabet $\mathcal{U}$ and a probability distribution $P_U$ on $\mc{U}$ defining a common randomness random variable $U$ that is revealed to both the transmitter and the receiver before the start of the transmission,\footnote{The realization $u$ of $U$ specifies the codebook.} 
	\item an encoding function $\mathsf{f}_n \colon \mc{U} \times [M] \to \mc{X}$, for each $n = 1, \ldots, n_L$, that assigns a codeword
	\begin{align}
	    \mathsf{f}(u, m)^{n_L} \triangleq (\mathsf{f}_1(u, m), \dots, \mathsf{f}_{n_L}(u, m))
	\end{align} to each message $m \in [M]$ and common randomness instance $u \in \mc{U}$,
    \item a non-negative integer-valued random stopping time $\tau \in \{n_1, \dots, n_L\}$ for the filtration generated by $\{U, Y^{n_i}\}_{i = 1}^L$ that satisfies an average decoding time constraint
	\begin{align}
	    \E{\tau} \leq N, \label{eq:averagetau}
	\end{align}
    \item and a decoding function
	$\mathsf{g}_{n_\ell} \colon \mc{U} \times \mc{Y}^{n_\ell} \to [M] \cup \{\ms e \}$ for each $\ell \in [L]$ (where $\ms{e}$ is the erasure symbol used to indicate that the receiver is not ready to decode), satisfying an average error probability constraint
   \begin{align}
        \Prob{\mathsf{g}_{\tau}(U, Y^{\tau}) \neq W} \leq \epsilon, \label{eq:averageerror}
    \end{align}
    where the message $W$ is equiprobably distributed on the set $[M]$, and $X^{\tau} = \mathsf f(U, W)^{\tau}$.
	
	\end{enumerate}
	\end{definition}
	
	Recall that  Definition~\ref{def:gaussvlf} with $L = 1$ recovers the fixed-length no-feedback codes in \cite{polyanskiy2010Channel}. As in \cite{polyanskiy2011feedback, yavas2020Random,trillingsgaard2018broadcast}, we here need common randomness because the traditional random-coding argument does not prove the existence of a single (deterministic) code that simultaneously satisfies conditions \eqref{eq:averagetau} and \eqref{eq:averageerror} on the code. Therefore, randomized codes are necessary for our achievability argument; here,  $|\mc{U}| \leq 2$ suffices \cite[Appendix~D]{yavas2020Random}.

We define the maximum achievable message set size $M^*(N, L, \epsilon)$ with $L$ decoding times, average decoding time $N$, and average error probability $\epsilon$ as
	\begin{align}
	    M^*(N, L, \epsilon) &\triangleq \max\{M \colon \text{ an } (N, L, M, \epsilon) \notag \\
	    &\text{ VLSF code exists}\}.\label{eq:Mstarmax}
	\end{align}
The maximum achievable message set size for VLSF codes with $L$ decoding times $n_1, \dots, n_L$ that are restricted to belong to a subset $\mc{N} \subseteq \mathbb{Z}_\geq$ is denoted by $M^*(N, L, \epsilon, \mc{N})$.

	\subsection{Related Work}
	The following discussion summarizes prior asymptotic expansions of the maximum achievable message set size for the DM-PPC.
	\begin{enumerate}[a)]
	\item $M^*(N, 1, \epsilon)$: For $L = 1$ and $\epsilon \in (0, 1/2)$, Polyanskiy \emph{et al.} \cite[Th.~49]{polyanskiy2010Channel} show that
    \begin{IEEEeqnarray}{rCl}
	   \log  M^*(N, 1, \epsilon) = N C - \sqrt{N V} Q^{-1}(\epsilon) + O(\log  N). \IEEEeqnarraynumspace \label{eq:PPCL1}
	\end{IEEEeqnarray}
	For $\epsilon \in [1/2, 1)$, the dispersion $V$ in \eqref{eq:dispersion} is replaced by the maximum dispersion  $V_{\max} \triangleq \max \limits_{P_X \colon \imath(X; Y) = C} V(X; Y)$. The $O(\log  N)$ term is lower bounded by $O(1)$ and upper bounded by $\frac{1}{2} \log  N + O(1)$. For nonsingular DM-PPCs, i.e., the channels that satisfy $\E{\Var{\imath(X; Y) | Y}} > 0$ for the distributions that achieve the capacity $C$ and the dispersion $V$, the $O(\log  N)$ term is equal to $\frac{1}{2} \log  N + O(1)$ \cite{tomamichel2013converse}. Moulin \cite{moulin2017log} derives lower and upper bounds on the $O(1)$ term in the asymptotic expansion when the channel is nonsingular with non-lattice information density. 
	
	\item $M^*(N, \infty, \epsilon)$: For VLSF codes with $L = n_{\max} = \infty$, Polyanskiy \emph{et al.}  \cite[Th.~2]{polyanskiy2011feedback} show that 
	for $\epsilon \in (0,1)$, 
    \begin{align}
	 \log  M^*(N, \infty, \epsilon) &\geq \frac{N C}{1 - \epsilon} - \log  N + O(1) 
	 \label{eq:feedb1}\\
	 \log  M^*(N, \infty, \epsilon) &\leq \frac{N C }{1-\epsilon} + \frac{ h_b(\epsilon)}{1-\epsilon}, \label{eq:feedb2}
	\end{align}
	where $h_b(\epsilon) \triangleq -\epsilon\log  \epsilon - (1-\epsilon) \log  (1-\epsilon)$ is the binary entropy function (in nats). The bounds in \eqref{eq:feedb1}--\eqref{eq:feedb2} indicate that the $\epsilon$-capacity (the first-order achievable term) is
	\begin{align}
	        \lim \inf_{N \to \infty} \frac{1}{N} \log  M^*(N, \infty, \epsilon) = \frac{C}{1-\epsilon}. \label{eq:epscaps}
	\end{align} 
	The achievable dispersion term is zero, i.e., the second-order term in the fundamental limit in \eqref{eq:feedb1}--\eqref{eq:feedb2} is $o(\sqrt{N})$. 
	\end{enumerate}
	
	\subsection{Our Achievability Bounds}\label{sec:main}
 \thmref{thm:nonasyPPC}, below, is our non-asymptotic achievability bound for VLSF codes with $L$ decoding times.

\begin{theorem}\label{thm:nonasyPPC}
	Fix a constant $\gamma$, decoding times $n_1 < \cdots < n_L$, and a positive integer $M$. For any positive number $N$ and $\epsilon \in (0, 1)$, there exists an $(N, L,M, \epsilon)$-VLSF code for the DM-PPC $(\mc{X}, P_{Y|X}, \mc{Y})$ with
	\begin{IEEEeqnarray}{rCl}
	    \epsilon &\leq& \Prob{\imath(X^{n_L}; Y^{n_L}) < \gamma} + (M-1) \exp\{-\gamma\}  \label{eq:boundeps}, \\
	    N &\leq& n_1 + \sum_{\ell = 1}^{L-1} (n_{\ell + 1} - n_\ell) \Prob{ \imath(X^{n_\ell}; Y^{n_\ell}) < \gamma}, \IEEEeqnarraynumspace \label{eq:boundN}
	\end{IEEEeqnarray}
	 where $P_{X^{n_L}}$ is a product of distributions of $L$ sub-vectors of lengths ${n_j - n_{j-1}}$, $j \in [L]$, i.e.,
	\begin{align}
	    P_{X^{n_L}}(x^{n_L}) = \prod_{j = 1}^L   P_{X^{n_{j-1}+1:n_j}}(x^{n_{j-1}+1:n_j}), \label{eq:probproduct}
	\end{align}
	where $n_0 = 0$.
	\end{theorem}
	\begin{IEEEproof}[Proof sketch]
    Polyanskiy \emph{et al.} \cite{polyanskiy2010Channel} interpret the information-density threshold test for a fixed-length code as a collection of hypothesis tests aimed at determining whether the channel output is $(H_0)$ or is not $(H_1)$ dependent on a given codeword. In our coding scheme, we use SHTs in a similar way. The strategy is as follows. 
	 
	 The VLSF decoder at each time $n_1, \dots, n_L$ runs $M$ SHTs between a hypothesis $H_0$ that the channel output results from transmission of the $m$-th codeword, $m \in [M]$, and the hypothesis $H_1$ that the channel output is drawn from the unconditional channel output distribution. The former indicates that the decoder hypothesizes that message $m$ is the sent message. The latter indicates that the decoder hypothesizes that message $m$ has not been sent and thus can be removed from the list of possible messages to decode. Transmission stops at the first time $n_i$ that hypothesis $H_0$ is accepted for some message $m$ or the first time $n_i$ that hypothesis $H_1$ is accepted for all $m$. If the latter happens, decoding fails and we declare an error. Transmission continues as long as one of the SHTs has not accepted either $H_0$ or $H_1$. If $H_0$ is declared for multiple messages at the same decoding time, then we stop and declare an error. Since $n_L$ is the last available decoding time, the SHTs are forced to decide between $H_0$ and $H_1$ at time $n_L$. Once $H_0$ or $H_1$ is decided for some message, the decision cannot be reversed at a later time. 

     The optimal SHT has the form of a two-sided information density threshold rule, where the thresholds depend on the individual decision times \cite[Th.~3.2.3]{sequentialbook}. 
     To simplify the analysis, we employ sub-optimal SHTs for which the upper threshold is set to a value $\gamma \in \mathbb{R}$ that is independent of the decoding times and the lower thresholds are set to $-\infty$ for $n_\ell < n_L$ and to $\gamma$ for $n_\ell = n_L$. That is, we declare $H_1$ for a message if and only if the corresponding information density never reaches $\gamma$ at any of decoding times $n_1, \dots, n_L$.
	\thmref{thm:nonasyPPC} analyzes the error probability and the average decoding time of the sub-optimal SHT-based decoder above, and extends the achievability bound in \cite[Th.~3]{polyanskiy2011feedback} that considers $L = \infty$ to the scenario where only a finite number of decoding times is allowed.
    The bound  on the average decoding time \eqref{eq:boundN} is obtained by expressing the bound on the average decoding time in \cite[eq.~(27)]{polyanskiy2011feedback} using the fact that the stopping time $\tau$ is in $\{n_1, \dots, n_L\}$.  When we compare \thmref{thm:nonasyPPC} with \cite[Th.~3]{polyanskiy2011feedback}, we see that the error probability bound in \eqref{eq:boundeps} has an extra term $\Prob{\imath(X^{n_L}; Y^{n_L}) < \gamma}$. This term appears since transmission always stops at or before time $n_L$. 

    \thmref{thm:nonasyPPC} is related to \cite[Lemma~1]{kim2015VLF}, which similarly treats $L < \infty$ but requires $n_{\ell + 1} - n_\ell = d$ for some constant $d \geq 1$, and~\cite[Cor.~2]{williamson2015VLConvolution}, where the transmitter retransmits the message if decoding attempts at times $n_1, \dots, n_L$ are unsuccessful. 
    
    See \appref{app:proofnonasy} for the proof details.

   \end{IEEEproof}
 
    \thmref{thm:mainPPC}, stated next, is our second-order achievability bound for VLSF codes with $L = O(1)$ decoding times over the DM-PPC. The proof of \thmref{thm:mainPPC} builds upon the non-asymptotic bound in \thmref{thm:nonasyPPC}.
    
	\begin{theorem}\label{thm:mainPPC}
	Fix an integer $L = O(1) \geq 2$ and real numbers $N > 0$ and $\epsilon \in (0, 1)$. For the DM-PPC with $V > 0$, the maximum message set size \eqref{eq:Mstarmax} achievable by $(N, L,M, \epsilon)$-VLSF codes satisfies
	\begin{align}
	\log  M^*\left(N, L, \epsilon\right) &\geq { \frac{N C}{1-\epsilon}} - \sqrt{N \log _{(L-1)} (N) \frac{V}{1-\epsilon}} \notag \\
	&+ \bigo{\sqrt{\frac{N}{\log _{(L-1)} (N)}}}. \label{eq:mainresultK}
	\end{align}
	The decoding times $\{n_1, \dots, n_L\}$ that achieve \eqref{eq:mainresultK} satisfy the equations
	\begin{align}
	&\log  M = n_\ell C - \sqrt{n_\ell  \log _{(L -\ell + 1)}(n_\ell) V}  - \log {n_\ell} + O(1) \label{eq:decodingeq}
	\end{align}
	for $\ell \in \{2, \dots, L\}$, and $n_1 = 0$. 
	\end{theorem}
	 \begin{IEEEproof}[Proof sketch]
     Inspired by \cite[Th.~2]{polyanskiy2011feedback}, the proof employs a time-sharing strategy between an $(N', L-1, M, \epsilon_N')$-VLSF code whose smallest decoding time is nonzero and a simple ``stop-at-time-zero'' procedure that does not involve any code and decodes an error at time 0. Specifically, we set the VLSF code as the one that achieves the bound in \thmref{thm:nonasyPPC}, and we use the VLSF code and the stop-at-time-zero procedure with probabilities $1-p$ and $p$, respectively, where $p$ and $\epsilon_N'$ satisfy
	 \begin{align}
	     \epsilon_N' &= \frac{1}{\sqrt{N' \log  N'}} \label{eq:epsNprimeset}\\
	     p &= \frac{\epsilon - \epsilon_N'}{1-\epsilon_N'} \label{eq:pset}
	 \end{align}
The error probability of the resulting code is bounded by $\epsilon$, and the average decoding time is 
	 \begin{align}
	     N = N' (1 - p) = N' (1-\epsilon) + \bigo{\sqrt{\frac{N'}{\log  N'}}}. \label{eq:Nset}
	 \end{align}
For the scenario where $L = \infty$, we again use time-sharing with the stop-at-time-zero procedure in the achievability bound in \cite[Th.~2]{polyanskiy2011feedback} with $\epsilon_N' = \frac{1}{N'}$ instead of \eqref{eq:epsNprimeset}. In the asymptotic regime $L = O(1)$, the choice in \eqref{eq:epsNprimeset} results in a better second-order term than that achieved by $\epsilon_N' = \frac{1}{N'}$.
  
  In the analysis of \thmref{thm:nonasyPPC}, we need to bound the probability $\Prob{\imath(X^{n_L}; Y^{n_L}) < \gamma} = \epsilon_N'(1 - o(1))$. Since this probability decays sub-exponentially to zero due to \eqref{eq:epsNprimeset}, 
  we use a moderate deviations result from \cite[Ch. 8]{petrov1975} to bound this probability. Such a tool was not needed in the proof of \cite[Th.~2]{polyanskiy2011feedback} for $L = \infty$ because when $n_L = \infty$, the term $\Prob{\imath(X^{n_L}; Y^{n_L}) < \gamma}$ disappears from \eqref{eq:boundeps}, and the average decoding time is bounded via martingale analysis instead of \eqref{eq:boundN}. Finally, we apply Karush-Kuhn-Tucker conditions to show that the decoding times in \eqref{eq:decodingeq} yield a value of $\log  M$ that is the maximal value achievable by the non-asymptotic bound up to terms of order $\bigo{\sqrt{\frac{N}{\log _{(L-1)} (N)}}}$. 
  The details of the proof appear in \appref{app:proofmainPPC}. 
  	 \end{IEEEproof}

\begin{figure*}[!htbp]
        \center
        \includegraphics[width=0.95\linewidth]{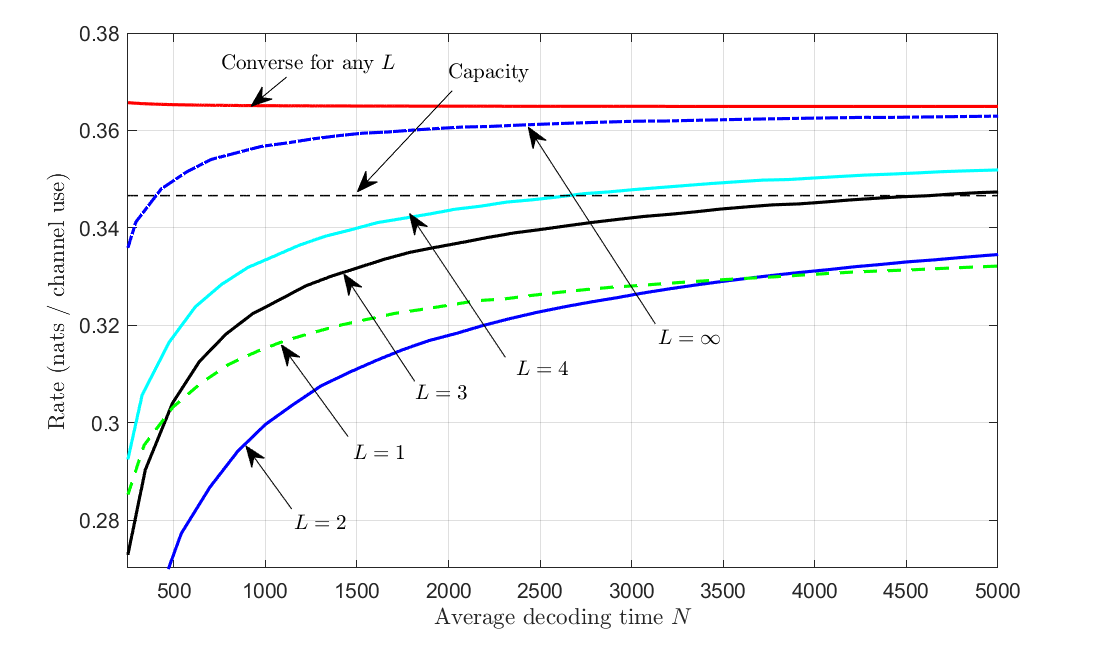}
        \caption{The non-asymptotic achievability bounds obtained from Theorems~\ref{thm:nonasyPPC} and \ref{thm:mainPPC} and the non-asymptotic converse bound \eqref{eq:feedb2} for the maximum achievable rate $\frac{\log  M^{*}(N, L, \epsilon)}{N}$ are shown for the BSC with crossover probability 0.11, $L \in \{1, 2, 3, 4, \infty\}$, and $\epsilon = 0.05$. The curves that $L = 1$ and $L = \infty$ are Polyanskiy \emph{et al.}'s achievability bounds from \cite[Th.~16]{polyanskiy2010Channel} and \cite[eq.~(102)]{polyanskiy2011feedback}, respectively.} 
        \label{fig:rates}
        \end{figure*}

  The non-asymptotic achievability bounds obtained from the coding scheme described in the proof sketch of \thmref{thm:mainPPC} are illustrated for the BSC in \figref{fig:rates}. For $L \in \{2, 3, 4\}$, the decoding times $n_1, \dots, n_L$ are chosen as described in \eqref{eq:decodingeq} with the $O(1)$ term ignored, and $\epsilon_N'$ in the stop-at-time-zero procedure is replaced with the right-hand side of \eqref{eq:boundeps}. For $L = 1$, \figref{fig:rates} shows the random coding union bound in \cite[Th.~16]{polyanskiy2010Channel}, which is a non-asymptotic achievability bound for fixed-length no-feedback codes. For $L = \infty$, \figref{fig:rates} shows the non-asymptotic bound in \cite[eq.~(102)]{polyanskiy2011feedback}. The curves for $L = 1$ and $L = 2$ cross because the choice of decoding times in \eqref{eq:decodingeq} requires $\epsilon \gg \frac{1}{\sqrt{N \log N}}$ and is optimal only as $N \to \infty$. 
 In \cite{yang2022}, Yang \emph{et al.} construct a computationally intensive integer program for the numerical optimization of the decoding times for finite $N$. If such a precise optimization is desired, our approximate decoding times in \eqref{eq:decodingeq} can be used as starting locations for that integer program.

    Replacing the information-density-based decoding rule in the proof sketch with the optimal SHT would improve the performance achieved on the right-hand side of \eqref{eq:mainresultK} by only $O(1)$.

	Since any $(N,L, M, \epsilon)$-VLSF code is also an $(N, \infty, M, \epsilon)$-VLSF code, \eqref{eq:feedb2} provides an upper bound on $\log  M^*(N, L, \epsilon)$ for an arbitrary $L$. 
	 The order of the second-order term, $-\sqrt{N \log _{(L-1)} (N) \frac{V}{1-\epsilon}}$, depends on the number of decoding times $L$. The larger $L$, the faster the achievable rate converges to the capacity. However, the dependence on $L$ is weak since $\log _{(L-1)} (N)$ grows very slowly in $N$ even if $L$ is small. For example, for $L =  4$ and $N = 1000$,  $\log _{(L-1)} (N) \approx 0.659$. For a finite $L$, this bound falls short of the $-\log  N$ achievability bound in \eqref{eq:feedb2} achievable with $L = \infty$. Whether the second-order term achieved in \thmref{thm:mainPPC} is tight remains an open problem. 
	
    The following theorem gives achievability and converse bounds for VLSF codes with decoding times uniformly spaced as $\{0, d_N, 2 d_N, \dots\}$.
    \begin{theorem} \label{thm:VLSFdN} Fix $\epsilon \in (0, 1)$.
    Let $d_N = o(N)$ with $d_N \to \infty$, and let $P_{Y|X}$ be any DM-PPC.  Then, it holds that
    \begin{align}
        \log  M^*(N, \infty, \epsilon, d_N \mathbb{Z}_\geq) &\geq \frac{N C}{1-\epsilon} - \frac{d_N C}{2} - \log  N + o(d_N). \label{eq:dNach}
    \end{align}
    If the DM-PPC $P_{Y|X}$ is a Cover--Thomas symmetric DM-PPC \cite[p.~190]{cover} i.e., the rows (and resp. the columns) of the transition probability matrix are permutations of each other, then
    \begin{align}
        \log  M^*(N, \infty, \epsilon, d_N \mathbb{Z}_\geq) &\leq \frac{N C}{1-\epsilon} - \frac{d_N C}{2} + o(d_N). \label{eq:dNconv}
    \end{align}
    \end{theorem}
    \begin{IEEEproof}[Proof Sketch]
    The achievability bound \eqref{eq:dNach} employs the sub-optimal SHT in the proof sketch of \thmref{thm:mainPPC}. To prove the converse in \eqref{eq:dNconv}, we first derive in \thmref{thm:metaconv}, in \appref{app:VLSFdN} below, the meta-converse bound for VLSF codes. The meta-converse bound in \thmref{thm:metaconv} bounds the error probability of any given VLSF code from below by the minimum achievable type-II error probability of the corresponding SHT; it is an extension and a tightening of Polyanskiy \emph{et al.}'s converse in \eqref{eq:feedb2} since for $d_N = 1$, weakening it by applying a loose bound on the performance of SHTs from \cite[Th.~3.2.2]{sequentialbook} recovers \eqref{eq:feedb2}. The Cover-Thomas symmetry assumption allows us to circumvent the maximization of that minimum type-II error probability over codes since the log-likelihood ratio $\log  \frac{P_{Y|X}(Y|x)}{P_Y(Y)}$ is the same regardless of the channel input $x$ for that channel class. In both bounds in \eqref{eq:dNach}--\eqref{eq:dNconv}, we use the expansions for the average stopping time and the type-II error probability from \cite[Ch.~2-3]{sequentialbook}. See \appref{app:VLSFdN} for details.
    \end{IEEEproof}
    
    \thmref{thm:VLSFdN} establishes that  when $\frac{d_N}{\log  N} \to \infty$, the second-order term of the logarithm of maximum achievable message set size among VLSF codes with uniformly spaced decoding times is $-\frac{d_N C}{2}$. \thmref{thm:VLSFdN} implies that in order to achieve the same performance  as achieved in \eqref{eq:mainresultK} with $L$ decoding times, one needs on average $\Omega\nB{\sqrt{\frac{N}{\log _{(L-1)}(N)}}}$ uniformly spaced stop-feedback instances, suggesting that the optimization of available decoding times considered in \thmref{thm:mainPPC} is crucial for attaining the second-order term in \eqref{eq:mainresultK}.
    
    The case where $d_N = \Omega(N)$ is not as interesting as the case where $d_N = o(N)$ since analyzing \thmref{thm:metaconv} using Chernoff bound would yield a bound on the probability that the optimal SHT makes a decision at times other than $n_1 = 0$ and $\frac{N}{1-\epsilon}(1 + o(1))$. Since that probability decays exponentially with $N$, the scenario where $L$ is unbounded and $d_N = \Omega(N)$ is asymptotically equivalent to $L = 2$. For example, for $d_N = \frac{1}{\ell} \frac{N}{1-\epsilon} \nB{1 + \bigo{\frac{1}{\sqrt{N \log  N}}}}$ for some $\ell \in \mathbb{Z}_+$, the right-hand side of \eqref{eq:mainresultK} is tight up to the second-order term.  
    
    \section{VLSF Codes for the DM-MAC} \label{sec:MAC}
    We begin by introducing the definitions used for the multi-transmitter setting.
    \subsection{Definitions}
    A $K$-transmitter DM-MAC is defined by a triple $\left( \prod_{k = 1}^K \mc{X}_k, P_{Y_K|X_{[K]}}, \mc{Y}_K\right)$, where $\mc{X}_k$ is the finite input alphabet for transmitter $k \in [K]$, $\mc{Y}_K$ is the finite output alphabet of the channel, and $P_{Y_K|X_{[K]}}$ is the channel transition probability.

    In what follows, the subscript and superscript indicate the corresponding transmitter indices and the codeword lengths, respectively.
    Let $P_{Y_K}$ denote the marginal output distribution induced by the input distribution $P_{X_{[K]}}$. The unconditional and conditional information densities are defined for each non-empty $\mc{A} \subseteq [K]$ as
\begin{align}
    \imath_{K}(x_{\mc{A}}; y) &\triangleq \log  \frac{P_{Y_K|X_{\mc{A}}}(y|x_{\mc{A}})}{P_{Y_K}(y)} \label{eq:ikA} \\
    \imath_{K}(x_{\mc{A}}; y | x_{\mc{A}^c}) &\triangleq \log  \frac{P_{Y_K|X_{[K]}}(y |x_{[K]})}{P_{Y_K | X_{\mc{A}^c}}(y | x_{\mc{A}^c})}, \label{eq:ikA2}
\end{align} 
where $\mc{A}^c = [K] \setminus \mc{A}$. Note that in \eqref{eq:ikA}--\eqref{eq:ikA2}, the information density functions depend on the transmitter set $\mc{A}$ unless further symmetry conditions are assumed (e.g., in some cases we assume that the components of $P_{X_{[K]}}$ are i.i.d., and $P_{Y_{K}|X_{[K]}}$ is invariant to permutations of the inputs $X_{[K]}$). 

The corresponding mutual informations under the input distribution $P_{X_{[K]}}$ and the channel transition probability $P_{Y_K|X_{[K]}}$ are defined as
\begin{align}
    I_K(X_{\mc{A}}; Y_K) &\triangleq \E{\imath_{K}(X_{\mc{A}}; Y_K)} \\
    I_K(X_{\mc{A}}; Y_K | {X_{\mc{A}^c}}) &\triangleq \E{\imath_{K}(X_{\mc{A}}; Y_K | X_{\mc{A}^c})}.
\end{align}
The dispersions are defined as
\begin{align}
    V_K(X_{\mc{A}}; Y_K) &\triangleq \Var{\imath_{K}(X_{\mc{A}}; Y_K)} \\
    V_K(X_{\mc{A}}; Y_K | {X_{\mc{A}^c}}) &\triangleq \Var{\imath_{K}(X_{\mc{A}}; Y_K | X_{\mc{A}^c})}.
\end{align}
For brevity, we define
\begin{align}
    I_K &\triangleq I_K(X_{[K]}; Y_K) \\
    V_K &\triangleq \Var{\imath_K(X_{[K]}; Y_K)}.
\end{align}

	A VLSF code for the MAC with $K$ transmitters is defined similarly to the VLSF code for the PPC. 
	\begin{definition}
    Fix $\epsilon \in (0, 1)$, $N \in (0, \infty)$, and positive integers $M_k, k \in [K]$. An $(N, L, M_{[K]}, \epsilon)$-VLSF code for the MAC comprises
    \begin{enumerate}
        \item non-negative integer-valued decoding times $n_1 < \cdots < n_L$,
        \item $K$ finite alphabets $\mc{U}_k$, $k \in [K]$, defining common randomness random variables $U_1, \dots, U_K$,
        \item $K$ sequences of encoding functions $\ms{f}_n^{(k)} \colon \mc{U}_k \times [M_k] \to \mc{X}_k$, $k \in [K]$,
        \item a stopping time $\tau \in \{n_1, \dots, n_L\}$ for the filtration generated by $\{U_1, \dots, U_K, Y_K^{n_\ell}\}_{\ell = 1}^L$, satisfying an average decoding time constraint \eqref{eq:averagetau}, and
        \item $L$ decoding functions
	$\mathsf{g}_{n_\ell} \colon \mc{U}_{[K]} \times \mc{Y}_K^{n_\ell} \to \prod \limits_{k = 1}^K [M_k] \cup \{\ms{e}\}$ for $\ell \in [L]$, satisfying an average error probability constraint
        \begin{align}
            \Prob{\mathsf{g}_{\tau}(U_{[K]}, Y_K^{\tau}) \neq W_{[K]}} \leq \epsilon, \label{eq:averageerrorMAC}
        \end{align}
        where the independent messages $W_1, \dots, W_K$ are uniformly distributed on the sets $[M_1], \dots, [M_K]$, respectively.
    \end{enumerate}
	\end{definition}

    \subsection{Our Achievability Bounds}	 
    Our main results are second-order achievability bounds for the rates approaching a point on the sum-rate boundary of the MAC achievable region expanded by a factor of $\frac{1}{1-\epsilon}$.
	 
    \thmref{thm:MACnonasy}, below, is a non-asymptotic achievability bound for any DM-MAC with $K$ transmitters and $L$ decoding times. 
	 \begin{theorem} 
	 \label{thm:MACnonasy}
		Fix constants $\epsilon \in (0, 1)$, $\gamma \in \mathbb{R}$, $\lambda^{(\mc{A})} > 0$ for $\mc{A} \in \mc{P}([K])$, integers $0 \leq n_1 < \cdots < n_L$, and distributions $P_{X_k}$, $k \in [K]$. For any DM-MAC with $K$ transmitters $(\prod_{k = 1}^K \mc{X}_k, P_{Y_K|X_{[K]}}, \mc{Y}_K)$, there exists an $(N, L, M_{[K]}, \epsilon)$-VLSF code with
	\begin{align}
	    &\epsilon \leq \Prob{\imath_{K}(X_{[K]}^{n_L}; Y_K^{n_L}) < \gamma} \label{eq:true} \\
	    &+ \prod_{k =1}^K (M_k-1) \exp\{-\gamma\} \label{eq:botherror}  \\
	    &+ \sum_{\ell = 1}^L \sum_{\mc{A} \in \mc{P}([K])} \Prob{\imath_{K}(X_{\mc{A}}^{n_\ell}; Y_K^{n_\ell}) > N (I_K(X_{\mc{A}}; Y) + \lambda^{(\mc{A})})} \label{eq:1errorS}\\
	    &+ \sum_{ \mc{A} \in \mc{P}([K])} \left(\prod_{k \in \mc{A}^{\mr{c}}} (M_k - 1) \right) \notag \\
	    &\quad \quad \quad \exp\{-\gamma + N I_K(X_{\mc{A}}; Y_K) + N\lambda^{(\mc{A})}\} \label{eq:1errorExp} \\
	    &N \leq n_1 + \sum_{\ell = 1}^{L-1} (n_{\ell + 1} - n_\ell) \Prob{\imath_{K}(X_{[K]}^{n_\ell}; Y_K^{n_\ell}) < \gamma}.  \label{eq:boundNMAC}
	\end{align}
	\end{theorem}
    \begin{IEEEproof}[Proof sketch]
        The proof of \thmref{thm:MACnonasy} uses a random coding argument that employs $K$ independent codebook ensembles, each with distribution $P_{X_k}^{n_L}$, $k \in [K]$. The receiver employs $L$ decoders that operate by comparing an information density $\imath_K(x_{[K]^{n_{\ell}}}; y^{n_{\ell}})$ for each possible transmitted codeword set to a threshold. At time $n_{\ell}$, decoder $g_{n_\ell}$ computes the information densities $\imath_{K}(X_{[K]}^{n_\ell}(m_{[K]}); Y_K^{n_{\ell}})$; if there exists a unique message vector $\hat{m}_{[K]}$ satisfying $\imath_{K}(X_{[K]}^{n_\ell}(\hat{m}_{[K]}); Y_K^{n_{\ell}}) > \gamma$, then the receiver decodes to the message vector $\hat{m}_{[K]}$; if there exists multiple such message vectors, then the receiver stops the transmission and decodes an error. If no such message vectors exist at time $n_{\ell}$, then the receiver emits output $\ms{e}$ and passes the decoding time $n_\ell$ without decoding if $n_\ell < n_L$ and decodes an error if $n_\ell = n_L$. The term \eqref{eq:true} bounds the probability that the information density corresponding to the true messages is below the threshold for all decoding times;  \eqref{eq:botherror} bounds the probability that all messages are decoded incorrectly; and \eqref{eq:1errorS}-\eqref{eq:1errorExp} bound the probability that the messages from the transmitter index set $\mc{A} \subseteq [K]$ are decoded incorrectly, and the messages from the index set $\mc{A}^{c}$ are decoded correctly. The proof of Theorem~\ref{thm:MACnonasy} appears in \appref{app:MACproofs}.
    \end{IEEEproof}

    Theorem \ref{thm:MAC}, below, is a second-order achievability bound in the asymptotic regime $L = O(1)$ for any DM-MAC. It follows from an application of \thmref{thm:MACnonasy}.
    
	 \begin{theorem} \label{thm:MAC}
	 Fix $\epsilon \in (0, 1)$, an integer ${{L = O(1) \geq 2}}$, and distributions $P_{X_k}$, $k \in [K]$. For any $K$-transmitter DM-MAC $(\prod_{k = 1}^K \mc{X}_k, P_{Y_K|X_{[K]}}, \mc{Y}_K)$, there exists a $K$-tuple $M_{[K]}$ and an $(N, L, M_{[K]}, \epsilon)$-VLSF code satisfying
	\begin{align}
	\sum_{k \in [K]} \log  M_k &= { \frac{N I_K}{1-\epsilon}} - \sqrt{N \log _{(L-1)} (N) \frac{V_K}{1-\epsilon}} \notag \\
	& \quad + \bigo{\sqrt{\frac{N}{\log _{(L-1)} (N)}}}. \label{eq:mainresultMAC}
	\end{align}
	 \end{theorem}
  \begin{IEEEproof}
      See \appref{app:MACproofs}.
  \end{IEEEproof}

    In the application of \thmref{thm:MACnonasy} to prove Theorem~\ref{thm:MAC}, we choose the parameters $\lambda^{(\mc{A})}$ and $\gamma$ so that the terms in \eqref{eq:1errorS}-\eqref{eq:1errorExp} decay exponentially with $N$, which become negligible compared to \eqref{eq:true} and \eqref{eq:botherror}. Between \eqref{eq:true} and \eqref{eq:botherror}, the term \eqref{eq:true} is dominant when $L$ does not grow with $N$, and \eqref{eq:botherror} is dominant when $L$ grows linearly with $N$. 
    
    Like the single-threshold rule from \cite{yavas2020Random} for the RAC, the single-threshold rule employed in the proof of \thmref{thm:MACnonasy} differs from the decoding rules employed in \cite{truong2018Journal} for VLSF codes over the Gaussian MAC with expected power constraints and in \cite{trillingsgaard2014VLF} for the DM-MAC. In both \cite{truong2018Journal} and \cite{trillingsgaard2014VLF}, $L = n_{\max} = \infty$, and the decoder employs $2^K - 1$ simultaneous threshold rules for each of the boundaries that define the achievable region of the MAC with $K$ transmitters. Those rules fix thresholds $\gamma^{(\mc{A})}$, $\mc{A} \in \mc{P}([K])$, and decode messages $m_{[K]}$ if for all $\mc{A} \in \mc{P}([K])$, the codeword for $m_{[K]}$ satisfies
    \begin{align}
        \imath_{K}(X_{\mc{A}}^{n_\ell}(m_{\mc{A}}); Y_K^{n_{\ell}} | X_{\mc{A}^{c}}^{n_\ell}(m_{\mc{A}^{c}})) &> \gamma^{(\mc{A})},\label{eq:thres1}
    \end{align}
    for some $\gamma^{(\mc{A})}$, $\mc{A} \in \mc{P}([K])$.  
    Our decoder can be viewed as a special case of \eqref{eq:thres1} obtained by setting $\gamma^{(\mc{A})} = -\infty$ for $\mc{A} \neq [K]$.
    
    Analyzing \thmref{thm:MACnonasy} in the asymptotic regime $L = \Omega(N)$, we determine that there exists a $K$-tuple $M_{[K]}$ and an $(N, \infty, M_{[K]}, \epsilon)$-VLSF code satisfying
	\begin{align}
	\sum_{k \in [K]} \log  M_k &= { \frac{N I_K}{1-\epsilon}} - \log  N + O(1). \label{eq:mainresultMACinf}
	\end{align}
	
	Both \eqref{eq:mainresultMAC} and \eqref{eq:mainresultMACinf} are achieved at rate points that approach a point on the sum-rate boundary of the $K$-MAC achievable region expanded by a factor of $\frac{1}{1-\epsilon}$.

	For any VLSF code, $L = \infty$ case can be treated as $L = \Omega(N)$ regardless of the number of transmitters since if we truncate an infinite-length code at time $n_{\max} = 2 N$, by Chernoff bound, the resulting penalty term added to the error probability decays exponentially with $N$, whose effect in \eqref{eq:mainresultMACinf} is $o(1)$. 
	See \appref{app:proofMACinf} for the proof of~\eqref{eq:mainresultMACinf}. 
	
    For $L = n_{\max} = \infty$, Trillingsgaard and Popovski  \cite{trillingsgaard2014VLF} numerically evaluate their non-asymptotic achievability bound for a DM-MAC while Truong and Tan \cite{truong2018Journal} provide an achievability bound with second-order term $-O(\sqrt{N})$ for the Gaussian MAC with average power constraints. Applying our single-threshold rule and analysis to the Gaussian MAC with average power constraints improves the second-order term in \cite{truong2018Journal} from $-O(\sqrt{N})$ to $-\log  N + O(1)$ for all non-corner points in the achievable region. The main challenge in \cite{truong2018Journal} is to derive a tight bound on the expected value of the maximum over $\mc{A} \subseteq [K]$ of stopping times $\tau^{(\mc{A})}$ for the corresponding threshold rules in \eqref{eq:thres1}. In our analysis, we avoid that challenge by employing a single-threshold decoder whose average decoding time is bounded by $\E{\tau^{([K])}}$.

    Under the same model and assumptions on $L$, to achieve non-corner rate points that do not lie on the sum-rate boundary, we modify our single-threshold rule to \eqref{eq:thres1}, where $\mc{A}$ is the transmitter index set corresponding to the capacity region's active sum-rate bound at the (non-corner) point of interest. Following steps similar to the proof of \eqref{eq:mainresultMACinf} gives second-order term $-\log  N + O(1)$ for those points as well. For corner points, more than one boundary is active\footnote{The capacity region of a $K$-transmitter MAC is characterized by the region bounded by $2^{K}-1$ planes. By definition of a corner point, at least two inequalities corresponding to these planes are active at a corner point.}; therefore, more than one threshold rule in \eqref{eq:thres1} is needed at the decoder. In this case, again for $L = \infty$, \cite{truong2018Journal} proves an achievability bound with a second-order term $-O(\sqrt{N})$. Whether this bound can be improved to $-\log  N + O(1)$ as in \eqref{eq:mainresultMACinf} remains an open problem.

    \section{VLSF Codes for the DM-RAC with at Most $K$ Transmitters} \label{sec:RAC}
    \begin{definition}[Yavas \emph{et al.} {{\cite[eq.~(1)]{yavas2020Random}}}]
A permutation-invariant, reducible DM-RAC for the maximal number of transmitters $K < \infty$ is defined by a family of DM-MACs $\left\{\left(\mc{X}^k, P_{Y_k|X_{[k]}}, \mc{Y}_k \right)\right\}_{k = 0}^K$, where the $k$-th DM-MAC defines the channel for $k$ active transmitters. 

By assumption, each of the DM-MACs satisfies the \emph{permutation-invariance} condition
\begin{align}
P_{Y_k | X_{[k]}}(y | x_{[k]}) = P_{Y_k | X_{[k]}}(y | x_{\pi[k]})
\label{eq:permutationinvariance}
\end{align}
for all permutations $\pi[k]$ of $[k]$, and $y \in \mathcal{Y}_k$, and
the \emph{reducibility} condition
\begin{align}
 P_{Y_s | X_{[s]}}(y|x_{[s]}) = P_{Y_k
| X_{[k]}}(y|x_{[s]},0^{k-s}) \quad 
\label{eq:reducibility}
\end{align}
for all $s < k$, $x_{[s]}\in\cX_{[s]}$, and $y\in \cY_s$, where $0 \in \mc{X}$ specifies a unique ``silence" symbol that is transmitted when a transmitter is silent.
\end{definition}

The permutation-invariance \eqref{eq:permutationinvariance} and reducibility \eqref{eq:reducibility} conditions simplify the presentation and enable us to show, using a single-threshold rule at the decoder \cite{yavas2020Random}, that the symmetrical rate point $(R, R, \dots, R)$ at which the code operates lies on the sum-rate boundary of each of the underlying DM-MACs,

		The VLSF RAC code defined here combines our rateless communication strategy from \cite{yavas2020Random} with the sparse feedback VLSF PPC and MAC codes with optimized average decoding times described above. Specifically, the decoder estimates the value of $k$ at time $n_0$. If the estimate $\hat{k}$ is not zero, it decodes at one of the $L$ decoding times $n_{\hat{k},1} < n_{\hat{k},2} < \dots < n_{\hat{k},L}$ (rather than just the single time $n_{\hat{k}}$ used in \cite{yavas2020Random, yavas2021Gaussian}).
		For every $k \in [K]$, the locations of the $L$ decoding times are optimized to attain the minimum average decoding delay. As in \cite{yavas2020Random}, we do not assume any probability distribution on the user activity pattern. We seek instead to optimize the rate-reliability trade-off simultaneously for all possible activity patterns. (By the permutation-invariance assumption, there are only $K$ distinguishable activity patterns to consider here indexed by the number of active transmitters.)
		If the decoder concludes that no transmitters are active, then it ends the transmission at time $n_0$ decoding no messages. At each time $n_{i, \ell}$, $i < \hat{k}$, the receiver broadcasts ``0'' to the transmitters, signaling that they should continue to transmit. At time $n_{\hat{k}, \ell}$, the receiver broadcasts feedback bit ``1" to the transmitters if it is able to decode $\hat{k}$ messages; otherwise, it outputs an erasure symbol ``$\ms{e}$" and sends feedback bit ``0", again signaling that decoding has not occurred and transmission should continue.  
	
	As in \cite{yavas2020Random, yavas2021Gaussian}, we assume that the transmitters know nothing about the set $\mc{A}$ except their own membership and the receiver's feedback at potential decoding times. We employ identical encoding \cite{polyanskiy2017perspective}, that is, all transmitters use the same codebook. This implies that the RAC code operates at the symmetrical rate point, i.e., $M_i = M$ for $i \in [K]$. As in \cite{polyanskiy2017perspective, yavas2020Random}, the decoder is required to decode the list of  messages transmitted by the active transmitters but not the identities of these transmitters.
 
    To deal with the scenario where the number of transmitters in the RAC grows linearly with the blocklength, i.e., $K = \Omega(N)$, \cite{polyanskiy2017perspective} employs the per-user error probability (PUPE) constraint rather than the joint error probability used here and in the analysis of the MAC (e.g., \cite{truong2018Journal, yavas2020Random, yavas2021Gaussian}). The PUPE is a weaker error probability constraint since, under PUPE, an error at one decoder does not count as an error at all other decoders.
    In \cite{yavas2020Random}, it is shown that when $K = O(1)$, PUPE and joint error probability constraints have the same second-order performance for random access coding. As a result, there is no advantage to using PUPE rather than the more stringent joint error criterion when $K = O(1)$. Therefore, we employ the joint error probability constraint throughout. 
    
    We formally define VLSF codes for the RAC as follows. 
	\begin{definition} \label{def:RAC}
	Fix $\epsilon_0, \dots, \epsilon_K \in (0, 1)$ and  $N_0, \dots, N_K \in (0, \infty)$. An $(\{N_k\}_{k = 0}^K, L, M, \{\epsilon_k\}_{k = 0}^K)$-VLSF code with identical encoders comprises
	\begin{enumerate}
	\item a set of integers $\mc{N} \triangleq \{n_0\} \cup \{n_{k, \ell} \colon k \in [K], \ell \in [L]\}$ (without loss of generality, we assume that $n_{K, L}$ is the largest available decoding time),
	\item a common randomness random variable $U$ on an alphabet~$\mc{U}$,
	\item a sequence of encoding functions $\mathsf{f}_n \colon \mc{U} \times [M] \to \mc{X}$, $n = 1, 2, \ldots, n_{K, L}$, defining $M$ length-$n_{K, L}$ codewords,
	 	\item $K$ non-negative integer-valued random stopping times $\tau_k \in \mc{N}$ for the filtration generated by $\{U, Y_k^{n}\}_{n \in \mc{N}}$, satisfying 
	\begin{align}
	    \E{\tau_k} \leq N_k \label{eq:averagetauRAC}
	\end{align}
	if $k \in \{0\} \cup [K]$ transmitters are active, and
	\item $KL + 1$ decoding functions $\ms{g}_{n_0} \colon \mc{U} \times \mc{Y}_0^{n_0} \to \{\emptyset\} \cup \{\ms{e}\}$ and
	$\mathsf{g}_{n_{k,  \ell}} \colon \mc{U} \times \mc{Y}_k^{n_{k,  \ell}} \to [M]^k \cup \{\ms{e}\}$, $k \in [K]$ and $\ell \in [L]$, satisfying an average error probability constraint
   \begin{align}
        \Prob{\mathsf{g}_{\tau_k}(U, Y_k^{\tau_k}) \stackrel{\pi}{\neq} W_{[k]}} \leq \epsilon_k \label{eq:averageerrorRAC}
    \end{align}
    when $k \in [K]$ messages $W_{[k]} = (W_1, \dots, W_k)$ are transmitted, where
    $W_1, \dots, W_k$ are independent and equiprobable on the set~$[M]$, and
    \begin{align}
        \Prob{\mathsf{g}_{\tau_0}(U, Y_0^{\tau_0}) \neq \emptyset} \leq \epsilon_0
    \end{align}
    when no transmitters are active.
	\end{enumerate}
	\end{definition}

	To guarantee that the symmetrical rate point arising from identical encoding lies on the sum-rate boundary for all $k \in [K]$, following \cite{yavas2020Random}, we assume that there exists an input distribution $P_X$ that satisfies the interference assumptions 
	\begin{align}
    P_{X_{[t]} | Y_k} \neq P_{X_{[s]}|Y_k} \, P_{X_{[s+1:t]}|Y_k} \quad \forall \,  s < t \leq k \leq K. \label{eq:interference}
    \end{align}
    Permutation-invariance \eqref{eq:permutationinvariance}, reducibility \eqref{eq:reducibility}, and interference \eqref{eq:interference} together imply that the mutual information per transmitter, $\frac{I_k}{k}$, strictly decreases with increasing $k$ (see {\cite[Lemma~1]{yavas2020Random}}). This property guarantees the existence of decoding times satisfying $n_{k_1, \ell_1} < n_{k_2, \ell_2}$ for any $k_1 < k_2$ and $\ell_1, \ell_2 \in [L]$.
    
    In order to be able to detect the number of active transmitters using the received symbols $Y^{n_{k, \ell}}$ but not the codewords themselves, we require that the input distribution $P_X$ satisfies the \emph{distinguishability} assumption
    \begin{align}
        P_{Y_{k_1}} \neq P_{Y_{k_2}} \quad \forall \, k_1 \neq k_2 \in \{0\} \cup [K], \label{eq:distinctoutput}
    \end{align}
    where $P_{Y_k}$ is the marginal output distribution under the $k$-transmitter DM-MAC with input distribution $P_{X_{[k]}} = (P_X)^k$.
    
	An example of a permutation-invariant and reducible DM-RAC that satisfies interference \eqref{eq:interference} and distinguishability~\eqref{eq:distinctoutput} assumptions is the adder-erasure RAC~in~\cite{ebrahimi2017coded, yavas2020Random}
    \begin{align}
    Y_k =
    \begin{cases}
    \sum_{i = 1}^{k} X_i, & \text{w.p. } 1 - \delta\\
    \mathsf e & \text{w.p. } \delta,
    \end{cases}
    \label{eq:addererasure}
    \end{align}
    where $X_i \in \{0, 1\}$, $Y_k \in \left\{ 0, \ldots, k\right\} \cup \{\mathsf e\}$, and $\delta \in (0, 1)$.

	\begin{theorem}\label{thm:RAC}
	Fix  $\epsilon \in (0, 1)$, finite integers $K \geq 1$ and $L \geq 2$, and a distribution $P_X$ satisfying~\eqref{eq:interference}--\eqref{eq:distinctoutput}. For any permutation-invariant~\eqref{eq:permutationinvariance} and reducible~\eqref{eq:reducibility} DM-RAC $\left\{(\mc{X}^k, P_{Y_k|X_{[k]}}, \mc{Y}_k)\right\}_{k = 0}^K$, there exists an $(\{N_k\}_{k = 0}^K, L, M, \{\epsilon_k\}_{k = 0}^K)$-VLSF code satisfying  
	\begin{align}
	k \log  M &= { \frac{N_k I_k}{1-\epsilon_k}} - \sqrt{N_k \log _{(L-1)} (N_k) \frac{V_k}{1-\epsilon_k}} \notag \\
	&\quad + \bigo{\sqrt{\frac{N_k}{\log _{(L)} (N_k)}}} \label{eq:mainresultRAC}
	\end{align}
	for $k \in [K]$,
	and
	\begin{align}
	    N_0 = c \log  N_1 + o(\log  N_1)
	\end{align}
	for some $c > 0$.
	
	\end{theorem}

\begin{IEEEproof}[Proof sketch]
The coding strategy to prove \thmref{thm:RAC} is as follows. The decoder applies a $(K+1)$-ary hypothesis test using the output sequence $Y^{n_0}$ and decides an estimate $\hat{k}$ of the number of active transmitters $k \in \{0, 1, \dots, K\}$. If the hypothesis test declares that $\hat{k} = 0$, then the receiver stops the transmission at time $n_0$, decoding no messages. If $\hat{k} \neq 0$, then the receiver decodes $\hat{k}$ messages at one of the times $n_{\hat{k}, 1}, \dots, n_{\hat{k}, L}$ using the VLSF code in \thmref{thm:MAC} for the $\hat{k}$-transmitter DM-MAC with $L$ decoding times. If the receiver decodes at time $n_{\hat{k}, \ell}$, then it sends feedback bit `0' at all previous decoding times $\{n \in \mc{N} \colon n < n_{\hat{k}, \ell}\}$ and feedback bit `1' at time $n_{\hat{k}, \ell}$.
	    Note that alternatively, the receiver can send its estimate $\hat{k}$ using $\lceil \log_2 (K+1) \rceil + L$ bits at time $n_0$, informing the transmitters that it will decode at some time $\{n_{\hat{k}, 1}, \dots, n_{\hat{k}, L}\}$; in this case, the number of feedback bits decreases from the worst-case $KL + 1$ that results from the strategy described above. The details of the proof appear in \appref{app:proof:RAC}.
\end{IEEEproof}

	     \section{VLSF Codes for the Gaussian PPC with Maximal Power Constraints} \label{sec:GaussianPPC}
	     \subsection{Gaussian PPC}
	     The output of a memoryless, Gaussian PPC of blocklength $n$ in response to the input $X^n \in \mathbb{R}^n$ is
        \begin{align}
            Y^n = X^n + Z^n, \label{eq:pointchannel}
        \end{align}
        where $Z_1, \ldots, Z_n$ are drawn i.i.d. from $\mathcal{N}(0, 1)$, independent of $X^n$.
        
        The channel's capacity $C(P)$ and dispersion $V(P)$ are
        \begin{align}
            C(P) &= \frac{1}{2} \log  (1 + P) \\
            V(P) &= \frac{P (P + 2)}{2 (1+P)^2}.
        \end{align}
    
    \subsection{Related Work on the Gaussian PPC}
    We first introduce the maximal and average power constraints on VLSF codes for the PPC. Given a VLSF code with $L$ decoding times $n_1, \dots, n_L$, the maximal power constraint requires that the length-$n$ prefixes, $n \in \{n_1, \dots, n_L\}$, of each codeword all satisfy a power constraint $P$, i.e.,
	\begin{IEEEeqnarray}{rCl}
	\norm{\ms{f}(u, m)^{n_\ell}}^2 \leq n_\ell P \,\text{ for all } m \in [M], u \in \mc{U}, \quad \ell \in [L]. \IEEEeqnarraynumspace  \label{eq:maxpower}.
	\end{IEEEeqnarray}
	The average power constraint on the length-$n_L$ codewords, as defined by \cite[Def.~1]{truong2018Journal}, is
    \begin{align}
	    \E{\norm{\mathsf{f}(U, W)^{n_L}}^2} 
	    &\leq N P. \label{eq:exppower}
	\end{align}
The definitions of $(N, L,M, \epsilon, P)_{\mr{max}}$ and $(N, L,M, \epsilon, P)_{\mr{ave}}$-VLSF codes for the Gaussian PPC are similar to Definition~\ref{def:gaussvlf} with the addition of maximal \eqref{eq:maxpower} and average \eqref{eq:exppower} power constraints, respectively. Similar to \eqref{eq:Mstarmax}, $M^*(N, L, \epsilon, P)_{\mathrm{max}}$ (resp.  $M^*(N, L, \epsilon, P)_{\mathrm{ave}}$) denotes the maximum achievable message set size with $L$ decoding times, average decoding time $N$, average error probability $\epsilon$, and maximal (resp. average) power constraint $P$.

	In the following, we discuss prior asymptotic expansions of $M^*(N, L, \epsilon, P)_{\mathrm{max}}$ and $M^*(N, L, \epsilon, P)_{\mathrm{ave}}$ for the Gaussian PPC, where $L \in \{1, \infty\}$.
	\begin{enumerate}[a)]
	    \item $M^*(N, 1, \epsilon, P)_{\mathrm{max}}$: For $L = 1$, $P > 0$, and $\epsilon \in (0, 1)$, Tan and Tomamichel {\cite[Th.~1]{tan2015Third}} and Polyanskiy \textit{et al.} {\cite[Th.~54]{polyanskiy2010Channel}} show that
	\begin{IEEEeqnarray}{rCl}
	    \IEEEeqnarraymulticol{3}{l}{\log  M^*(N, 1, \epsilon, P)_{\mr{max}}} \notag \\
	    &=& N C(P) - \sqrt{N V(P)} Q^{-1}(\epsilon) + \frac 1 2 \log  N + O(1). \IEEEeqnarraynumspace \label{eq:K1}
	\end{IEEEeqnarray}
	The converse for \eqref{eq:K1} is derived in \cite[Th.~54]{polyanskiy2010Channel} and the achievability for \eqref{eq:K1} in
  \cite[Th.~1]{tan2015Third}. The achievability scheme in \cite[Th.~1]{tan2015Third} generates i.i.d. codewords uniformly distributed on the $n$-dimensional sphere with radius $\sqrt{nP}$, and applies maximum likelihood (ML) decoding. These results imply that random codewords uniformly distributed on a sphere and ML decoding are, together, third-order optimal, meaning that the gap between the achievability and converse bounds in \eqref{eq:K1} is $O(1)$.
  \item $M^*(N, 1, \epsilon, P)_{\mathrm{ave}}$: For $L =1$ with an average-power-constraint, Yang \textit{et al.} show in \cite{yang2015Optimal} that
  \begin{align}
         &\log  M^*(N, 1, \epsilon, P)_{\mathrm{ave}} = N\, C\nB{\frac{P}{1-\epsilon}} - \notag \\
         &\quad - \sqrt{N \log  N\, V\nB{\frac{P}{1-\epsilon}}} + O(\sqrt{N}). \IEEEeqnarraynumspace\label{eq:Yang}
  \end{align}
  Yang \textit{et al.} use a power control argument to show the achievability of \eqref{eq:Yang}. They divide the messages into disjoint sets $\mathcal{A}$ and $[M] \setminus \mathcal{A}$, where $|\mathcal{A}| =  M (1-\epsilon)(1 - o(1))$. For the messages in $\mc{A}$, they use an $\left(N, 1, |\mc{A}|, \frac{2}{\sqrt{N \log  N}}, \frac{P}{1-\epsilon}(1 - o(1)) \right)$-VLSF code with a single decoding time $N$. The codewords are generated i.i.d. uniformly on the sphere with center at 0 and radius $\sqrt{N \frac{P}{1-\epsilon}(1 - o(1))}$. The messages in $[M] \setminus \mathcal{A}$ are assigned the all-zero codeword. The converse for \eqref{eq:Yang} follows from an application of the meta-converse \cite[Th.~26]{polyanskiy2010Channel}. 
   
	\item $M^*(N, \infty, \epsilon, P)_{\mathrm{ave}}$: For VLSF codes with $L =  n_{\max} = \infty$ and average power constraint \eqref{eq:exppower},  Truong and Tan show in \cite[Th.~1]{truong2016gaussian} that 
	for $\epsilon \in (0,1)$ and $P > 0$, 
    \begin{align}
	 \log  M^*(N, \infty, \epsilon,  P)_{\mathrm{ave}} &\geq \frac{N C(P)}{1 - \epsilon} - \log  N + O(1) 
	 \label{eq:truongach}\\
	 \log  M^*(N, \infty, \epsilon,  P)_{\mathrm{ave}} &\leq \frac{N C(P) }{1-\epsilon} + \frac{ h_b(\epsilon)}{1-\epsilon}, \label{eq:truongconv}
	\end{align}
	where $h_b$ is the binary entropy function. 
	The results in \eqref{eq:truongach}--\eqref{eq:truongconv} are analogous to the fundamental limits for DM-PPCs \eqref{eq:feedb1}--\eqref{eq:feedb2} and follow from arguments similar to those in \cite{polyanskiy2011feedback}. 
	Since the information density $\imath(X; Y)$ for the Gaussian channel is unbounded, bounding the expected value of the decoding time in the proof of
  \cite[Th.~1]{truong2016gaussian} requires different techniques from those applicable to DM-PPCs \cite{polyanskiy2011feedback}.
	
	\end{enumerate}

\begin{table*}[htbp!] \label{table:summary}
\renewcommand{\arraystretch}{2.2}
\caption{The performance of VLSF codes for the Gaussian channel in scenarios distinguished by the number of available decoding times $L$, the type of the power constraint, and the presence of feedback.}
\centering
\begin{tabular}{|c|c|c|c|c|c|}
\hline
\multicolumn{3}{|l|}{\multirow{2}{*}{}}                                                                                               & \multirow{2}{*}{First-order term}        & \multicolumn{2}{c|}{Second-order term}                                                                      					\\  \cline{5-6} 
\multicolumn{3}{|l|}{}                                                                                                                &                                          & Lower Bound                                          				& Upper Bound                                          \\  \hline \hline 
\multirow{4}{*}{\begin{tabular}[c|]{@{}c@{}}Fixed-length\\ $(L = 1)$\end{tabular}}         & \multirow{2}{*}{No Feedback} & Max. power & $N C(P)$                                 & $-\sqrt{N V(P)}Q^{-1}(\epsilon)$\quad(\hspace{1sp}\cite{tan2015Third, polyanskiy2010Channel})     & $-\sqrt{N V(P)}Q^{-1}(\epsilon)$\quad(\hspace{1sp}\cite{polyanskiy2010Channel})                     \\   \cline{3-6}  
                                                                                          &                              & Ave. power & $N C\left( \frac{P}{1-\epsilon} \right)$ & $-\sqrt{N \log  N V\left(\frac{P}{1-\epsilon}\right)}$\quad(\hspace{1sp}\cite{yang2015Optimal}) 				& $-\sqrt{N \log  N V\left(\frac{P}{1-\epsilon}\right)}$\quad(\hspace{1sp}\cite{yang2015Optimal}) \\  \cline{2-6}  
                                                                                          & \multirow{2}{*}{Feedback}    & Max. power & $N C(P)$                                 & $-\sqrt{N V(P)}Q^{-1}(\epsilon)$  \quad(\hspace{1sp}\cite{tan2015Third, polyanskiy2010Channel})    & $-\sqrt{N V(P)}Q^{-1}(\epsilon)$\quad(\hspace{1sp}\cite{fong2015feedbacknot})                    \\  \cline{3-6}  
                                                                                          &                              & Ave. power & $N C\left( \frac{P}{1-\epsilon} \right)$ & $-O(\log _{(L)}(N))$\quad(\hspace{1sp}\cite{truongfong2017logL})                                				        & $+\sqrt{N \log  N V\left(\frac{P}{1-\epsilon}\right)}$\quad(\hspace{1sp}\cite{truongfong2017logL}) \\  \hline \hline
\multirow{2}{*}{\begin{tabular}[c|]{@{}c@{}}Variable-length\\ $(L < \infty)$\end{tabular}} & \multicolumn{2}{c|}{Max. power}           & $\frac{ N C(P)}{1-\epsilon}$             &$-\sqrt{N \log _{(L-1)}(N) \frac{V(P)}{1-\epsilon}}$\quad(\hspace{1sp}\thmref{thm:Gaussian})  				&$+O(1)$\quad(\hspace{1sp}\cite{truong2016gaussian})                                          \\  \cline{2-6}  
                                                                                          & \multicolumn{2}{c|}{Ave. power}           & $\frac{ N C(P)}{1-\epsilon}$             &$-\sqrt{N \log _{(L-1)}(N) \frac{V(P)}{1-\epsilon}}$\quad(\hspace{1sp}\thmref{thm:Gaussian})   				& $+O(1)$\quad(\hspace{1sp}\cite{truong2016gaussian})                                            \\  \hline \hline 
\multirow{2}{*}{\begin{tabular}[c|]{@{}c@{}}Variable-length\\ $(L = n_{\max} = \infty)$\end{tabular}} & \multicolumn{2}{c|}{Max. power}           & $\frac{ N C(P)}{1-\epsilon}$             &$-O(\sqrt{N})$\quad(\hspace{1sp}\cite{yavas2021VLF})    				& $+O(1)$\quad(\hspace{1sp}\cite{truong2016gaussian})                                               \\  \cline{2-6} 
                                                                                          & \multicolumn{2}{c|}{Ave. power}           & $\frac{ N C(P)}{1-\epsilon}$             &$-\log  N$\quad\quad(\hspace{1sp}\cite{truong2016gaussian})                                               				& $+O(1)$\quad(\hspace{1sp}\cite{truong2016gaussian})                                              \\  \hline 	
\end{tabular}

\end{table*}

	Table II combines the $L = 1$ summary from \cite[Table~I]{truongfong2017logL} with the corresponding results for $L > 1$ to summarize the performance of VLSF codes for the Gaussian channel in different communication scenarios.
	
	\subsection{Main Result}
	The theorem below is our main result for the Gaussian PPC under the maximal power constraint \eqref{eq:maxpower}.
		\begin{theorem}\label{thm:Gaussian}
	Fix an integer $L = O(1) \geq 2$ and real numbers $P > 0$ and $\epsilon \in (0, 1)$. For the Gaussian channel with maximal power constraint \eqref{eq:maxpower}, the maximum message set size achievable by $(N, L,M, \epsilon, P)$-VLSF codes satisfies
	\begin{align}
	\log  M^*\left(N, L, \epsilon, P\right)_{\max} &\geq { \frac{N C(P)}{1-\epsilon}} - \sqrt{N \log _{(L-1)} (N) \frac{V(P)}{1-\epsilon}} \notag \\
	&+ \bigo{\sqrt{\frac{N}{\log _{(L-1)} (N)}}}. \label{eq:mainresGaussian}
	\end{align}
	The decoding times that achieve \eqref{eq:mainresGaussian} satisfy the equations
	\begin{align}
	&\log  M^*\left(N, L, \epsilon, P\right) \notag \\
	&= n_\ell C(P) - \sqrt{n_\ell  \log _{(L -\ell + 1)}(n_\ell) V(P)}  - \log {n_\ell} + O(1)
	\end{align}
	for $\ell \in \{2, \dots, L\}$, and $n_1 = 0$.
	\end{theorem}
	 \begin{IEEEproof}
	 See \appref{app:GaussianPPCproof}.
	 \end{IEEEproof}

  Note that the achievability bound in \thmref{thm:Gaussian} has the same form as the one in \thmref{thm:mainPPC} with $C$ and $V$ replaced with the Gaussian capacity $C(P)$ and the Gaussian dispersion $V(P)$, respectively. The bound in \eqref{eq:mainresGaussian} holds for the average power constraint as well since any code that satisfies the maximal power constraint also satisfies the average power constraint.

  From Shannon's work in \cite{shannon1959Probability}, it is known that for the Gaussian channel with a maximal power constraint, drawing i.i.d. Gaussian codewords yields a performance inferior to that achieved by the uniform distribution on the power sphere. As a result, almost all tight achievability bounds for the Gaussian channel in the fixed-length regime under a variety of settings (e.g., all four combinations of the maximal/average power constraint and feedback/no feedback \cite{tan2015Third, yang2015Optimal, fong2015feedbacknot, truongfong2017logL} in Table~I) employ random codewords drawn uniformly at random on the power sphere. A notable exception is Truong and Tan's result in \eqref{eq:truongach} \cite[Th.~1]{truong2016gaussian}, which considers VLSF codes with an average power constraint; that result employs i.i.d. Gaussian inputs. The Gaussian distribution works in this scenario because when $L =  \infty$, the usually dominant term $\Prob{\imath(X^{n_L}; Y^{n_L}) < \gamma}$ in \eqref{eq:boundeps} disappears. The second term $(M-1) \exp\{-\gamma\}$ in \eqref{eq:boundeps} is not affected by the input distribution. Unfortunately, the approach from \cite[Th.~1]{truong2016gaussian} does not work here since drawing codewords i.i.d. $\mc{N}(0, P)$ satisfies the average power constraint \eqref{eq:exppower} but not the maximal power constraint \eqref{eq:maxpower}. When $L = O(1)$ and the probability $\Prob{\imath(X^{n_L}; Y^{n_L}) < \gamma}$ dominates, using i.i.d. $\mc{N}(0, P)$ inputs achieves a worse second-order term in the asymptotic expansion \eqref{eq:mainresGaussian} of the maximum achievable message set size. For the case $L = O(1)$, we draw codewords according to the rule that the sub-codewords indexed from $n_{j-1}+1$ to $n_j$ are drawn uniformly on the $(n_j - n_{j-1})$-dimensional sphere of radius $\sqrt{(n_j - n_{j-1}) P}$ for $j \in [L]$, independently of each other. Note that this input distribution is dispersion-achieving for the fixed-length no-feedback case, i.e., $L = 1$ \cite{polyanskiy2010Channel} and 
  is superior to choosing codewords i.i.d. $\mathcal{N}(0, P)$, even under the average power constraint. In particular, i.i.d. $\mc{N}(0, P)$ inputs achieve \eqref{eq:mainresultK}, where the dispersion $V(P)$ is replaced by the variance
	        $\tilde{V}(P) = \frac{P}{1+P}$
	    of $\imath(X; Y)$ when $X \sim \mc{N}(0, P)$; here $\tilde{V}(P)$ is greater than the dispersion $V(P)$ for all $P > 0$ (see \cite[eq.~(2.56)]{molavianjaziThesis}). Whether or not our input distribution is optimal in the second-order term remains an open question.

\section{Conclusions} \label{sec:conclusion}
This paper investigates the maximum achievable message set size for sparse VLSF codes over the DM-PPC (\thmref{thm:mainPPC}), DM-MAC (\thmref{thm:MAC}), DM-RAC (\thmref{thm:RAC}), and Gaussian PPC (\thmref{thm:Gaussian}) in the asymptotic regime where the number of decoding times $L$ is constant as the average decoding time $N$ grows without bound. Under our second-order achievability bounds, the performance improvement due to adding more decoding time opportunities to our code quickly diminishes as $L$ increases. For example, for the BSC with crossover probability 0.11, at average decoding time $N = 2000$, our VLSF coding bound with only $L = 4$ decoding times achieves 96\% of the rate of Polyanskiy \emph{et al.}'s VLSF coding bound for $L = \infty$. Incremental redundancy automatic repeat request codes, which are some of the most common feedback codes, employ only a small number of decoding times and stop feedback. Our analysis shows that such a code design is not only practical but also has performance competitive with the best known dense feedback codes. 

In all channel types considered, the first-order term in our achievability bounds is $\frac{NC}{1-\epsilon}$, where $N$ is the average decoding time, $\epsilon$ is the error probability, and $C$ is the capacity (or the sum-rate capacity in the multi-transmitter case), and the second-order term is $\bigo{\sqrt{N \log _{(L-1)}(N)}}$. For DM-PPCs, there is a mismatch between the second-order term of our achievability bound for VLSF codes with $L = O(1)$ decoding times (\thmref{thm:mainPPC}) and the second-order term of the best known converse bound \eqref{eq:feedb2}; the latter applies to $L = \infty$, and therefore to any $L$. Towards closing the gap between the achievability and converse bounds, in \thmref{thm:metaconv} in \appref{app:VLSFdN}, below, we derive a non-asymptotic converse bound that links the error probability of a VLSF code with the minimum achievable type-II error probability of an SHT. However, since the threshold values of the optimal SHT with $L$ decoding times do not have a closed-form expression \cite[pp. 153-154]{sequentialbook}, analyzing the non-asymptotic converse bound in \thmref{thm:metaconv} is a difficult task. Whether the second-order term in \thmref{thm:mainPPC} is optimal is a question left to future work. 

In sparse VLSF codes, optimizing the values of $L$ available decoding times is important since to achieve the same performance as $L = O(1)$ optimized decoding times (\thmref{thm:mainPPC}), one needs $\Omega\nB{\sqrt{\frac{N}{\log _{(L-1)}(N)}}}$ uniformly spaced decoding times (\thmref{thm:VLSFdN}).

\begin{appendices} 
\renewcommand\thesubsection{\thesection.\Roman{subsection}}
\renewcommand\thesubsectiondis{\thesectiondis.\Roman{subsection}.}
\section{Proof of \thmref{thm:nonasyPPC}}
\label{app:proofnonasy} 

In this section, we derive an achievability bound based on a general SHT, which we use to prove \thmref{thm:nonasyPPC}.
\subsection{A General SHT-based Achievability Bound}
	\label{app:proofPPC}

\subsubsection{SHT definitions} \label{sec:SHTdef}
We begin by formally defining an SHT. We extend the definition in \cite[Ch.~3]{sequentialbook} to allow non-i.i.d. distributions and finitely many testing times.
Let $\{Z_i\}_{i = 1}^{n_L}$ be the observed sequence. 
Consider two hypotheses for the distribution of $Z^{n_L}$
    \begin{align}
        H_0 &\colon Z^{n_L} \sim P_0  \label{eq:H0Z}\\
        H_1 &\colon Z^{n_L} \sim P_1, \label{eq:H1Z}
    \end{align}
    where $P_0$ and $P_1$ are distributions on a common alphabet $\mc{Z}^{n_L}$. Let $\mc{N} \subseteq \{0, 1, 2, \dots, n_L\}$ be the set of times that the hypothesis is tested. Let $P_i^{(n_{\ell})}$ denote the marginal distribution of the first $n_\ell$ symbols in $P_i$, $i \in \{0, 1\}$. At time $n_{\ell} \in \mathcal{N}$, we either decide $H_0 \colon Z^{n_\ell} \sim P_0^{(n_{\ell})}$, $H_1 \colon Z^{n_{\ell}} \sim P_1^{(n_{\ell})}$, or we wait until the next available time $n_{\ell + 1}$ in $\mathcal{N}$.  
    Let $\tau$ be a stopping time adapted to the filtration $\{\mc{F}(X^n)\}_{n \in \mc{N}}$. Let $\delta$ be a $\{0, 1\}$-valued, $\mc{F}(\tau)$-measurable function. An SHT is a triple $(\delta, \tau, \mc{N})$, where $\delta$ is called the decision rule, $\tau$ is called the stopping time, and $\mc{N}$ is the set of available decision times. Type-I and type-II error probabilities are defined as
    \begin{align}
        \alpha &\triangleq \Prob{\delta = 1 | H_0} \\
        \beta &\triangleq \Prob{\delta = 0 | H_1}.
    \end{align}
Below, we derive an achievability using a general SHT.
   \subsubsection{Achievability Bound} \label{app:SHTach}
	Given some input distribution $P_{X^{n_L}}$, 
	define the common randomness random variable $U$ on $\mathbb{R}^{M n_L}$ with the distribution
	\begin{align}
	    P_U = \underbrace{P_{X^{n_L}} \times P_{X^{n_L}} \times \cdots \times P_{X^{n_L}}}_{M \text{times}}. \label{eq:U}
	\end{align}
	The realization of $U$ defines $M$ length-$n_L$ codewords $X^{n_L}(1), X^{n_L}(2), \dots, X^{n_L}(M)$. Denote the set of available decoding times by
	\begin{align}
	    \mc{N} \triangleq \{n_1, \dots, n_L\}.
	\end{align}
	
	Let $\{(\delta_m, \tilde{\tau}_m, \mc{N})\}_{m = 1}^{M}$ be $M$ copies of an SHT that distinguishes between the hypotheses 
        \begin{align}
        H_0 &\colon (X^{n_L}, Y^{n_L}) \sim P_{X^{n_L}} \times P_{Y|X}^{n_L} \label{eq:H0ach} \\
        H_1 &\colon (X^{n_L}, Y^{n_L}) \sim P_{X^{n_L}} \times P_{{Y}^{n_L}} \label{eq:H1ach}
        \end{align}
    for each message $m \in [M]$,
    where the type-I and type-II error probabilities are $\alpha$ and $\beta$, respectively. Define for $m \in [M]$ and $j \in \{0, 1\}$,
    \begin{align}
        \tau_m^{j} \triangleq \begin{cases} \tilde{\tau}_m &\text{if } \delta_m = j \\
        \infty &\text{otherwise}. \label{eq:taumjdef}
        \end{cases}
    \end{align}
    \thmref{thm:SHTnonasy}, below, is an achievability bound that employs an arbitrary SHT with $L$ decoding times. 
 
    \begin{theorem} \label{thm:SHTnonasy}
        Fix $L \leq \infty$, integers $M > 0$ and $0 \leq n_1 < n_2 < \dots < n_L \leq \infty$, a distribution $P_{X^{n_L}}$ as in \eqref{eq:U}, and $M$ copies of an SHT $\{(\delta_m, \tilde{\tau}_m, \{n_1, \dots n_L\})\}_{m = 1}^{M}$ as in \eqref{eq:H0ach}--\eqref{eq:taumjdef}. There exists an $(N, L, M, \epsilon)$-VLSF code for the DM-PPC $(\mc{X}, P_{Y|X}, \mc{Y})$ with
        \begin{align}
            \epsilon &\leq \alpha + (M-1) \beta \label{eq:epsboundSHT} \\
            N &\leq \E{ \min \cB{ \min \limits_{m \in [M]} \cB{\tau_m^{0}}, \max\limits_{m \in [M]} \cB{\tau_m^1}}}. \label{eq:NboundSHT}
        \end{align}
    \end{theorem}
    \begin{IEEEproof}
    We generate $M$ i.i.d. codewords according to \eqref{eq:U}. 
    For each of $M$ messages, we run the hypothesis test given in \eqref{eq:H0ach}--\eqref{eq:H1ach}. We decode at the earliest time that one of the following events happens
    \begin{itemize}
        \item $H_0$ is declared for some message $m \in [M]$, 
        \item $H_1$ is declared for all $m \in [M]$.
    \end{itemize}
    The decoding output is $m$ if $H_0$ is declared for $m$; if there exist more than one such $m$ or if there exists no such $m$, the decoder declares an error.
     
    Mathematically, the random decoding time of this code is expressed as
    \begin{align}
        \tau^* = \min \cB{ \min \limits_{m \in [M]} \cB{\tau_m^{0}}, \max\limits_{m \in [M]} \cB{\tau_m^1}}. \label{eq:tauSHT}
    \end{align}
    Note that $\tau^*$ is bounded by $n_L$ by construction.
    The average decoding time bound in \eqref{eq:NboundSHT} immediately follows from \eqref{eq:tauSHT}.
    The decoder output is
    \begin{align}
        \hat{W} \triangleq \begin{cases}
            m  &\text{if } \exists!\, m \in [M] \text{ s. t. }\tau^* = \tau_m^0 \\
            \text{error} &\text{otherwise.}
        \end{cases}
    \end{align}
    Since the messages are equiprobable, without loss of generality, assume that message $m = 1$ is transmitted. An error occurs if and only if $H_1$ is decided for $m = 1$ or if $H_0$ is decided for some $m \neq 1$, giving
    \begin{align}
        \epsilon = \Prob{ \{\delta_1 = 1\} \cup \cB{\bigcup_{m = 2}^M \{\delta_m = 0\}}}. \label{eq:epsilSHT}
    \end{align}
    Applying the union bound to \eqref{eq:epsilSHT} shows \eqref{eq:epsboundSHT}.
    \end{IEEEproof}
    
	\subsection{Proof of \thmref{thm:nonasyPPC}}
	
	\thmref{thm:nonasyPPC} particularizes the SHT in \thmref{thm:SHTnonasy} as an information density threshold rule.
	
    In addition to the random code design in \eqref{eq:U}, let $P_{X^{n_L}}$ satisfy \eqref{eq:probproduct}.  We here specify the stopping rule $\tau_m$ and the decision rule $\delta_m$ for the SHT in \eqref{eq:H0ach}--\eqref{eq:H1ach}. 
    
    Define the information density for message $m$ and decoding time $n_{\ell}$ as
	\begin{align}
	    S_{m, n_{\ell}} \triangleq \imath(X^{n_{\ell}}(m); Y^{n_{\ell}})  \text{ for } m \in [M], \ell \in [L]. \label{eq:SPPC}
	\end{align}
	Note that $S_{m, n_{\ell}}$ is the log-likelihood ratio between the distributions in hypotheses $H_0$ and $H_1$. We fix a threshold $\gamma \in \mathbb{R}$ and construct the SHTs
	\begin{align}
	    \tau_m &= \inf\{n_{\ell} \in \mc{N} \colon S_{m, n_{\ell}} \geq \gamma \} \label{eq:taum} \\
	    \tilde{\tau}_m &= \min\{\tau_m, n_L\} \label{eq:tautildemm}\\
	    \delta_m &= \begin{cases}
	    0 &\text{if } S_{m, \tilde{\tau}_m} \geq \gamma \\
	    1 &\text{if } S_{m, \tilde{\tau}_m} < \gamma
 	    \end{cases} \label{eq:deltam}
	\end{align}
	for all $m \in [M]$,
	that is, we decide $H_0$ for message $m$ at the first time $n_{\ell}$ that $S_{m, n_{\ell}}$ passes $\gamma$; if this never happens for $n_\ell \in \{n_1, \dots, n_L\}$, then we decide $H_1$ for $m$. Without loss of generality, assume that message 1 is transmitted.
	
	Bounding \eqref{eq:NboundSHT} from above, we get
	\begin{align}
	    N &\leq \E{\min\{\tau_1, n_L\}} \label{eq:taustar1}\\
	     &= \sum_{n = 0}^{\infty}  \Prob{\min\{\tau_1,  n_L\} > n } \\
	    &= n_1 + \sum_{\ell = 1}^{L - 1} (n_{\ell + 1} - n_\ell) \Prob{\tau_1 > n_\ell}. \label{eq:sumnprob}
	\end{align}
	The probability $\Prob{\tau_1 > n_\ell}$ is further bounded as
\begin{align}
    \Prob{\tau_1 > n_\ell} &= \Prob{\bigcap_{j = 1}^{\ell} \{\imath(X^{n_j}(1); Y^{n_j}) < \gamma\}} \\
    &\leq \Prob{ \imath(X^{n_\ell}(1); Y^{n_\ell}) < \gamma}. \label{eq:X1prob}
\end{align}
Combining \eqref{eq:sumnprob} and \eqref{eq:X1prob} proves \eqref{eq:boundN}.
	   
	We bound the type-I error probability of the given SHT as
	\begin{align}
	    \alpha &\triangleq \Prob{\delta_1 = 1} \label{eq:alphaPPC} \\
	    &= \Prob{\tau_1 = \infty} \\
	    &= \Prob{ \bigcap_{j = 1}^L \{\imath(X^{n_j}(1); Y^{n_j}) < \gamma\}} \label{eq:tauKplus1} \\
	   &\leq \Prob{\imath(X^{n_L}(1); Y^{n_L}) < \gamma}, \label{eq:tauK1bound}
	\end{align}
	where \eqref{eq:tauKplus1} uses the definition of the decision rule \eqref{eq:deltam}. 
	The type-II error probability is bounded as
	\begin{align}
	    \beta &\triangleq \Prob{\delta_2 = 0} \\
	    &\leq \Prob{\tau_2 < \infty}  \\
	    &= \E{\exp\{-\imath(X^{n_L}(1); Y^{n_L})\} 1\{\tau_1 < \infty\}} \label{eq:changemeasure}\\
	    &= \E{\exp\{-\imath(X^{\tau}(1); Y^{\tau})\} 1\{\tau_1 < \infty\}} \label{eq:doob}\\
	   &\leq \exp\{-\gamma\}, \label{eq:gammabound}
	\end{align}
	 where \eqref{eq:changemeasure} follows from changing measure from $P_{X^{n_L}(2) Y^{n_L}} = P_{X^{n_L}} P_{Y^{n_L}}$ to $P_{X^{n_L}(1), Y^{n_L}} = P_{X^{n_L}} P_{Y|X}^{n_L}$. Equality \eqref{eq:doob} uses the same arguments as in \cite[eq.~(111)-(118)]{polyanskiy2011feedback} and the fact that ${\{\exp\{-\imath(X^{n_{\ell}}(1); Y^{n_{\ell}})\} \colon n_{\ell} \in \mc{N}\}}$ is a martingale due to the product distribution in \eqref{eq:probproduct}. Inequality \eqref{eq:gammabound} follows from the definition of $\tau_1$ in \eqref{eq:taum}. Applying \eqref{eq:epsboundSHT} together with \eqref{eq:tauK1bound} and \eqref{eq:gammabound} proves \eqref{eq:boundeps}. \IEEEQEDhere

	    In his analysis of the error exponent regime, Forney \cite{forney68} uses a slightly different threshold rule than the one in \eqref{eq:taum}. Specifically, he uses a maximum a posteriori threshold rule, which can also be written as
	\begin{align}
	    \log  \frac{P_{Y^{n_\ell}|X^{n_\ell}}(Y^{n_\ell}|X^{n_\ell}(m))}{\frac{1}{M} \sum_{j = 1}^M P_{Y^{n_\ell}|X^{n_\ell}}(Y^{n_\ell}|X^{n_\ell}(j))} \geq \gamma, \label{eq:pdef}
	\end{align}
	whose denominator is the output distribution induced by the code rather than by the random codeword distribution $P_X^{n_\ell}$.

   \section{Proof of \thmref{thm:mainPPC}} \label{app:proofmainPPC}
  
    The proof uses an idea that is similar to that in \cite[Th.~2]{polyanskiy2010Channel}, which combines the achievability bound of a VLSF code with a sub-exponentially decaying error probability with the stop-at-time-zero procedure. The difference is that we set the sub-exponentially decaying error probability as $\epsilon_N' = \frac{1}{\sqrt{N' \log N'}}$ while \cite[Th.~2]{polyanskiy2010Channel} sets it to $\frac{1}{N'}$. This modification yields a better second-order term for finite $L$.

    Inverting \eqref{eq:Nset}, we get
	\begin{align}
	    N' = \frac{N}{1-\epsilon} \nB{1 + \bigo{\frac{1}{\sqrt{N \log  N}}}}. \label{eq:Nprime}
	\end{align}
    Next, we particularize the decision rules in the SHT at times $n_2, \dots, n_L$ to the information density threshold rule. \lemref{lem:VLFmoderate}, below, is an achievability bound for an $\left(N, L, M, \frac{1}{\sqrt{N \log  N}}\right)$-VLSF code that employs the information density threshold rule with the optimized decoding times and the threshold value.
	\begin{lemma}
	\label{lem:VLFmoderate}
	Fix an integer $L = O(1) \geq 1$. For the DM-PPC with $V > 0$, the maximum message set size \eqref{eq:Mstarmax} achievable by $\left(N, L, M, \frac{1}{\sqrt{N \log  N}}\right)$-VLSF codes satisfies
	\begin{align}
	\log  M^*\left(N, L, \frac{1}{\sqrt{N \log  N}}\right)
	&\geq {N C}
	- \sqrt{N \log _{(L)} (N) \, V} \notag \\
	&\quad + \bigo{\sqrt{\frac{N}{\log _{(L)} (N)}}}. \label{eq:achvanish}
	\end{align}
	The decoding times $n_1, \dots, n_L$ that achieve \eqref{eq:achvanish} satisfy the equations
	\begin{align}
	\log  M
	&= n_{\ell} C - \sqrt{n_{\ell}  \log _{(L -\ell + 1)}(n_{\ell}) V}  - \log {n_{\ell}} + O(1)
	\end{align}
	for $\ell \in [L]$.
	\end{lemma}
	\begin{IEEEproof}
	\lemref{lem:VLFmoderate} analyzes \thmref{thm:nonasyPPC}. See \appref{app:proofPPCMD}, below. For $L = O(1)$, the proof is significantly different than the corresponding result in \cite[eq.~(102)]{polyanskiy2011feedback} for $L = \infty$ because the dominant error probability term $\Prob{\imath(X^{n_L}; Y^{n_L}) < \gamma}$ in \eqref{eq:boundeps} disappears when $L = \infty$.
	\end{IEEEproof}
	
	We use the average decoding time $N$ and average error probability $\epsilon$ of a VLSF code in \lemref{lem:VLFmoderate} in the places of $N'$ and $\epsilon_N'$ in \eqref{eq:epsNprimeset}. By \lemref{lem:VLFmoderate}, there exists an $(N', L-1,  M, \epsilon_N')$-VLSF code with
	\begin{align}
	\log  M &= {N' C} - \sqrt{N' \log _{(L-1)} (N') \, V}  \notag \\
 &\quad + \bigo{\sqrt{\frac{N'}{\log _{(L-1)} (N')}}}. \label{eq:lnMN'}
	\end{align}
	Plugging \eqref{eq:Nprime} into \eqref{eq:lnMN'} and applying the necessary Taylor series expansions complete the proof.  \hspace*{\fill} \IEEEQEDhere
	
         \lemref{lem:VLFmoderate} is an achievability bound in the moderate deviations regime since the error probability $\frac{1}{\sqrt{N \log  N}}$ decays sub-exponentially to zero. The fixed-length scenario in \lemref{lem:VLFmoderate}, i.e., $L = 1$, is recovered by \cite{polyanskiy2010Moderate}, which investigates the moderate deviations regime in channel coding. A comparison between the right-hand side of \eqref{eq:achvanish} and \cite[Th.~2]{polyanskiy2010Moderate} highlights the benefit of using VLSF codes in the moderate deviations regime. The second-order rate achieved by a VLSF code with $L \geq 2$ decoding times, average decoding time $N$, and error probability $\frac{1}{\sqrt{N \log  N}}$ is achieved by a fixed-length code with blocklength $N$ and error probability $\frac{1}{\sqrt{\log _{(L-1)}(N) \log _{(L)}(N)}}$.	

         In the remainder of the appendix, we prove \lemref{lem:VLFmoderate} and show the second-order optimality of the parameters used in the proof of \thmref{thm:mainPPC} including the decoding times set in~\eqref{eq:decodingeq}.
        	\subsection{Proof of \lemref{lem:VLFmoderate}} \label{app:proofPPCMD}

	We first present two lemmas used in the proof of \lemref{lem:VLFmoderate} (step 1). We then choose the distribution $P_X^{n_L}$ of the random codewords (step 2) and the parameters $n_1, \dots, n_L, \gamma$ in \thmref{thm:nonasyPPC} (step 3). Finally, we analyze the bounds in \thmref{thm:nonasyPPC} using the supporting lemmas (step 4). 
	\subsubsection{Supporting lemmas}
	\lemref{lem:moderate}, below, is the moderate deviations result that bounds the probability that a sum of $n$ zero-mean i.i.d. random variables is above a function of $n$ that grows at most as quickly as $n^{2/3}$.

	\begin{lemma}[Petrov {\cite[Ch. 8, Th.~ 4]{petrov1975}}]\label{lem:moderate}
	Let $Z_1, \dots, Z_n$ be i.i.d. random variables. Let $\E{Z_1} = 0$, $\sigma^2 = \Var{Z_1}$, and $\mu_3 = \E{Z_1^3}$. Suppose that the moment generating function $\E{\exp\{tZ\}}$ is finite in the neighborhood of $t = 0$. (This condition is known as Cram\'er's condition.) Let $0 \leq z_n = O(n^{1/6})$. As $n \to \infty$, it holds that
	\begin{align}
	    &\Prob{\sum\limits_{i = 1}^n Z_i \geq z_n \sigma \sqrt{n}} \notag \\
	    &\quad = Q(z_n) \exp\left\{\frac{z_n^3 \mu_3}{6 \sqrt{n} \sigma^3} \right\} +  O\left(\frac{1}{\sqrt{n}} \exp\left\{-\frac{z_n^2}{2}\right\} \right)
	\end{align}
	\end{lemma}

	\lemref{lem:asympxy}, below, gives the asymptotic expansion of the root of an equation. We use \lemref{lem:asympxy} to find the asymptotic expansion for the gap between two consecutive decoding times $n_\ell$ and $n_{\ell+1}$.
	
	\begin{lemma}\label{lem:asympxy} Let $f(x)$ be a differentiable increasing function that satisfies $f'(x) \to 0$ as $x \to \infty$. Suppose that 
	\begin{align}
	    x + f(x) = y. \label{eq:equationf}
	\end{align}
	Then, as $x \to \infty$ it holds that
	\begin{align}
	    x = y - f(y) \nB{1- o(1)}.
	\end{align}
	\end{lemma}
	\begin{IEEEproof}[Proof of \lemref{lem:asympxy}]
	Define the function $F(x) \triangleq x + f(x) - y$. Applying Newton's method with the starting point $x_0 = y$ yields
	\begin{align}
	    x_1 &= x_0 -  \frac{F(x_0)}{F'(x_0)} \\
	    &= y - \frac{f(y)}{1 + f'(y)} \\
	    &= y - f(y)(1 - f'(y) + O(f'(y)^2). \label{eq:fprimetaylor}
	\end{align}
	Recall that $f'(y) = o(1)$ by assumption. Equality  \eqref{eq:fprimetaylor} follows from the Taylor series expansion of the function $\frac{1}{1 + x}$ around $x = 0$. Let
	\begin{align}
	    x^{\star} = y - f(y) (1 - o(1)). \label{eq:xstar}
	\end{align}
	From Taylor's theorem, it follows that
	\begin{align}
	    f(x^{\star}) = f(y) - f'(y_0) f(y) (1 - o(1)), \label{eq:fstar}
	\end{align}
	for some $y_0 \in [y - f(y)(1-o(1)), y]$. Therefore, $f'(y_0) = o(1)$, and $f(x^{\star}) = f(y) (1 - o(1))$. Putting \eqref{eq:xstar}--\eqref{eq:fstar} in \eqref{eq:equationf}, we see that $x^\star$ is a solution to the equality in \eqref{eq:equationf}. 
	
	\end{IEEEproof}
	
	   \subsubsection{Random encoder design} \label{sec:randomencoder}
	    We set the distribution of the random codewords $P_{X^{n_L}}$ as the product of $P_X^*$'s, where $P_X^*$ is the capacity-achieving distribution with minimum dispersion, i.e.,
	    \begin{align}
	        P_{X^{n_L}} &= (P_X^*)^{n_L} \\
	        P_X^* &= \arg \min_{P_X} \{V(X; Y)\colon I(X; Y) = C\}.
	    \end{align}

\subsubsection{Choosing the threshold $\gamma$ and decoding times $n_1, \dots, n_L$} \label{sec:choosedec}
We choose $\gamma, n_1, \dots, n_L$ so that the equalities
\begin{align}
    \gamma  = n_{\ell} C - \sqrt{n_{\ell}  \log _{(L -\ell + 1)}(n_{\ell}) V} \label{eq:setnk}
\end{align}
hold for all $\ell \in [L]$. This choice minimizes our upper bound \eqref{eq:boundN} on the average decoding time up to the second-order term in the asymptotic expansion. See \appref{app:optimalityproof} for the proof. Applying \lemref{lem:asympxy} with
\begin{align}
    x &= n_{\ell+1} \\
    y &= n_{\ell} - \frac{1}{C} \sqrt{n_{\ell}  \log _{(L-\ell + 1)}(n_{\ell}) V} \\
    f(x) &= - \frac{1}{C} \sqrt{n_{\ell+1}  \log _{(L-\ell)} (n_{\ell+1}) V}
\end{align}
for $\ell \in \{1, \dots, L-1\}$, gives the following gaps between consecutive decoding times
\begin{align}
    n_{\ell+1} - n_{\ell} &= \frac{1}{C} \Big( \sqrt{n_{\ell}  \log _{(L-\ell)} (n_{\ell}) V} \notag \\
    &- \sqrt{n_{i}  \log _{(L- \ell + 1)} (n_{\ell}) V} \Big) (1 + o(1)). \label{eq:ngap}
\end{align}

\subsubsection{Analyzing the bounds in \thmref{thm:nonasyPPC}}
The information density $\imath(X; Y)$ of a DM-PPC is a bounded random variable. Therefore, $\imath(X; Y)$ satisfies Cram\'er's condition (see \lemref{lem:moderate}).\footnote{Here, Cram\'er's condition is the bottleneck that determines whether our proof technique applies to a specific DM-PPC. For DM-PPCs with infinite input or output alphabets, Cram\'er's condition may or may not be satisfied. Our proof technique applies to any DM-PPC for which the information density satisfies Cram\'er's condition.} For each $\ell \in [L]$, applying \lemref{lem:moderate} with $\gamma, n_1, \dots, n_L$ satisfying \eqref{eq:setnk} gives
\begin{IEEEeqnarray}{rCl}
     \IEEEeqnarraymulticol{3}{l}{\Prob{\imath(X^{n_{\ell}}; Y^{n_{\ell}}) < \gamma}} \notag\\
     &\leq& Q\left(\sqrt{\log _{(L-\ell+1)}(n_{\ell})}\right) \exp\left\{\frac{-(\log _{(L-\ell+1)}(n_{\ell}))^{3/2} \mu_3}{6 \sqrt{n} V^{3/2}} \right\} \notag \\
     && + O\left(\frac{1}{\sqrt{n}} \exp\left\{-\frac{\log _{(L-\ell+1)} (n_{\ell})}{2}\right\} \right) \label{eq:PetrovQ} \\
    &\leq& \frac{1}{\sqrt{2 \pi}}  \frac{1}{\sqrt{\log _{(L- \ell)}(n_{\ell})}} \frac{1}{\sqrt{\log _{(L-\ell+1)}(n_{\ell})}} \, \notag \\
    && \cdot \nB{1 + \bigo{\frac{(\log _{(L-\ell+1)}(n_{\ell}))^{(3/2)}}{\sqrt{n_{\ell}}}}} \label{eq:pboundQ}
\end{IEEEeqnarray}
for $\ell < L$, where 
\begin{align}
    \mu_3 \triangleq \E{(\imath(X;Y) - C)^3} < \infty,
\end{align}
and
\eqref{eq:pboundQ} follows from the Taylor series expansion $\exp\{x\} = 1 + x + O(x^2)$ as $x \to 0$, and the well-known bound (e.g., \cite[Ch. 8, eq.~(2.46)]{petrov1975})
\begin{align}
    Q(x) \leq \frac{1}{\sqrt{2 \pi}} \frac{1}{x} \exp\left\{-\frac{x^2}{2}\right\} \quad \text{for } x > 0.
\end{align}
For $\ell = L$, \lemref{lem:moderate} gives
\begin{IEEEeqnarray}{rCl}
    \IEEEeqnarraymulticol{3}{l}{\Prob{\imath(X^{n_L}; Y^{n_L}) < \gamma}} \notag \\ 
    &\leq& \frac{1}{\sqrt{2 \pi}} \frac{1}{\sqrt{n_L}} \frac{1}{\sqrt{\log  n_L}} \nB{1 + \bigo{\frac{(\log  n_L)^{(3/2)}}{\sqrt{n_L}}}}. \IEEEeqnarraynumspace \label{eq:pboundQK1}
\end{IEEEeqnarray}

By \thmref{thm:nonasyPPC}, there exists a VLSF code with $L$ decoding times $n_1 < n_2 < \cdots < n_L$ such that the expected decoding time is bounded as
	\begin{align}
	    N \leq n_1 + \sum_{\ell = 1}^{L-1} (n_{\ell+1} - n_\ell) \Prob{\imath(X^{n_{\ell}}; Y^{n_{\ell}}) < \gamma}.   \label{eq:Nbound}
	\end{align}
By \eqref{eq:ngap}, we have $n_{\ell + 1} = n_{\ell}(1 + o(1))$ for $\ell \in [L-1]$. Multiplying these asymptotic equations, we get
\begin{align}
    n_{\ell} = n_1 (1 + o(1)) \label{eq:nkn1}
\end{align}
for $\ell \in [L]$. Plugging \eqref{eq:ngap}, \eqref{eq:pboundQ}, and \eqref{eq:nkn1} into \eqref{eq:Nbound}, we get
\begin{align}
    N \leq n_1 +  \frac{\sqrt{V}}{\sqrt{2 \pi} \,C} \frac{\sqrt{n_1}}{\sqrt{\log _{(L)} (n_1)}} (1 + o(1)).\label{eq:Nn1}
\end{align}
Applying \lemref{lem:asympxy} to \eqref{eq:Nn1}, we get 
\begin{align}
    n_1 \geq N - \frac{\sqrt{V}}{\sqrt{2 \pi} \,C} \frac{\sqrt{N}}{\sqrt{\log _{(L)} (N)}}(1 + o(1)). \label{eq:n1great}
\end{align}

Comparing \eqref{eq:n1great} and \eqref{eq:ngap}, we observe that for $n_1$ large enough, 
\begin{align}
    n_1 < N < n_2 < \dots < n_L. \label{eq:norder}
\end{align}
Further, from \eqref{eq:setnk} and \eqref{eq:Nn1}, we have
\begin{align}
    n_L = N \left(1 + O\left(\sqrt{ \frac{\log  N}{N}}\right) \right). \label{eq:nLN}
\end{align}
Finally, we set message set size $M$ such that
\begin{align}
    \log  M = \gamma - \log  N. \label{eq:Mgamma}
\end{align}
Plugging \eqref{eq:pboundQK1} and \eqref{eq:Mgamma} into \eqref{eq:boundeps}, we bound the error probability as
 \begin{IEEEeqnarray}{rCl}
       \IEEEeqnarraymulticol{3}{l}{ \Prob{\mathsf{g}_{\tau^*}(U, Y^{\tau^*}) \neq W}} \notag \\
       &\leq& \Prob{\imath(X^{n_L}; Y^{n_L}) < \gamma} + (M-1) \exp\{-\gamma\} \\
    &\leq& \frac{1}{\sqrt{2\pi}} \frac{1}{\sqrt{n_L}} \frac{1}{\sqrt{\log  n_L}} (1 + o(1)) + \frac{1}{N} \label{eq:nklarge}\\
    &\leq& \frac{1}{\sqrt{2\pi}} \frac{1}{\sqrt{N}} \frac{1}{\sqrt{\log  N}} (1 + o(1)) + \frac{1}{N}, \IEEEeqnarraynumspace \label{eq:Nnkrel} 
\end{IEEEeqnarray}
where \eqref{eq:Nnkrel} follows from \eqref{eq:norder}. Inequality \eqref{eq:Nnkrel} implies that the error probability is bounded by $\frac{1}{\sqrt{N \log  N}}$ for $N$ large enough. Plugging \eqref{eq:n1great} and \eqref{eq:Mgamma} into \eqref{eq:setnk} with $\ell = 1$, we conclude that there exists an $\left(N, L, M, \frac{1}{\sqrt{N \log  N}}\right)$-VLSF code with
\begin{align}
    \log  M &\geq N C - \sqrt{N \log _{(L)}(N) V } \notag \\
    &\quad - \frac{1}{\sqrt{2 \pi}} \sqrt{\frac{N V}{\log _{(L)} (N)}} (1 + o(1)) -\log  N, \label{eq:lnMfinal}
\end{align}
which completes the proof. \hspace*{\fill} \IEEEQEDhere

	       \subsection{Second-order optimality of the decoding times in \thmref{thm:mainPPC}} 
    \label{app:optimalityproof}
    	From the code construction described in Theorems~\ref{thm:nonasyPPC}--\ref{thm:mainPPC}, the average decoding time is 
	\begin{align}
	    N(n_2, \dots, n_L, \gamma) = N' (1-\epsilon) \frac{1}{1-\epsilon_N'}, \label{eq:Nexp}
	\end{align}
	where
	\begin{align}
	    N' &= n_2 + \sum_{i = 2}^{L-1} (n_{i+1} - n_i) \Prob{\imath(X^{n_{i}}; Y^{n_{i}}) < \gamma} \label{eq:epsNprimen2}\\
	    \epsilon_N' &= \Prob{\imath(X^{n_L}; Y^{n_L}) < \gamma} + M \exp\{-\gamma\}. \label{eq:Nprimeexp}
	\end{align}
	
	We here show that given a fixed $M$, the parameters $n_2, n_3, \dots, n_L, \gamma$ chosen according to \eqref{eq:setnk} and \eqref{eq:Mgamma} (and also the error value $\epsilon_N'$ chosen in \eqref{eq:epsNprimeset} since $\epsilon_N'$ is a function of $(n_L, \gamma$)) minimize the average decoding time in \eqref{eq:Nexp} in the sense that the second-order expansion of $\log  M$ in terms of $N$ is maximized. That is, our parameter choice optimizes our bound on our code construction.
    
\subsubsection{Optimality of $n_2, \dots, n_{L-1}$}   

We first set $n_L$ and $\gamma$ to satisfy the equations
\begin{align}
    \frac{1}{\sqrt{n_L \log  n_L}} &= \Prob{\imath(X^{n_L}; Y^{n_L}) < \gamma} + (M-1) \exp\{-\gamma\} \label{eq:errorapprox} \\
    \log  M &= \gamma - \log  n_L, \label{eq:nkapprox}
\end{align}
and optimize the values of $n_2, \dots, n_{L-1}$ under \eqref{eq:errorapprox}--\eqref{eq:nkapprox}.
\secref{sec:optgammank} proves the optimality of the choices in \eqref{eq:errorapprox}--\eqref{eq:nkapprox}.

Under \eqref{eq:errorapprox}--\eqref{eq:nkapprox}, the optimization problem in \eqref{eq:Nexp}--\eqref{eq:Nprimeexp} reduces to
\begin{equation}
\begin{aligned}
    \min \quad & N'(n_2, \dots, n_{L-1})  \\
    &= n_2 + \sum_{i = 2}^{L-1} (n_{i+1} - n_i) \Prob{\imath(X^{n_{i}}; Y^{n_{i}}) < \gamma} \\
    \textrm{s.t.} \quad  & \frac{1}{\sqrt{n_L \log  n_L}} = \Prob{\imath(X^{n_L}; Y^{n_L}) < \gamma} \\
    &\quad \quad \quad \quad \quad \quad + (M-1) \exp\{-\gamma\}. \label{eq:problem}
\end{aligned}
\end{equation}

Next, we define the functions 
\begin{align}
    g(n) &\triangleq \frac{n C - \gamma}{\sqrt{n V}} \label{eq:gn}\\
    F(n) &\triangleq Q(-g(n)) = 1 - Q(g(n)) \\
    f(n) &\triangleq F'(n) = \frac{1}{\sqrt{2 \pi}} \exp\left\{-\frac{g(n)^2}{2}\right\} \, g'(n).
\end{align}
Assume that $\gamma = \gamma_n$ is such that $g(n) = O(n^{1/6})$, and $\lim \limits_{n \to \infty} g(n) = \infty$. Then by \lemref{lem:moderate}, $\Prob{\imath(X^{n}; Y^{n}) < \gamma}$, which is a step-wise constant function of $n$, is approximated by differentiable function $1 - F(n)$ as
\begin{align}
     \Prob{\imath(X^{n}; Y^{n}) < \gamma} = (1-F(n))(1 + o(1)). \label{eq:probasymp}
\end{align}
Taylor series expansions give
\begin{align}
    1 - F(n) &= \frac{1}{g(n)} \label{eq:Fnap} \frac{1}{\sqrt{2 \pi}} \exp\left\{-\frac{g(n)^2}{2}\right\} (1 + o(1)) \\
    f(n) &= (1 - F(n)) g(n) g'(n) (1 + o(1))  \label{eq:fnap}\\
    g'(n) &= \frac{C}{\sqrt{n V}} (1 + o(1)). \label{eq:gnap}
\end{align}

Let ${\mathbf{n}^{*} = (n_2^*, \dots, n_{L-1}^*)}$ denote the solution to the optimization problem in \eqref{eq:problem} with $\Prob{\imath(X^{n}; Y^{n}) < \gamma}$ replaced by $ (1-F(n))$. Then $\mathbf{n}^{*}$ must satisfy the Karush-Kuhn-Tucker conditions $\nabla N' (\mathbf{n}^*) = \mathbf{0}$, giving
\begin{align}
    \left.\frac{\partial N'}{\partial n_2} \right\vert_{\mathbf{n} = \mathbf{n}^*} &= F(n_2^*) - (n_3^* - n_2^*) f(n_2^*) = 0 \label{eq:optcond1}\\
    \left.\frac{\partial N'}{\partial n_{\ell}}\right\vert_{\mathbf{n} = \mathbf{n}^*} &= F(n_{\ell}^*) - F(n_{L-1}^*) - (n_{\ell+1}^* - n_{\ell}^*) f(n_{\ell}^*) = 0 \label{eq:optcond2} 
\end{align}
for $\ell = 3, \dots, L-1$.

Let $\tilde{\mathbf{n}} = (\tilde{n}_2, \dots, \tilde{n}_{L-1})$ be the decoding times chosen in \eqref{eq:setnk}. We evaluate 
\begin{align}
    g(\tilde{n}_\ell) &= \sqrt{\log _{(L-\ell+1)}(\tilde{n}_\ell)}(1 + o(1)) \label{eq:gtildeni}\\
    1 - F(\tilde{n}_\ell) &= \frac{1}{\sqrt{2\pi}} \frac{1}{g(\tilde{n}_{\ell + 1}) g(\tilde{n}_\ell)} (1 + o(1)) \\
    f(\tilde{n}_\ell) &= \frac{1}{\sqrt{2 \pi}} \frac{g'(\tilde{n}_\ell)}{g(\tilde{n}_{\ell + 1})} \\
    \tilde{n}_{\ell + 1} - \tilde{n}_\ell &= \frac{g(\tilde{n}_{\ell+1})}{g'(\tilde{n}_\ell)} (1 + o(1)) \label{eq:ntildegap}
\end{align}

for $\ell = 2, \dots, L-1$. Plugging \eqref{eq:gtildeni}--\eqref{eq:ntildegap} into \eqref{eq:optcond1}--\eqref{eq:optcond2} for $\left.\frac{\partial N'}{\partial n_\ell} \right\vert_{\mathbf{n} = \mathbf{\tilde{n}}}$, we get
\begin{align}
    \nabla N' (\tilde{\mathbf{n}}) &= \left(1-\frac{1}{\sqrt{2 \pi}}, -\frac{1}{\sqrt{2 \pi}}, -\frac{1}{\sqrt{2 \pi}}, \dots, -\frac{1}{\sqrt{2 \pi}}\right) \notag \\
    &\quad (1 + o(1)).
\end{align}

Our goal is to find a vector $\Delta \mathbf{n} = (\Delta n_2, \dots, \Delta n_{L-1})$ such that
\begin{align}
    \nabla N'(\tilde{\mathbf{n}} + \Delta \mathbf{n}) = \mathbf{0}, \label{eq:gradientDelta}
\end{align}
Assume that $\Delta n = O(\sqrt{n})$. By plugging $n + \Delta n$ into \eqref{eq:Fnap}--\eqref{eq:gnap} and using the Taylor series expansion of $g(n + \Delta n)$, we get
\begin{align}
    1 - F(n + \Delta n) &= (1 - F(n)) \notag \\
    &\quad \cdot \exp\{-\Delta n g(n) g'(n)\} (1 + o(1)) \label{eq:Fdelta}\\
    f(n + \Delta n) &= f(n) \exp\{-\Delta n g(n) g'(n)\} (1 + o(1)). \label{eq:fdelta}
\end{align}
Using \eqref{eq:Fdelta}--\eqref{eq:fdelta}, and putting $\tilde{\mathbf{n}} + \Delta \mathbf{n}$ in \eqref{eq:optcond1}--\eqref{eq:optcond2}, we solve \eqref{eq:gradientDelta} as
\begin{align}
    \Delta n_2 &= - \frac{\log  \sqrt{2 \pi}}{g(\tilde{n}_2) g'(\tilde{n}_2)} (1 + o(1)) \\
    &= -\frac{\sqrt{V} \, \log  {\sqrt{2 \pi}}}{C} \frac{ \sqrt{\tilde{n}_2}}{\sqrt{\log _{(L-1)}(\tilde{n}_2)}} (1 + o(1)) \\
    \Delta n_i &= \frac{1}{2} \frac{g(\tilde{n}_{i-1})^2}{g(\tilde{n}_i) g'(\tilde{n}_i)} = o(\Delta n_2) (1 + o(1)) 
\end{align}
for $i = 3, \dots, L-1$.
Hence, $\tilde{\mathbf{n}} + \Delta{\mathbf{n}}$ satisfies the optimality criterion, and $\mathbf{n}^* = \tilde{\mathbf{n}} + \Delta{\mathbf{n}}$.

It remains only to evaluate the gap $N'(\mathbf{n}^*) - N'(\tilde{\mathbf{n}})$. We have
\begin{align}
    &N'(\mathbf{n}^*) - N'(\tilde{\mathbf{n}}) \notag \\
    &= \bigg(\Delta n_2 + \sum_{i = 2}^{L-1} (\tilde{n}_{i+1} - \tilde{n}_i) Q(g(\tilde{n}_i)) \notag \\
    &\quad \cdot \left( \exp\{-\Delta n_i g(\tilde{n}_i) g'(\tilde{n}_i)\} - 1 \right) \bigg) (1 + o(1)) \label{eq:NstarNgap1}\\
    &= \left(\Delta n_2 + \left(1 - \frac{1}{\sqrt{2 \pi}}\right) \frac{1}{g(\tilde{n}_1) g'(\tilde{n}_i)} - \sum_{i = 3}^{L-1} \Delta n_i \right) \notag\\
    &\quad \cdot (1 + o(1)) \\
    &= - B \frac{\sqrt{\tilde{n}_2}}{\sqrt{\log _{(L-1)}(\tilde{n}_2)}}  (1 + o(1)), \label{eq:NstarN}
\end{align}
where $B = \left(\log  \sqrt{2 \pi} + \frac{1}{\sqrt{2 \pi}} - 1 \right) \frac{\sqrt{V}}{C}$ is a positive constant. From the relationship between $n_{\ell}$ and $n_2$ in \eqref{eq:nkn1} and the equality \eqref{eq:NstarN}, we get
\begin{align}
    N'(\tilde{\mathbf{n}}) = N'(\mathbf{n}^*) + B \frac{\sqrt{N'(\mathbf{n}^*)}}{\sqrt{\log _{(L-1)}(N'(\mathbf{n}^*))}} (1 + o(1)). \label{eq:NstarNrel}
\end{align}
Plugging \eqref{eq:NstarNrel} into our VLSF achievability bound \eqref{eq:lnMfinal} gives
\begin{align}
    \log  M &\geq N'(\mathbf{n}^*) C - \sqrt{N'(\mathbf{n}^*) \log _{(L-1)}(N'(\mathbf{n}^*)) V } \notag \\
    &\quad - O\left( \sqrt{\frac{N'(\mathbf{n}^*) }{\log _{(L-1)} (N'(\mathbf{n}^*))}} \right). \label{eq:lnMfinalstar}
\end{align}
Comparing \eqref{eq:lnMfinalstar} and \eqref{eq:lnMfinal}, note that the decoding times chosen in \eqref{eq:setnk} have the optimal second-order term in the asymptotic expansion of the maximum achievable message set size within our code construction. Moreover, the order of the third-order term in \eqref{eq:lnMfinalstar} is the same as the order of the third-order term in \eqref{eq:lnMfinal}. \IEEEQEDhere

The method of approximating the probability $\Prob{\imath(X^{n}; Y^{n}) \geq \gamma}$, which is an upper bound for $\Prob{\tau \leq n}$ (see \eqref{eq:sumnprob}), by a differentiable function $F(n)$ is introduced in \cite[Sec.~III]{vakilinia2016} for the optimization problem in \eqref{eq:problem}. In \cite{vakilinia2016}, Vakilinia \emph{et al.} approximate the distribution of the random stopping time $\tau$ by the Gaussian distribution, where $\E{\tau}$ and $\Var{\tau}$ are found empirically. They derive the Karush-Kuhn-Tucker conditions in \eqref{eq:optcond1}--\eqref{eq:optcond2}, which is known as the SDO algorithm. A similar analysis appears in \cite{heidarzadeh2019Systematic} for the binary erasure channel.
The analyses in \cite{vakilinia2016, heidarzadeh2019Systematic} use the SDO algorithm to numerically solve the equations \eqref{eq:optcond1}--\eqref{eq:optcond2} for a fixed $L$, $M$, and $\epsilon$. Unlike \cite{vakilinia2016, heidarzadeh2019Systematic}, we find the analytic solution to  \eqref{eq:optcond1}--\eqref{eq:optcond2} as decoding times $n_2, \dots, n_L$ approach infinity, and we derive the achievable rate in \thmref{thm:mainPPC} as a function of $L$.

	\subsubsection{Optimality of $n_L$ and $\gamma$} \label{sec:optgammank}

	Let $(\mathbf{n}^*, \gamma^*) = (n_2^*, \dots, n_L^*, \gamma^*)$ be the solution to $\nabla N(\mathbf{n}^*, \gamma^*) = \mathbf{0}$ in \eqref{eq:Nexp}.
	Section A finds  the values of $n_2^*, n_3^*, \dots, n_{L-1}^*$ that minimize $N'$. Minimizing $N'$ also minimizes $N$ in \eqref{eq:Nexp} since $\epsilon_N'$ depends only on $n_L$ and $\gamma$, and $\epsilon$ is a constant. Therefore, to minimize $N$, it only remains to find $(n_L^*, \gamma^*)$ such that
	\begin{align}
	    \left.\frac{\partial N}{\partial n_L}\right\vert_{(\mathbf{n}, \gamma) = (\mathbf{n}^*, \gamma^*)}&= 0 \label{eq:derivnK} \\
	    \left.\frac{\partial N}{\partial \gamma}\right\vert_{(\mathbf{n}, \gamma) = (\mathbf{n}^*, \gamma^*)} &= 0. \label{eq:derivgamma}
	\end{align}
	Consider the case where $L > 2$. Solving \eqref{eq:derivnK} and \eqref{eq:derivgamma} using \eqref{eq:optcond1}--\eqref{eq:ntildegap} gives
	\begin{align}
	    g(n_L^*) &= \sqrt{\log  n_L^* + \log _{(2)}(n_L^*) + \log _{(3)}{(n_L^*)} + O(1)} \label{eq:eq1deriv}\\
	    0 &= c_0 + N' \bigg(\frac{1}{\sqrt{2 \pi n_L^*}} \exp\left\{-\frac{g(n_L^*)^2}{2}\right\} (1 + o(1)) \notag \\
	    &\quad - M \exp\{-\gamma^*\} \bigg), \label{eq:eq2deriv}
	\end{align}
	where $c_0$ is a positive constant.
	Solving \eqref{eq:eq1deriv}--\eqref{eq:eq2deriv} simultaneously, we obtain
	\begin{align}
	    \log  M &= \gamma^* - \log  n_L^* + O(1). \label{eq:gammasol}
	\end{align}
    Plugging \eqref{eq:eq1deriv} and \eqref{eq:gammasol} into \eqref{eq:Nprimeexp}, we get
    \begin{align}
        \epsilon_{N}'^* = \frac{c_1}{\sqrt{n_L^* \log _{(2)}(n_L^*)} \log  n_L^*} (1 + o(1)), \label{eq:optpes}
    \end{align}
    where $c_1$ is a constant. Let $(\tilde{\mathbf{n}}, \tilde{\gamma}) = (\tilde{n}_2, \dots, \tilde{n}_{K}, \tilde{\gamma})$ be the parameters chosen in \eqref{eq:setnk} and \eqref{eq:Mgamma}. Note that $ \epsilon_{N}'^*$ is order-wise different than $\epsilon_N'$ in \eqref{eq:epsNprimeset}. Following steps similar to \eqref{eq:NstarNgap1}--\eqref{eq:NstarN}, we compute
    \begin{align}
        N(\mathbf{n}^*, \gamma^*) - N(\tilde{\mathbf{n}}, \tilde{\gamma}) = - O\left(\sqrt{\frac{n_L^*}{\log  n_L^*}}\right). \label{eq:Noptgap}
    \end{align}
    Plugging \eqref{eq:Noptgap} into \eqref{eq:mainresultK} gives
    \begin{align}
        \log  M &=  { \frac{N(\mathbf{n}^*, \gamma^*) C}{1-\epsilon}} \notag \\
        &- \sqrt{N(\mathbf{n}^*, \gamma^*) \log _{(L-1)} (N(\mathbf{n}^*, \gamma^*)) \frac{V}{1-\epsilon}} \notag \\
	&+ \bigo{\sqrt{\frac{N(\mathbf{n}^*, \gamma^*)}{\log _{(L-1)} (N(\mathbf{n}^*, \gamma^*))}}}. \label{eq:perf}
    \end{align}
    Comparing \eqref{eq:mainresultK} and \eqref{eq:perf}, we see that although \eqref{eq:epsNprimeset} and \eqref{eq:optpes} are different, the parameters $(\tilde{\mathbf{n}}, \tilde{\gamma})$ chosen in \eqref{eq:setnk} and \eqref{eq:Mgamma} have the same second-order term in the asymptotic expansion of the maximum achievable message set size as the parameters $(\mathbf{n}^*, \gamma^*)$ that minimize the average decoding time in the achievability bound in \thmref{thm:nonasyPPC}.
    
    For $L =  2$, the summation term in \eqref{eq:epsNprimen2} disappears; in this case, the solution to \eqref{eq:derivnK} gives 
    \begin{align}
        \epsilon_N'^* = \frac{c_2}{\sqrt{n_L^* \log  n_L^*}} (1 + o(1))
    \end{align}
    for some constant $c_2$. Following the steps in \eqref{eq:Noptgap}--\eqref{eq:perf}, we conclude that the parameter choices in \eqref{eq:setnk} and \eqref{eq:Mgamma} are second-order optimal for $L =  2$ as well. 
       
    \section{Proof of \thmref{thm:VLSFdN}} \label{app:VLSFdN}
    
    Let $P_0$ and $P_1$ be two distributions. Let $Z \triangleq \log  \frac{\mathrm{d} P_0}{\mathrm{d} P_1}$ be the log-likelihood ratio, and let
    \begin{align}
        S_n = \sum_{i = 1}^n Z_i, \label{eq:Snsum}
    \end{align}
    where $Z_i$'s are i.i.d. and have the same distribution as $Z$. For $i \in \{0, 1\}$, we denote the probability measures and expectations under distribution $P_i$ by $\mathbb{P}_i$ and $\mathbb{E}_i$, respectively.  Given a threshold $a_0 \in \mathbb{R}$, define the stopping time 
    \begin{align}
        T &\triangleq \inf\{n \geq 1 \colon S_n \geq a_0\} \label{eq:Teq}
    \end{align}
    and the overshoot 
    \begin{align}
        \xi_0 = S_T - a_0.
    \end{align}
    The following lemma from \cite{sequentialbook}, which gives the refined asymptotics for the stopping time $T$, is the main tool to prove our bounds.  
    \begin{lemma}[{\cite[Cor.~2.3.1, Th.~2.3.3, Th.~2.5.3, Lemma~3.1.1]{sequentialbook}}] \label{lem:SHT}
    Suppose that $\Ez{(Z_1^+)^2} < \infty$, and $Z_1$ is non-arithmetic. Then, as $a \to \infty$, it holds that
    \begin{align}
        \Ez{T} &= \frac{1}{D(P_0 \| P_1)} (a_0 + \Ez{\xi_0}) \\
        &= \frac{1}{D(P_0 \| P_1)} \Bigg(a_0 + \frac{\Ez{Z_1^2}}{2 D(P_0 \| P_1)} \notag \\
        &\quad -\sum_{n = 1}^\infty   \frac{1}{n} \Ez{S_n^-} + o(1)\Bigg), \label{eq:E0T}
    \end{align}
    and
    \begin{align}
        \Pz{T < \infty} &= 1 \\
        \Po{T < \infty} &= e^{-a_0} \Ez{e^{-\xi_0}} \label{eq:Ponebound} \\
        \Ez{e^{-\lambda \xi_0}} &= \frac{1 + o(1)}{\lambda D(P_0 \| P_1)} \exp\cB{-\sum_{n = 1}^{\infty} \frac{1}{n} \Ez{e^{-\lambda S_n^+}}}. \label{eq:lambdaexp}
    \end{align}
    \end{lemma}

    \subsection{Achievability Proof}
    Let $P_X$ be a capacity-achieving distribution of the DM-PPC.
    Define the hypotheses
    \begin{align}
        H_0 &\colon (X^{d_N}, Y^{d_N})^{\infty} \sim P_0^{\infty} = ((P_X \times P_{Y|X})^{d_N})^{\infty} \\
        H_1 &\colon (X^{d_N}, Y^{d_N})^{\infty} \sim P_1^{\infty} = ((P_X \times P_Y)^{d_N})^{\infty},
    \end{align}
    and the random variables
    \begin{align}
    W_i &\triangleq \frac{1}{d_N} \log  \frac{\mathrm{d} P_0^{d_N}}{\mathrm{d} P_1^{d_N}}\left(X^{(i-1) d_N + 1:i d_N}, Y^{(i-1) d_N + 1 : i d_N} \right) \\
    &= \frac{1}{d_N} \imath(X^{(i-1) d_N + 1: i d_N}; Y^{(i-1) d_N + 1 : i d_N})
    \end{align}
    for $i \in \mathbb{N}.$
    Note that under $H_0$, $\Ez{W_i} = C$. 
    Here, we use $W_i$ in the place of $Z_i$ in \eqref{eq:Snsum}.
    Define 
    \begin{align}
    S_n \triangleq \sum_{i = 1}^n W_i, \label{eq:Sndef}
    \end{align}
    and
    \begin{align}
        \tau &\triangleq \inf\{k \geq 1 \colon S_k \geq a_0 / d_N\} \\
        T &\triangleq d_N \, \tau. \label{eq:tauT}
    \end{align}
    
    We employ the stop-at-time-zero procedure described in the proof sketch of \thmref{thm:mainPPC} with $\epsilon_N' = \frac{1}{\Ez{T}}$ and the information density threshold rule \eqref{eq:SPPC}--\eqref{eq:deltam} from the proof of \thmref{thm:nonasyPPC}, where the threshold $\gamma$ is set to $a_0$. Here, $T$ is as in \eqref{eq:Teq}.
    We set $M$ and $a_0$ so that
    \begin{align}
        M \Po{T < \infty}  &\leq M e^{-a_0} = \epsilon_N' = \frac{1}{\Ez{T}}, \label{eq:Ma0}
    \end{align}
    where the inequality follows from \eqref{eq:Ponebound}.
    Following steps identical to \eqref{eq:Nprime}, and noting that $\Pz{T = \infty} = 0$, we get
    \begin{align}
        N = (1- \epsilon) \Ez{T} + O(1), \label{eq:ETN}
    \end{align}
    and the average error probability of the code is bounded by $\epsilon$. 
    
    To evaluate $\Ez{T}$, we use \lemref{lem:SHT} with $W_i$ in place of $Z_i$. A straightforward calculation yields 
    \begin{align}
        \Ez{W_1^2} &= \Ez{W_1}^2 - \Var{\frac{1}{d_N}\sum_{i = 1}^{d_N} \imath(X_i; Y_i)} \\
        &= 
        C^2 - \frac{1}{d_N} \Var{\imath(X_1; Y_1)}. \label{eq:W0sq}
    \end{align}
    Next, we have that
    \begin{align}
        \Ez{S_n^-} = - n d_N \E{\frac{1}{n d_N} S_n 1\cB{ \frac{1}{n d_N} S_n \leq 0}},
    \end{align}
    where $S_n = \sum_{j = 1}^{n d_N} \imath(X_j; Y_j)$. Applying the saddlepoint approximation (e.g., \cite[eq.~(1.2)]{butler2007book}) to $\frac{1}{n d_N} S_n$, we get
    \begin{align}
        \Ez{S_n^-} = n d_N \int_{-\infty}^0 c(x) \sqrt{n d_N} e^{-  n d_N g(x) + \log  x} \mathrm{d} x, \label{eq:saddlep}
    \end{align}
    where $c(x)$ and $g(x)$ are bounded from below a positive constant for all $x \in (-\infty, 0]$. Applying the Laplace's integral \cite[eq.~(2.5)]{butler2007book} to \eqref{eq:saddlep}, we get
    \begin{align}
        \Ez{S_n^-} = e^{- n d_N c_n + o(n d_N)} \label{eq:ESbound}
    \end{align}
    for all $n \in \mathbb{Z}_+$, where each $c_n$ is a positive constant depending on $n$. Putting \eqref{eq:W0sq} and \eqref{eq:ESbound} into \eqref{eq:E0T} and \eqref{eq:tauT}, we get
    \begin{align}
        \Ez{T} = \frac{a_0}{C} + \frac{d_N}{2} + o(d_N). \label{eq:Ez1} 
    \end{align}
    From \eqref{eq:Ma0}--\eqref{eq:ETN}, we get
    \begin{align}
        \Ez{T} &= \frac{N}{1-\epsilon} + \bigo{1} \\
        \log  M &= a_0 - \log  N.  \label{eq:Ma}
    \end{align}
    Putting \eqref{eq:Ez1}--\eqref{eq:Ma} together completes the proof of \eqref{eq:dNach}.
    
    \subsection{Converse Proof}
    Recall the definition of an SHT $(\delta, \tau, \mc{N})$ from \appref{sec:SHTdef} that tests the hypotheses
    \begin{align}
        H_0 &\colon Z^{\infty} \sim P_0 \\
        H_1 &\colon Z^{\infty} \sim P_1,
    \end{align}
    where $P_0$ and $P_1$ are distributions on a common alphabet $\mc{Z}^{\infty}$. Here, $Z^{\infty} \triangleq (Z_1, Z_2, \dots)$ denotes a vector of infinite length whose joint distribution is either $P_0$ or $P_1$, which need not be product distributions in general.
    We define the minimum achievable type-II error probability, subject to a type-I error probability bound and a maximal expected decoding time constraint, with decision times restricted to the set $\mc{N}$ as 
    \begin{align}
        \beta_{(\epsilon, N, \mc{N})}(P_0, P_1) \triangleq \min_{\substack{(\delta, \tau, \mc{N}) \colon \Pz{\delta = 1} \leq {\epsilon}, \\ \max\{\mathbb{E}_0[\tau], \mathbb{E}_1[\tau]\} \leq N }}{\Po{\delta = 0}}, \label{eq:SHTL}
    \end{align}
    which is the SHT version of the $\beta_{\alpha}$-function defined for the fixed-length binary hypothesis test \cite{polyanskiy2010Channel}.
    
    The following theorem extends the meta-converse bound \cite[Th.~27]{polyanskiy2010Channel}, which is a fundamental theorem used to show converse results in fixed-length channel coding without feedback and many other applications.

    \begin{theorem} \label{thm:metaconv}
    Fix any set $\mc{N} \subseteq \mathbb{Z}_\geq$, a real number $N > 0$, and a DM-PPC $P_{Y|X}$. Then, it holds that
    \begin{align}
        &\log  M^*(N, |\mc{N}|, \epsilon, \mc{N}) \notag \\
        &\quad \leq \sup_{P_{X^{\infty}}} \inf_{Q_{Y^{\infty}}} - \log  \beta_{(\epsilon, N, \mc{N})}(P_{X^{\infty}} \times P_{Y|X}^{\infty}, P_{X^{\infty}} \times Q_{Y^{\infty}}).
    \end{align}
    \end{theorem}
    \begin{IEEEproof}
    The proof is similar to that in \cite{polyanskiy2010Channel}.
    Let $W$ denote a message equiprobably distributed on $[M]$, and let $\hat{W}$ be its reconstruction. Given any VLSF code with the set of available decoding times $\mc{N}$, average decoding time $N$, error probability $\epsilon$, and codebook size $M$, let $\hat{P}_{X^{\infty}}$ denote the input distribution induced by the code's codebook. The code operation creates a Markov chain $W \to X^{\infty} \to Y^{\infty} \to \hat{W}$. As full feedback breaks this Markov chain, stop feedback does not since the channel inputs are conditionally independent of the channel outputs given the message $W$. Fix an arbitrary output distribution $Q_{Y^{\infty}}$, and consider the SHT 
    \begin{align}
        H_0 &\colon (X^{\infty}, Y^{\infty}) \sim \hat{P}_{X^{\infty}} \times P_{Y|X}^{\infty} \\
        H_1 &\colon (X^{\infty}, Y^{\infty}) \sim \hat{P}_{X^{\infty}} \times Q_{Y^{\infty}}
    \end{align}
    with a test $\delta = 1\{\hat{W} \neq W\}$, where $(W, \hat{W})$ are generated by the (potentially random) encoder-decoder pair of the VLSF code. The type-I and type-II error probabilities of this code-induced SHT are
    \begin{align}
        \alpha &= \Pz{\delta = 1} \label{eq:WW} = \Prob{\hat{W} \neq W} \leq \epsilon\\
        \beta &= \Po{\delta = 0} =  \frac{1}{M}, \label{eq:1M}
    \end{align}
    where \eqref{eq:1M} follows since the sequence $Y^{\infty}$ is independent of $X^{\infty}$ under $H_1$. The stopping time of this SHT under $H_0$ or $H_1$ is bounded by $N$ by the definition of a VLSF code. Since the error probabilities in \eqref{eq:WW}--\eqref{eq:1M} cannot be better than that of the optimal SHT, it holds that
    \begin{align}
        &\log  M \notag \\
        &\leq - \log  \beta_{(\epsilon, N, \mc{N})}(\hat{P}_{X^{\infty}} \times P_{Y|X}^{\infty}, \hat{P}_{X^{\infty}}\times Q_{Y^{\infty}}) \\  &\leq \inf_{Q_{Y^{\infty}}} - \log  \beta_{(\epsilon, N, \mc{N})}(\hat{P}_{X^{\infty}} \times P_{Y|X}^{\infty}, \hat{P}_{X^{\infty}}\times Q_{Y^{\infty}}) \label{eq:infstep}\\
        &\leq \sup_{P_{X^{\infty}}} \inf_{Q_{Y^{\infty}}} - \log  \beta_{(\epsilon, N, \mc{N})}(P_{X^{\infty}} \times P_{Y|X}^{\infty}, P_{X^{\infty}} \times Q_{Y^{\infty}}), \label{eq:suptheorem}
    \end{align}
    where \eqref{eq:infstep} follows since the choice $Q_{Y^{\infty}}$ is arbitrary. 
    \end{IEEEproof}
    
    To prove \eqref{eq:dNconv}, we apply \thmref{thm:metaconv} and get
    \begin{align}
        \log  M \leq - \log  \beta_{(\epsilon, N, \mc{N})}(P_{Y|X}^{\infty}, P_Y^{\infty}), \label{eq:redconv}
    \end{align}
    where $P_Y$ is the capacity-achieving output distribution, and $\mc{N} = \{0, d_N, 2 d_N, \dots\}$. The reduction from \thmref{thm:metaconv} to \eqref{eq:redconv} follows since $\log  \frac{P_{Y|X}(Y|x)}{P_Y(Y)}$ has the same distribution for all $x \in \mc{X}$ for Cover--Thomas symmetric channels \cite[p.~190]{cover}. In the remainder of the proof, we derive an upper bound for the right-hand side of \eqref{eq:redconv}. 
    
    Consider any SHT $(\delta, \tau, \mc{N})$ with $\Ez{\tau} \leq N$ and $\Eo{\tau} \leq N$. Our definition in \eqref{eq:SHTL} is slightly different than the classical SHT definition from \cite{wald} since our definition allows one to make a decision at time 0. Notice that at time 0, any test has three choices: decide $H_0$, decide $H_1$, or decide to start taking samples. When the test decides to start taking samples, the remainder of the procedure becomes a classical SHT.  From this observation, any test satisfies
    \begin{align}
        \epsilon \geq \alpha &= \epsilon_0 + (1- \epsilon_0 - \epsilon_1) \alpha' \geq \epsilon_0 \label{eq:alp} \\
        \beta &= \epsilon_1 + (1-\epsilon_0 - \epsilon_1) \beta' \geq (1-\epsilon_0)  \beta', \label{eq:bet}
    \end{align}
    where at time 0, the test decides $H_i$ with probability $\epsilon_{1-i}$, and $\alpha'$ and $\beta'$ are the type-I and type-II error probabilities conditioned on the event that the test decides to take samples at time 0, which occurs with probability $1-\epsilon_0 - \epsilon_1$.
    
    Let $\tau'$ denote the average stopping time of the test with error probabilities $(\alpha', \beta')$. We have
    \begin{align}
        \Ez{\tau} &= (1-\epsilon_0 - \epsilon_1) \Ez{\tau'} \label{eq:tb3} \\
        &= (1-\epsilon_0) (\Ez{\tau'} + e^{-O(N)}) \leq N \\
        \Eo{\tau} &= (1-\epsilon_0 - \epsilon_1) \Eo{\tau'} \\
        &= (1-\epsilon_0) (\Eo{\tau'} + e^{-O(N)}) \leq N \label{eq:tb4}
    \end{align}
    since $\beta$ decays exponentially with $\Ez{\tau}$.
    
    The following argument is similar to that in \cite[Sec. V-C]{li2020sht}. Set an arbitrary $\nu > 0$ and the thresholds
    \begin{align}
        \tilde{a}_0 &= C \nB{\frac{N}{1-\epsilon_0} - \frac{d_N}{2} - o(d_N) + {\nu}} \label{eq:tildea0} \\
        \tilde{a}_1 &= D(P_Y \| P_{Y|X = x}) \nB{\frac{N}{1-\epsilon_0} - \frac{d_N}{2} - o(d_N) + {\nu}}, 
    \end{align}
    where $x \in \mc{X}$ is arbitrary, and let $(\tilde{\delta}, \tilde{\tau}, \mc{N})$ be the SPRT associated with the thresholds $(-\tilde{a}_1, \tilde{a}_0)$, and type-I and type-II error probabilities $\tilde{\alpha}$ and $\tilde{\beta}$.
    
    Applying \cite[eq.~(3.56)]{sequentialbook} to \eqref{eq:Ez1}, we get
    \begin{align}
        \Ez{\tilde{\tau}} &= \frac{\tilde{a}_0}{C} + \frac{d_N}{2} + o(d_N) \\
        \Eo{\tilde{\tau}} &= \frac{\tilde{a}_1}{D(P_Y \| P_{Y|X = x})} + \frac{d_N}{2} + o(d_N). \label{eq:E1tilde}
    \end{align}
    Combining \eqref{eq:tildea0}--\eqref{eq:E1tilde} gives
    \begin{align}
        \Ez{\tilde{\tau}} &\geq \frac{N}{1-\epsilon_0} + {\nu} \label{eq:tb1} \\
        \Eo{\tilde{\tau}} &\geq \frac{N}{1-\epsilon_0} + {\nu}. \label{eq:tb2}
    \end{align}
    Letting $\nu = \bigo{\frac{1}{N}}$, it follows from \eqref{eq:tb3}--\eqref{eq:tb4} and \eqref{eq:tb1}--\eqref{eq:tb2} that
    \begin{align}
        \Ez{\tilde{\tau}} &\geq \Ez{\tau'} \\
        \Eo{\tilde{\tau}} &\geq \Eo{\tau'}
    \end{align}
    for a large enough $N$.
    Using Wald and Wolfowitz's SPRT optimality result \cite{waldwolfowitz}, we get
    \begin{align}
        \alpha' &\geq \tilde{\alpha} \\
        \beta' &\geq \tilde{\beta}. \label{eq:betprim}
    \end{align}
    Now it only remains to lower bound $\tilde{\beta}$. Applying \cite[Th. 3.1.2, 3.1.3]{sequentialbook} and \eqref{eq:lambdaexp} gives
    \begin{align}
        \tilde{\beta} = \tilde{\zeta} e^{-\tilde{a}_0} (1 + o(1)), 
    \end{align}
    where
   \begin{align}
        \tilde{\zeta} 
        &= \frac{1}{d_N C} \left(\exp\cB{- \sum_{n = 1}^{\infty} \frac{1}{n} \Pz{S_n < 0} + \Po{S_n > 0}}  \right), \label{eq:kappabound}
    \end{align}
    and $S_n$ is as in \eqref{eq:Sndef}. Since $S_n$ is a sum of $n d_N \to \infty$ i.i.d. random variables, where the summands have a non-zero mean, the Chernoff bound implies that each of the probabilities in \eqref{eq:kappabound} decays exponentially with $d_N$. Thus,
    \begin{align}
         \tilde{\zeta}  = \frac{1}{d_N C} (1 + o(1)). \label{eq:zeta}
    \end{align}
    From \eqref{eq:tildea0} and \eqref{eq:zeta}, we get
    \begin{align}
        -\log  \tilde{\beta} = C \nB{\frac{N}{1-\epsilon_0} - \frac{d_N}{2} - o(d_N) + o(\log  d_N)} \\
       \leq C \nB{\frac{N}{1-\epsilon} - \frac{d_N}{2} - o(d_N) + o(\log  d_N)},
        \label{eq:lnbound}
    \end{align}
    where \eqref{eq:lnbound} follows from \eqref{eq:alp}. 
    Inequalities \eqref{eq:bet}, \eqref{eq:betprim}, and \eqref{eq:lnbound} imply \eqref{eq:dNconv}.

	\section{Proofs for the DM-MAC} \label{app:MACproofs}
	 In this section, we prove our main results for the DM-MAC, beginning with Theorem~\ref{thm:MACnonasy}, which is used to prove \thmref{thm:MAC}. 
	
	\subsection{Proof of \thmref{thm:MACnonasy}} \label{app:proofnonasymMAC}
	For each transmitter $k$, $k \in [K]$, we generate $M_k$ $n_L$-dimensional  codewords i.i.d. from $P_{X_k}^{n_L}$. Codewords for distinct transmitters  are drawn independently of each other. Denote the codeword for transmitter $k$ and message $m_k$ by $X_k^{n_L}(m_k)$ for $k \in [K]$ and $m_k \in [M_k]$. The proof extends the DM-PPC achievability bound in \thmref{thm:nonasyPPC} that is based on a sub-optimal SHT to the DM-MAC. Below, we explain the differences.
    
    Without loss of generality, assume that $m_{[K]} = \bs{1}$ is transmitted.	
	The hypothesis test in \eqref{eq:H0ach}--\eqref{eq:H1ach} is replaced by
        \begin{align}
        H_0 &\colon (X_{[K]}^{n_L}, Y_K^{n_L}) \sim \left(\prod_{k = 1}^K P_{X_k}^{n_L}\right) \times P_{Y_K|X_{[K]}}^{n_L} \label{eq:H0achMAC} \\
        H_1 &\colon (X_{[K]}^{n_L}, Y_K^{n_L}) \sim \left(\prod_{k = 1}^K P_{X_k}^{n_L} \right) \times P_{Y_K}^{n_L}, \label{eq:H1achMAC}
        \end{align}
    which is run for every message tuple $m_{[K]} \in \prod \limits_{k = 1}^K [M_k]$. The information density \eqref{eq:SPPC}, the stopping times \eqref{eq:taum}--\eqref{eq:tautildemm}, and the decision rule \eqref{eq:deltam} are extended to the DM-MAC as
    \begin{align}
        S_{m_{[K]}, n_{\ell}} &\triangleq \imath_K(X_{[K]}^{n_\ell}(m_{[K]}); Y_K^{n_\ell}) \\
        \tau_{m_{[K]}} &\triangleq \inf\{n_{\ell} \in \mc{N} \colon S_{m_{[K]}, n_{\ell}} \geq \gamma\} \label{eq:taumMAC} \\
        \tilde{\tau}_{m_{[K]}}  &\triangleq \min\{\tau_{m_{[K]}} , n_L\} \\
	    \delta_{m_{[K]}} &\triangleq \begin{cases}
	    0 &\text{if } S_{m_{[K]}, n_{\ell}} \geq \gamma \\
	    1 &\text{if } S_{m_{[K]}, n_{\ell}} < \gamma
 	    \end{cases} \label{eq:deltamMAC}
    \end{align}
    for every message tuple $m_{[K]}$ and decoding time $n_\ell$.
	For brevity, 
 let $(X_{[K]}^{n_\ell}, Y_K^{n_\ell}, \bar{X}_{[K]}^{n_\ell})$ be drawn i.i.d. according to the joint distribution
	\begin{align}
	    &P_{X_{[K]}, Y_K, \bar{X}_{[K]}}(x_{[K]}, y, \bar{x}_{[K]}) \notag \\
	    &= P_{Y_K|X_{[K]}}(y|x_{[K]}) \prod_{k = 1}^K P_{X_k}(x_k)  P_{X_k} (\bar{x}_k). \label{eq:dist}
	\end{align}
    
    \emph{Expected decoding time analysis:} Following steps identical to \eqref{eq:taustar1}--\eqref{eq:sumnprob}, we get \eqref{eq:boundNMAC}.
	
	\emph{Error probability analysis:} 
	The following error analysis extends the PPC bounds in \eqref{eq:epsilSHT} and \eqref{eq:alphaPPC}--\eqref{eq:gammabound} to the DM-MAC. 
	
	In the analysis below, for brevity, we write $m_{\mc{A}} \neq 1$ to denote that $m_i \neq 1$ for $i \in \mc{A}$. The error probability is bounded as
	\begin{align}
	    \epsilon &\leq \mathbb{P}\bigg[ \bigcup_{m_{[K]} \neq \mb{1}} \{\tau_{m_{[K]}} \leq \tau_{\mb{1}} < \infty\} \bigcup \{\tau_{\mb{1}} = \infty\}  \bigg]  \label{eq:W1W2eq}\\
	      &\leq   \Prob{\tau_{\mb{1}} = \infty } + \Prob{ \bigcup_{\substack{m_{[K]} \neq 1}} \{\tau_{m_{[K]}} < \infty \}  }  \label{eq:allwrong} \\
	      &\quad + \Prob{ \bigcup_{\substack{m_{[K]} \neq \mb{1} \colon \exists \, i \in [K] \\ m_i = 1}} \{\tau_{m_{[K]}} < \infty \}  } \label{eq:2ndwrong},
	    \end{align}
	    where \eqref{eq:allwrong}--\eqref{eq:2ndwrong} apply the union bound to separate the probabilities of the following error events:
	    \begin{enumerate}
	        \item the information density of the true message tuple does not satisfy the threshold test for any available decoding time;
	        \item the information density of a message tuple in which all messages are incorrect satisfies the threshold test for some decoding time;
	        \item the information density of a message tuple in which the messages from some transmitters are correct and the messages from the other transmitters are incorrect satisfies the threshold test for some decoding time. 
	    \end{enumerate}
	    
	   The terms in \eqref{eq:allwrong} are bounded using steps identical to \eqref{eq:alphaPPC}--\eqref{eq:gammabound} as
	    \begin{align}
	        \Prob{\tau_{\mb{1}} = \infty} &\leq \Prob{\imath_K(X_{[K]}^{n_L}; Y_K^{n_L}) < \gamma} \label{eq:bothwrongerror} \\
	      \Prob{ \bigcup_{\substack{m_{[K]} \neq 1}} \{\tau_{m_{[K]}} < \infty \} }&\leq \prod_{k = 1}^K (M_k - 1) \exp\{-\gamma\}. \label{eq:taubartruong}
	    \end{align}
	 
	    For the cases where at least one message is decoded correctly, we delay the application of the union bound. Let $\mc{A} \in \mc{P}([K])$ be the set of transmitters whose messages are decoded correctly. Define 
	    \begin{align}
	        \mc{M}^{(\mc{A})} &\triangleq \{m_{[K]} \in [M]^K \colon m_k = 1 \text{ for } k \in \mc{A}, \notag \\
	        &\quad \quad m_k \neq 1 \text{ for } k \in \mc{A}^{\mr{c}} \} \\ \tilde{\mc{M}}^{(\mc{A})} &\triangleq \{m_{\mc{A}} \in [M]^{|\mc{A}|} \colon m_k \neq 1 \text{ for } k \in \mc{A} \}.
	    \end{align}
	    
	    We first bound the probability term in \eqref{eq:2ndwrong} by applying the union bound according to which subset $\mc{A}$ of the transmitter set $[K]$ is decoded correctly, and get 
	    \begin{align}
	        &\Prob{ \bigcup_{\substack{m_{[K]} \neq \mb{1} \colon \exists \, i \in [K] \\ m_i = 1}} \{\tau_{m_{[K]}} < \infty \} } \notag \\
	        &\leq \sum_{\mc{A} \in \mc{P}([K])} \Prob{ \bigcup_{m_{[K]} \in \mc{M}^{(\mc{A})}} \{\tau_{m_{[K]}} < \infty \} } \\
	        &= \sum_{\mc{A} \in \mc{P}([K])} \Prob{ \bigcup_{\substack{m_{\mc{A}^c} \in \tilde{\mc{M}}^{(\mc{A}^c)} \\ n_{\ell} \in \mc{N}}} \left\{\imath_K(\bar{X}_{\mc{A}^{c}}^{n_\ell}(m_{\mc{A}^{c}}), X_{\mc{A}}^{n_\ell}; Y_K^{n_\ell}) \geq \gamma \right \}}, \label{eq:unionbar}
	    \end{align}
	    where $\bar{X}_{\mc{A}^{c}}^{n_\ell}(m_{\mc{A}^{c}})$ refers to the random sample from the codebooks of transmitter set $\mc{A}^{c}$, independent from the codewords $X_{{\mc{A}^{c}}}^{n_\ell}$ transmitted by the transmitters $\mc{A}^{c}$ and the received output~$Y^{n_\ell}$. 
	 
        We bound the right-hand side of \eqref{eq:unionbar} using the same method as in \cite[eq.~(65)--(66)]{yavas2020Random}. This step is crucial in enabling the single-threshold rule for the rate vectors approaching a point on the sum-rate boundary. 
        Set an arbitrary $\lambda^{(\mc{A})} > 0$. Define two events
        \begin{align}
            \mc{E}(\mc{A}) &\triangleq \bigcup_{n_\ell \in \mc{N}} \left\{ \imath_K(X_{\mc{A}}^{n_\ell}; Y_K^{n_\ell}) > N I_K(X_{\mc{A}}; Y_K) + N \lambda^{(\mc{A})} \right\} \\
            \mc{F}(\mc{A}) &\triangleq \bigcup_{\substack{m_{\mc{A}^{c}} \in \tilde{\mc{M}}^{(\mc{A}^c)} \\ n_\ell \in \mc{N}}} \left\{\imath_K(\bar{X}_{\mc{A}^{c}}^{n_\ell}(m_{{\mc{A}}^{c}}), X_{\mc{A}}^{n_\ell}; Y_K^{n_\ell}) \geq \gamma \right \}.
        \end{align}
        Define the threshold 
        \begin{align}
            \bar{\gamma}^{(\mc{A})} \triangleq \gamma  - N I_K(X_{\mc{A}}; Y_K) - N \lambda^{(\mc{A})}.
        \end{align}
        We have
        \begin{align}
            &\Prob{\mc{F}(\mc{A})} \notag \\
            &= \Prob{\mc{F}(\mc{A}) \cap \mc{E}(\mc{A})} + \Prob{\mc{F}(\mc{A}) \cap \mc{E}(\mc{A})^{c}} \label{eq:F}\\
            &\leq \Prob{\mc{E}(\mc{A})} \notag \\
            &+ \mathbb{P} \Biggm[ \bigcup_{\substack{\substack{m_{\mc{A}^{c}} \in \tilde{\mc{M}}^{(\mc{A}^c)}}\\ n_\ell \in \mc{N}}} \bigg\{\imath_K(\bar{X}_{\mc{A}^{\mr{c}}}^{n_\ell}(m_{\mc{A}^{\mr{c}}}); Y_K^{n_\ell} | X_{\mc{A}}^{n_\ell}) \geq \bar{\gamma}^{(\mc{A})} \bigg\} \Biggm] \label{eq:probdelay} \\
            &\leq  \sum_{n_\ell \in \mc{N}}  \Prob{\imath_K(X_{\mc{A}}^{n_\ell}; Y_K^{n_\ell}) > N I_K(X_{\mc{A}}; Y_K) + N \lambda^{(\mc{A})} }  \notag \\
            &+ \prod_{k \in \mc{A}^{\mr{c}}} (M_k-1)
            \Prob{\bigcup_{n_\ell \in \mc{N}} \left\{\imath_K(\bar{X}_{\mc{A}^{\mr{c}}}^{n_\ell}; Y_K^{n_\ell} | X_{\mc{A}}^{n_\ell}) \geq \bar{\gamma}^{(\mc{A})} \right\}} \label{eq:unionE} \\
            &\leq \sum_{n_\ell \in \mc{N}} \Prob{\imath_K(X_{\mc{A}}^{n_\ell}; Y_K^{n_\ell}) > N I_K(X_{\mc{A}}; Y_K) + N \lambda^{(\mc{A})} } \notag \\
            &\quad + \prod_{k \in \mc{A}^{\mr{c}}} (M_k - 1) \exp\{-\bar{\gamma}^{(\mc{A})}\}  \label{eq:fbound},
        \end{align}
        where inequality \eqref{eq:probdelay} uses the chain rule for mutual information, \eqref{eq:unionE} applies the union bound, and \eqref{eq:fbound} follows from \cite[eq.~(88)]{truong2018Journal}. 
        
        Applying the bound in \eqref{eq:fbound} to each of the probabilities in \eqref{eq:unionbar} and plugging \eqref{eq:bothwrongerror}, \eqref{eq:taubartruong}, and \eqref{eq:unionbar} into \eqref{eq:allwrong}--\eqref{eq:2ndwrong}, we complete the proof of \thmref{thm:MACnonasy}.
        
        \subsection{Proof of \thmref{thm:MAC}} \label{app:proofMACasy}
        We employ the stop-at-time-zero procedure in the proof sketch of \thmref{thm:mainPPC} with $\epsilon_N' = \frac{1}{\sqrt{N' \log  N'}}$. Therefore, we first show that there exists an $(N, L, M_{[K]}, 1/\sqrt{N \log  N})$-VLSF code satisfying
	\begin{align}
       \sum_{k = 1}^K \log  M_k &= N I_K - \sqrt{N \log _{(L)}(N) V_K } \notag \\
       &\quad + O\left(\sqrt{\frac{N V_K}{\log _{(L)} (N)}}\right). \label{eq:MACeq} 
     \end{align}
     
     We set the parameters
     \begin{align}
    \gamma  &= n_\ell I_K - \sqrt{n_\ell  \log _{(L - \ell + 1)}(n_\ell) V_K} \quad \forall \, \ell \in [L]\label{eq:setnellMAC} \\
            &= \sum_{k = 1}^K \log  M_k + \log  N \label{eq:gammaM1M2}\\
    \lambda^{(\mc{A})} &= \frac{N I_K(X_{\mc{A}^{c}}; Y_K|X_{\mc{A}}) - \sum_{k \in \mc{A}^{c}}\log  M_k}{2 N} \label{eq:lambda1_2}, \quad \mc{A} \in \mc{P}([K]).
    \end{align}
    Note that $\lambda^{(\mc{A})}$'s are bounded below by a positive constant for rate points lying on the sum-rate boundary.
    
    By  \thmref{thm:MACnonasy}, there exists a VLSF code with $L$ decoding times $n_1 < n_2 < \cdots < n_L$ such that the average decoding time is bounded as
	\begin{align}
	    N \leq n_1 + \sum_{\ell = 1}^{L-1} (n_{\ell+1} - n_\ell) \Prob{\imath_{K}(X_{[K]}^{n_\ell}; Y_K^{n_\ell}) < \gamma}.  \label{eq:NboundMAC}
	\end{align}
Following the analysis in \eqref{eq:Nn1}--\eqref{eq:nLN}, we conclude that 
\begin{align}
n_\ell = N(1 + o(1)) \label{eq:Nn1MAC}
\end{align}
for all $\ell \in [L]$.
Applying the Chernoff bound to the probability terms in \eqref{eq:1errorS}--\eqref{eq:1errorExp} using  \eqref{eq:setnellMAC} and \eqref{eq:Nn1MAC}, we get that the sum of the terms in \eqref{eq:1errorS}--\eqref{eq:1errorExp} is bounded by $\exp\{-N E\}$ for some $E > 0$.  

Applying \lemref{lem:moderate} to the probability in \eqref{eq:true} with \eqref{eq:setnellMAC} gives 
\begin{align}
    \Prob{\imath_{K}(X_{[K]}^{n_L}; Y_K^{n_L}) < \gamma} &\leq \frac{1}{\sqrt{2 \pi}} \frac{1}{\sqrt{n_{L}}} \frac{1}{\sqrt{\log  n_{L}}} \notag \\
    &\quad \cdot \nB{1 + \bigo{\frac{(\log  n_{L})^{(3/2)}}{\sqrt{n_L}}}}. \label{eq:Lbound}
\end{align}

Applying \thmref{thm:MACnonasy} with \eqref{eq:gammaM1M2}, \eqref{eq:Lbound}, and the exponential bound on the sum of the terms in \eqref{eq:1errorS}--\eqref{eq:1errorExp}, we bound the error probability as
\begin{align}
    &\Prob{\mathsf{g}_{\tau^*}(U, Y^{\tau^*}) \neq W_{[K]}} \notag \\
    &\leq \frac{1}{\sqrt{2 \pi}} \frac{1}{\sqrt{N}} \frac{1}{\sqrt{\log  N}} \cdot \nB{1 + \bigo{\frac{(\log  N)^{(3/2)}}{\sqrt{N}}}} \notag \\
    &\quad + \frac{1}{N} + \exp\{-N E \},
\end{align}
which is further bounded by $\frac{1}{\sqrt{N \log  N}}$ for $N$ large enough. Following steps identical to \eqref{eq:Nn1}--\eqref{eq:lnMfinal}, we prove the existence of a VLSF code that satisfies \eqref{eq:MACeq} for the DM-MAC with $L$ decoding times and error probability $\frac{1}{\sqrt{N \log  N}}$. 

Finally, invoking \eqref{eq:MACeq} with $L$ replaced by $L-1$ and the stop-at-time-zero procedure in the proof sketch of \thmref{thm:mainPPC} with $\epsilon_N' = \frac{1}{\sqrt{N' \log  N'}}$, we complete the proof of \thmref{thm:MAC}.

       \subsection{Proof of \eqref{eq:mainresultMACinf}} \label{app:proofMACinf}
       The proof of \eqref{eq:mainresultMACinf} follows steps similar to those in the proof of \cite[Th.~2]{polyanskiy2011feedback}. Below, we detail the differences between the proofs of \eqref{eq:mainresultMACinf}, \thmref{thm:MAC}, and \cite[Th.~2]{polyanskiy2011feedback}.
        \begin{enumerate}
            \item In \eqref{eq:mainresultMACinf}, we choose $cN + 1$ decoding times as $n_i = i-1$ for $i = 1, \dots, cN + 1$ 
            for a sufficiently large constant $c > 1$. This differs from  \thmref{thm:MAC} where $L$ does not grow with $N$ ($L = O(1)$) and the gaps between consecutive decoding times can differ. In \cite[Th.~2]{polyanskiy2011feedback}, any integer time is available for decoding, giving $L = n_{\max} = \infty$. 
            \item We here set the parameter $\gamma$ differently from how it was set in \eqref{eq:setnellMAC} and \eqref{eq:gammaM1M2}. The difference accounts for the error event that some of the messages are decoded incorrectly and some of the messages are decoded correctly. Specifically, we set
            \begin{align}
                \gamma &= N I_K - a \label{eq:gammai12}\\
            &= \sum_{k = 1}^K \log  M_k + \log  N + b, \label{eq:gamma2}
            \end{align}
            where $a$ is an upper bound on the information density $\imath_K(X_{[K]}; Y_K)$, and $b$ is a positive constant to be determined later. Since the number of decoding times $L$ grows linearly with $N$ and $c > 1$, applying the Chernoff bound gives
            \begin{align}
                \eqref{eq:true} +  \eqref{eq:1errorS} + \eqref{eq:1errorExp} \leq \exp\{-NE\} \label{eq:NEbound}
            \end{align}
            for some $E > 0$ and $N$ large enough. Hence, the error probability $\epsilon$ in \thmref{thm:MACnonasy} is bounded by $\frac{\exp\{-b\}}{N} + \exp\{-NE\}$, which can be further bounded by $\frac{1}{N}$ by appropriately choosing the constant~$b$.

            The term \eqref{eq:true} disappears in \cite[Th.~2]{polyanskiy2011feedback} because $n_L = \infty$; the terms \eqref{eq:1errorS} and  \eqref{eq:1errorExp} disappear in \cite[Th.~2]{polyanskiy2011feedback} because the channel is point-to-point. Therefore, $b$ is set to 0 in \cite[Th.~2]{polyanskiy2011feedback}.
            \item We bound the average decoding time $\E{\tau^*}$ as
            \begin{align}
            \E{\tau^*} &\leq \frac{\gamma + a}{I_K} = N\label{eq:aveMACinf}
        \end{align} 
        using Doob's optional stopping theorem
            as used in\cite[eq.~(106)-(107)]{polyanskiy2011feedback}
            while $\E{\tau^*}$ in the proof of \thmref{thm:MAC} is bounded by bounding the tail probability of information density. 
            
            The steps above show the achievability of an $(N, cN, M_{[K]}, 1/N)$ code with
        \begin{align}
            \sum_{k = 1}^K \log  M_k = N I_K - \log  N + O(1).
        \end{align}
        
        \item Lastly, as in \cite[Th.~2]{polyanskiy2011feedback}, we invoke the stop-at-time-zero procedure from the proof sketch of \thmref{thm:mainPPC} with $\epsilon_N' = \frac{1}{N'}$. 
            
        \end{enumerate}

\section{Proof of \thmref{thm:RAC}}
\label{app:proof:RAC}
In \thmref{thm:RAC}, we employ a multiple hypothesis test at some early time $n_0$ to estimate the number of active transmitters followed by a VLSF MAC coding. Since VLSF MAC codeword design employed in \thmref{thm:MAC} is unchanged (up to the coding dimension), the VLSF MAC code employs a single nested codebook, as described in the proof below.
If the test decides that the number of active transmitters is $\hat{k} = 0$, then the decoder declares no active transmitters at time $n_0$ and stops the transmission at that time. If the estimated number of active transmitters is $\hat{k} \neq 0$, then the decoder decides to decode at one of the available times $n_{\hat{k}, 1}$, \dots, $n_{\hat{k}, L}$ using the decoder for the MAC with $\hat{k}$ transmitters.

\subsection{Encoding and decoding}
\textbf{Encoding}: As in the DM-PPC and DM-MAC cases, the codewords are generated i.i.d. from the distribution $P_X^{n_{K, L}}$.

\textbf{Decoding}: The decoder combines a $(K+1)$-ary hypothesis test and the threshold test that is used for the DM-MAC. 

\emph{Multiple hypothesis test}: Given distributions $P_{Y_k}$, $k \in \{0, \dots, K\}$ where $\mc{Y}_K$ is the common alphabet, we test the hypotheses
\begin{align}
    H_k \colon Y^{n_0} \sim P_{Y_k}^{n_0}, \quad k \in \{0, \dots, K\}.
\end{align}
The error probability constraints of our test are
\begin{align}
    \Prob{\text{Decide } H_s \text{ where } s \neq 0 | H_0} &\leq \epsilon_0 \label{eq:ht0}\\
    \Prob{\text{Decide } H_s \text{ where } s \neq k | H_k} &\leq \exp\{-n_0 E + o(n_0)\} \label{eq:htk}
\end{align}
for $k \in [K]$, where $E > 0$ is a constant.

Due to the asymmetry in \eqref{eq:ht0}--\eqref{eq:htk}, we employ a composite hypothesis testing to decide whether $H_0$ is true; that is, the test declares $H_0$ if 
\begin{align}
    \log  \frac{P_{Y_0}^{n_0}(y^{n_0})}{P_{Y_s}^{n_0}(y^{n_0})} \geq \eta_s \label{eq:0test}
\end{align}
for all $s \in [K]$, where the threshold values $\eta_s$, $s \in [K]$, are chosen to satisfy \eqref{eq:ht0}. If the condition in \eqref{eq:0test} is not satisfied, then the test applies the maximum likelihood decoding rule, i.e., the output is $H_s$, where
\begin{align}
    s = \arg \max_{s \in [K]} P_{Y_s}^{n_0}(y^{n_0}). 
\end{align}

From the asymptotics of the error probability bound for composite hypothesis testing in \cite[Th. 5]{yavas2020Random}, the maximum type-II error of the composite hypothesis test is bounded as
\begin{align}
    &\max_{k \in [K]} \Prob{\text{Decide } H_0 | H_k} \notag \\
    &\quad \leq \exp\left\{-n_0 \min_{k \in [K]} D(P_{Y_0} \| P_{Y_k}) + O(\sqrt{n_0})\right\}. \label{eq:E0test}
\end{align}
If $P_{Y_0}$ is not absolutely continuous with respect to $P_{Y_k}$, \eqref{eq:E0test} remains valid when $D(P_{Y_0} \| P_{Y_k}) = \infty$ since in that case, we can achieve arbitrarily large type-II error probability exponent (see \cite[Lemmas~57-58]{polyanskiy2010Channel}.)

From \cite{leang97}, the maximum likelihood test yields
\begin{align}
    \max_{(k, s) \in [K]^2} \Prob{\text{Decide } H_s | H_k} \leq \exp\{-n E_C + o(n)\},
\end{align}
where
\begin{align}
    E_C = \min_{k, s} \min_{\lambda \in (0, 1)} \log  \sum_{y \in \mc{Y}_K} P_{Y_k}(y)^{1-\lambda} P_{Y_s}(y)^{\lambda} \label{eq:EC}
\end{align}
is the minimum Chernoff distance between the pairs $(P_{Y_k}, P_{Y_s})$, $k \neq s \in [K]$. Combining \eqref{eq:E0test} and \eqref{eq:EC}, the conditions in \eqref{eq:ht0}--\eqref{eq:htk} are satisfied with 
\begin{align}
    E = \min \left\{ \min_{k \in [K]} D(P_{Y_0} \| P_{Y_k}), E_C \right\} > 0.
\end{align}

If the hypothesis test declares the hypothesis $H_{\hat{k}}$, $\hat{k} \neq 0$, then the receiver decides to decode $\hat{k}$ messages at one of the decoding times in $\{n_{\hat{k}, 1}, \dots, n_{\hat{k}, L}\}$ using the VLSF code in \secref{app:MACproofs} for the $\hat{k}$-MAC, where $n_{\hat{k}, 1}$ is set to $n_0$.

\subsection{Error analysis}
We here bound the probability of error for the RAC code in Definition \ref{def:RAC}. 

\emph{No active transmitters}: For $k = 0$, the only error event is that the composite hypothesis test at time $n_0$ does not declare $H_0$ given that $H_0$ is true. By \eqref{eq:ht0}, the probability of this event is bounded by $\epsilon_0$.

\emph{$k \geq 1$ active transmitters}: When there is at least one active transmitter, there is an error if and only if at least one of the following events occurs:

\begin{itemize}
\item $\mc{E}_{\hat{k} \neq k}$: The number of active transmitters is estimated incorrectly at time $n_0$, i.e., $\hat{k} \neq k$, which results in decoding of $\hat{k}$ messages instead of $k$ messages.
\item $\mc{E}_{\textnormal{mes}}$: A list of messages $m_{[k]}' \neq m_{[k]}$ is decoded at one of the times in $\{n_{k, 1}, \dots, n_{k, L}\}$.
\end{itemize}
In the following discussion, we bound the probability of these events separately, and apply the union bound to combine them.

Since the encoders are identical, the event $\mc{E}_{\textnormal{rep}} = \{W_i = W_j \text{ for some } i \neq j \}$, where two or more transmitters send the same yields a dependence with transmitted codewords. Since transmitted codewords are usually independent, treating $\mc{E}_{\mr{rep}}$ as an error simplifies the analysis.
By the union bound, we have
\begin{align}
\Prob{\mc{E}_{\textnormal{rep}}} \leq \frac{k (k-1)}{2 M}. \label{eq:rep}
\end{align}

Applying the union bound, we bound the error probability as
\begin{IEEEeqnarray}{rCl}
\epsilon_k 
&\leq& \Prob{\mc{E}_{\textnormal{rep}}}  + \Prob{\mc{E}^{\mr{c}}_{\textnormal{rep}}} \Prob{\mc{E}_{\hat{k} \neq k} \cup \mc{E}_{\textnormal{mes}} \middle| \mc{E}^{\mr{c}}_{\textnormal{rep}}  } \notag \\
&\leq&  \Prob{\mc{E}_{\textnormal{rep}}}  +  \Prob{\mc{E}_{\hat{k} \neq k} \middle| \mc{E}^{\mr{c}}_{\textnormal{rep}} } + \Prob{\mc{E}_{\textnormal{mes}} \middle| \mc{E}^{\mr{c}}_{\textnormal{rep}} \cap \mc{E}^{\mr{c}}_{\hat{k} \neq k}}. \label{eq:probb} \IEEEeqnarraynumspace 
\end{IEEEeqnarray}
By \eqref{eq:htk}, the probability $ \Prob{\mc{E}_{\hat{k} \neq k} \middle| \mc{E}^{\mr{c}}_{\textnormal{rep}} } $ is bounded as
\begin{align}
     \Prob{\mc{E}_{\hat{k} \neq k} \middle| \mc{E}^{\mr{c}}_{\textnormal{rep}} } \leq \exp\{-n_0 E + o(n_0)\}. \label{eq:activebound}
\end{align}

Since the number of active transmitters $k$ is not available at the decoder at time 0, we here slightly modify the stop-at-time-zero procedure from the proof sketch of \thmref{thm:mainPPC}. We set the smallest decoding time $n_{j, 1}$ to $n_0 \neq 0$ for all $j \in [K]$. Given the estimate $\hat{k}$ of the number of active transmitters $k$, we employ the stop-at-time-zero procedure with the triple $(N', \epsilon, \epsilon_{N}')$ replaced by  $(N_{\hat{k}}', \epsilon_{\hat{k}}, \epsilon_{N_{\hat{k}}'})$.

Let $\mc{E}_{\textnormal{stop}}$ denote the event that the decoder chooses to stop at time $n_{k, 1} = n_0$. We bound $\Prob{\mc{E}_{\textnormal{mes}} \middle| \mc{E}^{\mr{c}}_{\textnormal{rep}} \cap \mc{E}^{\mr{c}}_{\hat{k} \neq k}}$ as
\begin{align}
    &\Prob{\mc{E}_{\textnormal{mes}} \middle| \mc{E}^{\mr{c}}_{\textnormal{rep}} \cap \mc{E}^{\mr{c}}_{\hat{k} \neq k}} 
    \leq \Prob{\mc{E}_{\textnormal{stop}} | \mc{E}^{\mr{c}}_{\textnormal{rep}} \cap \mc{E}^{\mr{c}}_{\hat{k} \neq k}} \notag \\
    &\quad +  \Prob{\mc{E}_{\textnormal{stop}}^{\mr{c}} | \mc{E}^{\mr{c}}_{\textnormal{rep}} \cap \mc{E}^{\mr{c}}_{\hat{k} \neq k}} \Prob{\mc{E}_{\textnormal{mes}} \middle| \mc{E}^{\mr{c}}_{\textnormal{rep}} \cap \mc{E}^{\mr{c}}_{\hat{k} \neq k} \cap \mc{E}^{\mr{c}}_{\textnormal{stop}}}. \label{eq:stoptrick}
\end{align}

By \thmref{thm:MACnonasy}, when the RAC decoder decodes decode a list of $k$ messages from $[M]$ at time $n_{k, \ell}$, we get
\begin{align}
    &\Prob{\mc{E}_{\textnormal{mes}} \middle| \mc{E}^{\mr{c}}_{\textnormal{rep}} \cap \mc{E}^{\mr{c}}_{\hat{k} \neq k} \cap \mc{E}^{\mr{c}}_{\textnormal{stop}}}  \\
    &\quad \leq \Prob{\imath_{k}(X_{[k]}^{n_{k, L}}; Y_k^{n_{k, L}}) < \gamma_k} \\
	    &\quad \quad + \binom{M-k}{k} \exp\{-\gamma_k\}  \\
	    &\quad \quad  + \sum_{\ell = 2}^L \sum_{\mc{A} \in \mc{P}([k])} \notag \\
	    &\quad \quad \quad  \Prob{\imath_{k}(X_{\mc{A}}^{n_{k,\ell}}; Y_k^{n_{k, \ell}}) > N_k' I_k(X_{\mc{A}}; Y_k) + N_k' \lambda^{(k, \mc{A})}} \\
	    &\quad \quad  + \sum_{ \mc{A} \in \mc{P}([k])} \binom{M-k}{|\mc{A}|} \notag \\
	    &\quad \quad \quad \exp\{-\gamma + N_k' I_k(X_{\mc{A}}; Y_k) + N_k \lambda^{(k, \mc{A})}\},
\end{align}
where $N_k'$ is the average decoding time given $\mc{E}_{\textnormal{stop}}^c$, and $ \gamma_k$ and $\lambda^{(k, \mc{A})}$ are constants chosen to satisfy the equations
\begin{align}
    \gamma_k  &= n_{k, \ell} I_k - \sqrt{n_{k, \ell}  \log _{(L - \ell + 1 )}(n_{k, \ell}) V_k} \label{eq:gammakvalue}\\
            &= k \log  M + \log  N_k' + O(1) \label{eq:gammaMRAC}
\end{align}
for all $\ell \in \{2, \dots, L\}$, and
\begin{align}
    \lambda^{(k, \mc{A})} &= \frac{N_k' I_k(X_{\mc{A}^{c}}; Y_k|X_{\mc{A}}) - |\mc{A}^{c}|\log  M}{2 N_k'} , \quad \mc{A} \in \mc{P}([k]).
\end{align}
The fact that each $\lambda^{(k, \mc{A})}$ is bounded below by a positive constant follows from \eqref{eq:gammaMRAC}, \cite[Lemma~1]{yavas2020Random}, and the symmetry assumptions on the RAC.

Following the analysis in \appref{app:proofMACasy}, we conclude that
	\begin{align}
       &k \log  M = N_k' I_k - \sqrt{N_k' \log _{(L-1)}(N_k') V_k } \notag \\
       &\quad + O\left(\sqrt{\frac{N_k' V_k}{\log _{(L-1)} (N_k')}}\right) \label{eq:maineq}\\
      &\Prob{\mc{E}_{\textnormal{mes}} \middle| \mc{E}^{\mr{c}}_{\textnormal{rep}} \cap \mc{E}^{\mr{c}}_{\hat{k} \neq k} \cap \mc{E}^{\mr{c}}_{\textnormal{stop}}} \leq \frac{1}{\sqrt{N_k' \log  N_k'}}. \label{eq:Estop}
     \end{align}
     Note that by \eqref{eq:rep} and \eqref{eq:maineq}, the bound on $\Prob{\mc{E}_{\textnormal{rep}}}$ decays exponentially with $N_k$. 
     A consequence of \eqref{eq:gammakvalue} and \eqref{eq:maineq} is that 
     \begin{align}
         N_k' = n_{k, \ell} (1 + o(1)) \label{eq:Nko}
     \end{align} 
     for all $\ell \geq 2$ and $k \in [K]$.

Note that from \eqref{eq:maineq}, the right-hand side of \eqref{eq:rep} is bounded by $\frac{1}{N_k'}$ for $N_k'$ large enough. We set the time $n_0$ so that the right-hand side of \eqref{eq:activebound} is bounded by $\frac{1}{4 \sqrt{N_k' \log  N_k'}}$ for all $k \in [K]$. This condition is satisfied if
\begin{align}
    n_0 \geq \frac{1}{2 E} \log  N_k' + o(\log  N_k').
\end{align}
The above arguments imply that
\begin{align}
    \Prob{\mc{E}_{\textnormal{rep}}}  +  \Prob{\mc{E}_{\hat{k} \neq k} \middle| \mc{E}^{\mr{c}}_{\textnormal{rep}} } \leq \frac{1}{2 \sqrt{N_k' \log  N_k'}} \label{eq:sumrep}
\end{align}
for $N_k'$ large enough.
As in the DM-MAC case, we set 
\begin{align}
    p \triangleq \Prob{\mc{E}_{\textnormal{stop}} | \mc{E}^{\mr{c}}_{\textnormal{rep}} \cap \mc{E}^{\mr{c}}_{\hat{k} \neq k}} = \frac{\epsilon_k' - \frac{1}{\sqrt{N_k' \log  N_k'}}}{1 - \frac{1}{\sqrt{N_k' \log  N_k'}}} \label{eq:pRAC}
\end{align}
where 
\begin{align}
    \epsilon_k' = \epsilon_k - \frac{1}{2 \sqrt{N_k' \log  N_k'}}. \label{eq:epskRAC}
\end{align}
Combining \eqref{eq:probb},  \eqref{eq:stoptrick}, \eqref{eq:Estop}, and \eqref{eq:sumrep}--\eqref{eq:pRAC}, the error probability of the RAC code is bounded by
\begin{align}
    &\Prob{\mc{E}_{\textnormal{rep}}}  +  \Prob{\mc{E}_{\hat{k} \neq k} \middle| \mc{E}^{\mr{c}}_{\textnormal{rep}} } + \Prob{\mc{E}_{\textnormal{stop}} | \mc{E}^{\mr{c}}_{\textnormal{rep}} \cap \mc{E}^{\mr{c}}_{\hat{k} \neq k}} \notag \\
    &+  \Prob{\mc{E}_{\textnormal{stop}}^{\mr{c}} | \mc{E}^{\mr{c}}_{\textnormal{rep}} \cap \mc{E}^{\mr{c}}_{\hat{k} \neq k}} \Prob{\mc{E}_{\textnormal{mes}} \middle| \mc{E}^{\mr{c}}_{\textnormal{rep}} \cap \mc{E}^{\mr{c}}_{\hat{k} \neq k} \cap \mc{E}^{\mr{c}}_{\textnormal{stop}}} \\
    &\quad \leq \frac{1}{2 \sqrt{N_k' \log  N_k'}} + p + (1-p)  \frac{1}{\sqrt{N_k' \log  N_k'}} \\
    &\quad = \epsilon_k.
\end{align}

The average decoding time of the code is bounded as
\begin{align}
    N_k &\leq \E{\tau_k^* | \mc{E}_{\hat{k} \neq k} \cup \mc{E}_{\textnormal{rep}}} \Prob{\mc{E}_{\hat{k} \neq k} \cup \mc{E}_{\textnormal{rep}} } \notag \\
    &\quad + \E{\tau_k^* | \mc{E}_{\hat{k} \neq k}^{\mr{c}} \cap \mc{E}_{\textnormal{rep}}^{\mr{c}}} \Prob{\mc{E}_{\hat{k} \neq k}^{\mr{c}} \cap \mc{E}_{\hat{k} \neq k}^{\mr{c}}} \\
    &\leq \frac{N_{K, L}}{2 \sqrt{N_k' \log  N_k'}} + n_0 p + N_k' (1-p). 
\end{align}
From \eqref{eq:Nko} and \eqref{eq:pRAC}--\eqref{eq:epskRAC}, we get
\begin{align}
	    N_k' = \frac{N_k}{1-\epsilon_k'} \nB{1 + \bigo{\frac{1}{\sqrt{N_k \log  N_k}}}}.  \label{eq:Nkprime}
\end{align}
Plugging \eqref{eq:Nkprime} into \eqref{eq:maineq} completes the proof. 

	\section{Proof of \thmref{thm:Gaussian}}
	\label{app:GaussianPPCproof}
	    The non-asymptotic achievability bound in \thmref{thm:nonasyPPC} applies to the Gaussian PPC with maximal power constraint $P$ \eqref{eq:maxpower} with the modification that the error probability \eqref{eq:boundeps} has an additional term for power constraint violations
	    \begin{align}
    \Prob{\bigcup_{\ell = 1}^L \left\{\norm{X^{n_\ell}}^2 > n_\ell P \right\}}. \label{eq:powerviolation}
\end{align}
        
        The proof follows similarly to the proof of \thmref{thm:mainPPC} as we employ the stop-at-time-zero procedure in the proof sketch of \thmref{thm:mainPPC}. We extend \lemref{lem:VLFmoderate} to the Gaussian PPC, showing
        \begin{align}
	    &\log  M^*\left(N, L, \frac{1}{\sqrt{N \log  N}}, P\right) \notag \\
	&\geq {N C(P)}
	- \sqrt{N \log _{(L)} (N) \, V(P)}  + \bigo{\sqrt{\frac{N}{\log _{(L)} (N)}}}. \label{eq:achvanishGaussian}
	\end{align}
	    The input distribution $P_{X^{n_L}}$ used in the proof of \eqref{eq:achvanishGaussian} differs from the one used in the proof of \lemref{lem:VLFmoderate}, causing changes in  the analysis on the probability $\Prob{\imath(X^{n_L}; Y^{n_L}) < \gamma}$ and the threshold $\gamma$ in \eqref{eq:setnk}. Below, we detail these differences.
	    
		\subsubsection{The input distribution $P_{X^{n_L}}$} We choose the distribution of the random codewords, $P_{X^{n_L}}$, in \thmref{thm:nonasyPPC} 
	    as follows. Set $n_0 = 0$. For each codeword, independently draw sub-codewords $X^{n_{j-1}+1:n_j}$, $j \in [L]$ from the uniform distribution on the $(n_j - n_{j-1})$-dimensional sphere of radius $\sqrt{(n_j - n_{j-1}) P}$. Let $P_{X^{n_L}}$ denote the distribution of the length-$n_L$ random codewords described above. 
    Since codewords chosen under $P_{X^{n_L}}$ never violate the power constraint \eqref{eq:maxpower}, the power violation probability in \eqref{eq:powerviolation} is 0.
Furthermore, the power constraint is satisfied with equality at each of the dimensions $n_1, \dots, n_L$;
	    our analysis in \cite{yavas2021Gaussian} shows that for any finite $L$, and sufficiently large increments $n_\ell - n_{\ell-1}$ for all $\ell \in [L]$, using this restricted subset instead of the entire $n_L$-dimensional power sphere results in no change in the asymptotic expansion \eqref{eq:K1} for the fixed-length no-feedback codes up to the third-order term.

\subsubsection{Bounding the probability of the information density random variable}
We begin by bounding the probability 
\begin{align}
    \Prob{\imath(X^{n_{\ell}}; Y^{n_{\ell}}) < \gamma}, \quad \ell \in [L], \label{eq:probimath}
\end{align}
that appears in \thmref{thm:nonasyPPC}
under the input distribution described above. Under this choice of $P_{X^{n_L}}$, the random variable $\imath(X^{n_{\ell}}; Y^{n_{\ell}})$ is not a sum of $n_{\ell}$ i.i.d. random variables. We wish to apply the moderate deviations result in \lemref{lem:moderate}. To do this, we first
introduce the following lemma from \cite{molavianjazi2015second}, which uniformly bounds the Radon-Nikodym derivative of the channel output distribution in response to the uniform distribution on a sphere as compared to the channel output distribution in response to i.i.d. Gaussian inputs.
	\begin{lemma}[MolavianJazi and Laneman {\cite[Prop.~2]{molavianjazi2015second}}] \label{lem:molavianJazi}
		Let $X^n$ be distributed uniformly over the $n$-dimensional sphere of radius $\sqrt{nP}$. Let $\tilde{X}^n \sim \mathcal{N}(\mathbf{0}, P \mathsf{I}_n)$. Let $P_{Y^n}$ and $P_{\tilde{Y}^n}$ denote the channel output distributions in response to $P_{X^n}$ and $P_{\tilde{X}^n}$, respectively, where $P_{Y^n|X^n}$ is the point-to-point Gaussian channel \eqref{eq:pointchannel}. Then there exists an $n_0 \in \mathbb{N}$ such that for all $n \geq n_0$ and $y^n \in \mathbb{R}^{n}$, it holds that
		\begin{align}
		\frac{\mathrm{d} P_{Y^n}(y^n)}{\mathrm{d} P_{\tilde{Y}^n}(y^n)} &\leq J(P) \triangleq {27}{\sqrt{\frac{\pi}{8}}}\frac{1 + P}{\sqrt{1 + 2 P}} \label{eq:molavianjazi}.
		\end{align}
	\end{lemma}

Let $P_{\tilde{Y}}^{n_{\ell}}$ be $\mc{N}(\mathbf{0}, (1 + P)\mathsf{I}_{n_{\ell}})$. By \lemref{lem:molavianJazi}, we bound $\eqref{eq:probimath}$ as
\begin{IEEEeqnarray}{rCl}
    \IEEEeqnarraymulticol{3}{l}{\Prob{\imath(X^{n_{\ell}}; Y^{n_{\ell}}) < \gamma}} \notag\\
    &=& \Prob{\log  \frac{\mathrm{d} P_{Y^{n_{\ell}}|X^{n_{\ell}}}(Y^{n_{\ell}}|X^{n_{\ell}})}{\mathrm{d} P_{\tilde{Y}^{n_{\ell}}}(Y^{n_{\ell}})} < \gamma + \log  \frac{\mathrm{d} P_{Y^{n_{\ell}}}(Y^{n_{\ell}})}{\mathrm{d} P_{\tilde{Y}^{n_{\ell}}}(Y^{n_{\ell}})}} \IEEEeqnarraynumspace\\
    &\leq& \Prob{\log  \frac{\mathrm{d} P_{Y^{n_{\ell}}|X^{n_{\ell}}}(Y^{n_{\ell}}|X^{n_{\ell}})}{\mathrm{d} P_{\tilde{Y}^{n_{\ell}}}(Y^{n_{\ell}})} < \gamma + \ell \log  J(P)}, \label{eq:molused}
\end{IEEEeqnarray}
where $J(P)$ is the constant given in \eqref{eq:molavianjazi}, and \eqref{eq:molused} follows from the fact that $P_{Y^{n_{\ell}}}$ is the product of $\ell$ output distributions of dimensions $n_j - n_{j-1}, j \in [\ell]$, each induced by a uniform distribution over a sphere of the corresponding radius. As argued in \cite{polyanskiy2010Channel, tan2015Third, molavianjazi2015second, yavas2021Gaussian}, by spherical symmetry, the distribution of the random variable 
\begin{align}
    \log  \frac{\mathrm{d} P_{Y^{n_{\ell}}|X^{n_{\ell}}}(Y^{n_{\ell}}|X^{n_{\ell}})}{\mathrm{d} P_{\tilde{Y}^{n_{\ell}}}(Y^{n_{\ell}})} \label{eq:lnP}
\end{align}
depends on $X^{n_{\ell}}$ only through its norm $\norm{X^{n_{\ell}}}$. Since $\norm{X^{n_{\ell}}}^2 = n_{\ell} P$ with probability 1, any choice of $x^{n_{\ell}}$ such that $\norm{x^{n_i}}^2 = n_i P$ for $i \in [\ell]$ gives
\begin{align}
    &\Prob{\log  \frac{\mathrm{d} P_{Y^{n_{\ell}}|X^{n_{\ell}}}(Y^{n_{\ell}}|X^{n_{\ell}})}{\mathrm{d} P_{\tilde{Y}^{n_{\ell}}}(Y^{n_{\ell}})} < \gamma + \ell \log  J(P)} = \notag \\
    &
    \Prob{\log  \frac{\mathrm{d} P_{Y^{n_{\ell}}|X^{n_{\ell}}}(Y^{n_{\ell}}|X^{n_{\ell}})}{\mathrm{d} P_{\tilde{Y}^{n_{\ell}}}(Y^{n_{\ell}})} < \gamma + \ell \log  J(P) \middle| X^{n_{\ell}} = x^{n_{\ell}}}. \IEEEeqnarraynumspace \label{eq:probcondx}
\end{align}
We set $x^{n_{\ell}} = (\sqrt{P}, \sqrt{P}, \dots, \sqrt{P}) = \sqrt{P} \mathbf{1}$ to obtain an i.i.d. sum in \eqref{eq:probcondx}. Given $X^{n_{\ell}} =  \sqrt{P} \mathbf{1}$, the distribution of \eqref{eq:lnP} is the same as the distribution of the sum 
\begin{align}
    \sum_{i = 1}^{n_{\ell}} A_i \label{eq:Ai}
\end{align}
of $n_{\ell}$ i.i.d. random variables
\begin{align}
    A_i = C(P) + \frac{P}{2 (1 + P)}\nB{1 - Z_i^2 + \frac{2}{\sqrt{P}} Z_i}, \quad i \in [n_{\ell}], \label{eq:Ainotsum}
\end{align}
where $Z_1, \dots, Z_{n_{\ell}}$ are drawn independently from $\mc{N}(0, 1)$ (see e.g., \cite[eq.~(205)]{polyanskiy2010Channel}). The mean and variance of $A_1$ are 
\begin{align}
    \E{A_1} &= C(P) \label{eq:EA1}\\
    \Var{A_1} &= V(P).
\end{align}
From \eqref{eq:molused}--\eqref{eq:Ai}, we get
\begin{align}
    \Prob{\imath(X^{n_{\ell}}; Y^{n_{\ell}}) < \gamma} \leq  \Prob{\sum_{i = 1}^{n_{\ell}} A_i < \gamma + \ell \log  J(P)}. \label{eq:gammaJ}
\end{align}
To verify that \lemref{lem:moderate} is applicable to the right-hand side of \eqref{eq:gammaJ}, it only remains to show that $\E{(A_1 - C(P))^3}$ is finite, and $A_1 - C(P)$ satisfies Cram\'er's condition, that is, there exists some $t_0 > 0$ such that $\E{\exp\{t(A_1 - C(P))\}} < \infty$ for all $|t| < t_0$. From \eqref{eq:Ainotsum}, $(A_1 - C(P))^3$ has the same distribution as a 6-degree polynomial of the Gaussian random variable $Z \sim \mc{N}(0, 1)$. This polynomial has a finite mean since all moments of $Z$ are finite. Let $c \triangleq \frac{P}{2 (1 + P)}$, $f \triangleq \frac{2}{\sqrt{P}}$, and $t' \triangleq tc$. To show that Cram\'er's condition holds, we compute
\begin{align}
    &\E{\exp\{t(A_1 - C(P))\}} \notag \\
    &=\E{\exp\{t'(1 - Z^2 + fZ)\}} \\
    &= \int_{-\infty}^{\infty} \frac{1}{\sqrt{2 \pi}} \exp\left\{-\frac{x^2}{2} + t'(1 - x^2 + fx)\right\} \mathrm{d}x \\
    &= \frac{1}{\sqrt{1 + 2t'}} \exp\left\{t' + \frac{t' f}{2(1 + 2t')}\right\}.
\end{align}
Thus, $\E{\exp\{t(A_1 - C(P))\}} < \infty$
for $t' > -\frac{1}{2}$, and $t_0 = \frac{1}{2c} > 0$ satisfies Cram\'er's condition.

\subsubsection{The threshold $\gamma$}
We set $\gamma, n_1, \dots, n_L$ so that the equalities
\begin{align}
    \gamma  = n_{\ell} C(P) - \sqrt{n_{\ell}  \log _{(L -\ell + 1)}(n_{\ell}) V(P)} - \ell \log  J(P) \label{eq:setGaussian}
\end{align}
hold for all $\ell \in [L]$.

The rest of the proof follows identically to \eqref{eq:PetrovQ}--\eqref{eq:lnMfinal} with $C$ and $V$ replaced by $C(P)$ and $V(P)$, respectively, giving 
\begin{align}
    \log  M &\geq N C(P) - \sqrt{N \log _{(L)}(N) V(P) } \notag \\
    &- \frac{1}{\sqrt{2 \pi}} \sqrt{\frac{N V(P)}{\log _{(L)} (N)}}(1 + o(1)) -\log  N - L \log  J(P), \label{eq:lnMfinalGaussian}
\end{align}
which completes the proof.

\end{appendices}
	
\bibliographystyle{IEEEtran}
\bibliography{mac} 

\begin{IEEEbiographynophoto}{Recep Can Yavas}
(S'18--M'22) received the B.S. degree (Hons.) in electrical engineering from Bilkent University, Ankara, Turkey, in 2016. He received the M.S. and Ph.D. degrees in electrical engineering from the California Institute of Technology (Caltech) in 2017 and 2023, respectively. He is currently a research fellow at CNRS at CREATE, Singapore. His research interests include information theory, probability theory, and multi-armed bandits.
\end{IEEEbiographynophoto}

\begin{IEEEbiographynophoto}{Victoria Kostina}
    (S'12--M'14--SM'22)
    is a professor of electrical engineering and of computing and mathematical sciences at Caltech. She received the bachelor's degree from Moscow Institute of Physics and Technology (MIPT) in 2004, the master's degree from University of Ottawa in 2006, and the Ph.D. degree from Princeton University in 2013.  During her studies at MIPT, she was affiliated with the Institute for Information Transmission Problems of the Russian Academy of Sciences. 
 
 Her research interests lie in information theory, coding, communications, learning, and control.
She has served as an Associate Editor for IEEE Transactions of Information Theory, and as a Guest Editor for the IEEE Journal on Selected Areas in Information Theory.  She received the Natural Sciences and Engineering Research Council of Canada postgraduate scholarship during 2009--2012, Princeton Electrical Engineering Best Dissertation Award in 2013, Simons-Berkeley research fellowship in 2015 and the NSF CAREER award in 2017.  
 \end{IEEEbiographynophoto}


\begin{IEEEbiographynophoto}{Michelle Effros}
(S'93--M'95--SM'03--F'09) is the George Van Osdol
Professor of Electrical Engineering and Vice Provost at the California
Institute of Technology.  She was a co-founder of Code On Technologies, a
technology licensing firm, which was sold in 2016.  Dr. Effros is a fellow
of the IEEE and has received a number of awards including Stanford’s
Frederick Emmons Terman Engineering Scholastic Award (for excellence in
engineering), the Hughes Masters Full-Study Fellowship, the National
Science Foundation Graduate Fellowship, the AT\&T Ph.D. Scholarship, the
NSF CAREER Award, the Charles Lee Powell Foundation Award, the Richard
Feynman-Hughes Fellowship, an Okawa Research Grant, and the Communications
Society and Information Theory Society Joint Paper Award.  She was cited
by Technology Review as one of the world’s top 100 young innovators in
2002, became a fellow of the IEEE in 2009, and is a member of Tau Beta Pi,
Phi Beta Kappa, and Sigma Xi.  She received the B.S.~(with distinction),
M.S., and Ph.D.~degrees in electrical engineering from Stanford
University.  Her research interests include information theory (with a
focus on source, channel, and network coding for multi-node networks)
and theoretical neuroscience (with a focus on neurostability and memory).

Dr. Effros served as the Editor of the IEEE Information Theory Society
Newsletter from 1995 to 1998 and as a Member of the Board of Governors of
the IEEE Information Theory Society from 1998 to 2003 and from 2008 to
2017. She served as President of the IEEE Information Theory Society in
2015 and as Executive Director for the film ``The Bit Player,'' a movie
about Claude Shannon, which came out in 2018.  She was a member of the
Advisory Committee and the Committee of Visitors for the Computer and
Information Science and Engineering (CISE) Directorate at the National
Science Foundation from 2009 to 2012 and in 2014, respectively. She served
on the IEEE Signal Processing Society Image and Multi-Dimensional Signal
Processing (IMDSP) Technical Committee from 2001 to 2007 and on ISAT from
2006 to 2009. She served as Associate Editor for the joint special issue
on Networking and Information Theory in the IEEE Transactions on
Information Theory and the IEEE/ACM Transactions on Networking, as
Associate Editor for the special issue honoring the scientific legacy of
Ralf Koetter in the IEEE Transactions on Information Theory and, from 2004
to 2007 served as Associate Editor for Source Coding for the IEEE
Transactions on Information Theory.  She has served on numerous technical
program committees and review boards, including serving as general
co-chair for the 2009 Network Coding Workshop and technical program
committee co-chair for the 2012 IEEE International Symposium on
Information Theory and the 2023 IEEE Information Theory Workshop.
\end{IEEEbiographynophoto}

	\end{document}